\documentclass[12pt]{article}
\usepackage{amsmath,amsthm}
\usepackage{eufrak}
\newcommand{\ar}{\renewcommand{\arraystretch}{1}} % 1.0 % 0.6
\DeclareMathAlphabet{\bb}{U}{msb}{m}{n}
\gdef\C{\bb C}
\gdef\dZ{\bb Z}

\gdef\dS{\bb S}
\gdef\R{\bb R}

\gdef\BH{\bb H}

\DeclareMathOperator{\spin}{{\bf Spin}}

\DeclareMathOperator{\fD}{\mathfrak{D}}

\DeclareMathOperator{\Sym}{Sym}

\DeclareMathOperator{\sign}{sign}
\DeclareMathOperator{\Tr}{Tr}

\newcommand{\s}{\!}

\newcommand{\re}{\mbox{\rm Re}\,}
\newcommand{\im}{\mbox{\rm Im}\,}

\newcommand{\cP}{{\cal P}}
\newcommand{\cL}{\mathcal{L}}
\newcommand{\cM}{{\cal M}}
\newcommand{\cH}{{\cal H}}

\newcommand{\sA}{{\sf A}}
\newcommand{\sB}{{\sf B}}
\newcommand{\sI}{{\sf I}}

\newcommand{\sX}{{\sf X}}
\newcommand{\sY}{{\sf Y}}

\newcommand{\sa}{{\sf a}}
\newcommand{\sfb}{{\sf b}}
\newcommand{\sfc}{{\sf c}}
\newcommand{\sz}{{\sf z}}
\newcommand{\sL}{\Lambda}

\newcommand{\bk}{{\bf k}}
\newcommand{\bp}{{\bf p}}
\newcommand{\bx}{{\bf x}}
\newcommand{\by}{{\bf y}}
\newcommand{\bz}{{\bf z}}
\newcommand{\bB}{{\bf B}}

\newcommand{\bE}{{\bf E}}

\newcommand{\fM}{\mathfrak{M}}
\newcommand{\fG}{\mathfrak{G}}

\newcommand{\fH}{\mathfrak{H}}

\newcommand{\fP}{\mathfrak{P}}
\newcommand{\fL}{\mathfrak{L}}
\newcommand{\fT}{\mathfrak{M}}
\newcommand{\fS}{\mathfrak{S}}

\newcommand{\fg}{\mathfrak{g}}
\newcommand{\fz}{\mathfrak{z}}

\newcommand{\balpha}{\boldsymbol{\alpha}}
\newcommand{\cl}{C\kern -0.2em \ell}

\newcommand{\e}{\mbox{\bf e}}

\newcommand{\hypergeom}[5]{\mbox{$
_#1 F_#2\left.
\!\!
\left(
\!\!
\begin{array}{c}
\multicolumn{1}{c}{\begin{array}{c}
#3
\end{array}}\\[1mm]
\multicolumn{1}{c}{\begin{array}{c}
#4
\end{array}}\end{array}
\!\!
\right|\displaystyle{#5}\right)
$}
}
\newcommand{\ld}{\left[}
\newcommand{\rd}{\right]}
\newcommand{\lf}{\left\{}
\newcommand{\rf}{\right\}}
\newcommand{\tg}{\tan}
\newcommand{\ch}{\cosh}
\newcommand{\sh}{\sinh}
\newcommand{\tnh}{\tanh}
\newcommand{\cth}{\coth}
\newcommand{\ctg}{\cot}
\newcommand{\coupone}{
{}_{\raisebox{-0.45mm}{$\scriptscriptstyle\lfloor$}}\!
\underline{\boldsymbol{\psi}(\boldsymbol{\alpha})
\overline{\boldsymbol{\psi}}}\!{}_{\raisebox{-0.44mm}
{$\scriptscriptstyle\rfloor$}}(\boldsymbol{\alpha}^\prime)
}
\newcommand{\couptwo}{
{}_{\raisebox{-0.45mm}{$\scriptscriptstyle\lfloor$}}\!
\underline{\overline{\boldsymbol{\psi}}(\boldsymbol{\alpha})
\boldsymbol{\psi}}\!{}_{\raisebox{-0.44mm}
{$\scriptscriptstyle\rfloor$}}(\boldsymbol{\alpha}^\prime)
}
\newcommand{\coupthree}{
{}_{\raisebox{-0.45mm}{$\scriptscriptstyle\lfloor$}}\!
\underline{\boldsymbol{\psi}(\boldsymbol{\alpha})
\boldsymbol{\psi}}\!{}_{\raisebox{-0.44mm}
{$\scriptscriptstyle\rfloor$}}(\boldsymbol{\alpha}^\prime)
}
\newcommand{\coupfour}{
{}_{\raisebox{-0.45mm}{$\scriptscriptstyle\lfloor$}}\!
\underline{\overline{\boldsymbol{\psi}}(\boldsymbol{\alpha})
\overline{\boldsymbol{\psi}}}\!{}_{\raisebox{-0.44mm}
{$\scriptscriptstyle\rfloor$}}(\boldsymbol{\alpha}^\prime)
}
\newcommand{\coupfife}{
{}^{\raisebox{0.88mm}{$\scriptscriptstyle\lceil$}}\!
\overline{\boldsymbol{\psi}(\boldsymbol{\alpha})
\overline{\boldsymbol{\psi}}}\!
{}^{\raisebox{0.88mm}{$\scriptscriptstyle\rceil$}}
(\boldsymbol{\alpha}^\prime)
}
\newcommand{\coupsix}{
{}^{\raisebox{0.88mm}{$\scriptscriptstyle\lceil$}}\!
\overline{\boldsymbol{\psi}(\boldsymbol{\alpha}^\prime)
\overline{\boldsymbol{\psi}}}\!
{}^{\raisebox{0.88mm}{$\scriptscriptstyle\rceil$}}
(\boldsymbol{\alpha})
}
\newcommand{\coupseven}{
{}^{\raisebox{0.45mm}{$\scriptscriptstyle\lceil$}}\!
\overline{\boldsymbol{\psi}(\boldsymbol{\alpha})
\boldsymbol{\psi}}\!
{}^{\raisebox{0.45mm}{$\scriptscriptstyle\rceil$}}
(\boldsymbol{\alpha}^\prime)
}
\newcommand{\coupeight}{
{}^{\raisebox{0.88mm}{$\scriptscriptstyle\lceil$}}\!
\overline{\overline{\boldsymbol{\psi}}(\boldsymbol{\alpha})
\overline{\boldsymbol{\psi}}}\!
{}^{\raisebox{0.88mm}{$\scriptscriptstyle\rceil$}}
(\boldsymbol{\alpha}^\prime)
}
\newcommand{\coupnine}{
{}_{\raisebox{-0.45mm}{$\scriptscriptstyle\lfloor$}}\!
\underline{\overline{\boldsymbol{\psi}}(\boldsymbol{\alpha}^\prime)
\boldsymbol{\psi}}\!{}_{\raisebox{-0.44mm}
{$\scriptscriptstyle\rfloor$}}(\boldsymbol{\alpha})
}
\newcommand{\coupten}{
{}^{\raisebox{0.88mm}{$\scriptscriptstyle\lceil$}}\!
\overline{\overline{\boldsymbol{\psi}}(\boldsymbol{\alpha}^\prime)
\boldsymbol{\psi}}\!
{}^{\raisebox{0.88mm}{$\scriptscriptstyle\rceil$}}
(\boldsymbol{\alpha})
}
\newcommand{\coupeleven}{
{}_{\raisebox{-0.45mm}{$\scriptscriptstyle\lfloor$}}\!
\underline{\boldsymbol{\phi}(\boldsymbol{\alpha})
\boldsymbol{\phi}}\!{}_{\raisebox{-0.44mm}
{$\scriptscriptstyle\rfloor$}}(\boldsymbol{\alpha}^\prime)
}
\newcommand{\couptwelve}{
{}^{\raisebox{0.45mm}{$\scriptscriptstyle\lceil$}}\!
\overline{\boldsymbol{\phi}(\boldsymbol{\alpha})
\boldsymbol{\phi}}\!
{}^{\raisebox{0.45mm}{$\scriptscriptstyle\rceil$}}
(\boldsymbol{\alpha}^\prime)
}
\newcommand{\coupthirteen}{
{}_{\raisebox{-0.45mm}{$\scriptscriptstyle\lfloor$}}\!
\underline{\boldsymbol{\phi}(\boldsymbol{\alpha}^\prime)
\boldsymbol{\phi}}\!{}_{\raisebox{-0.44mm}
{$\scriptscriptstyle\rfloor$}}(\boldsymbol{\alpha})
}

\begin{document}
\title{Towards the Quantum Electrodynamics on the Poincar\'{e} Group}
\author{V. V. Varlamov\\
{\small\it Department of Mathematics, Siberia State
University of Industry,}\\
{\small\it Kirova 42, Novokuznetsk 654007, Russia}}
\date{}
\maketitle
\begin{abstract}
A general scheme of construction and analysis of physical fields on the
various homogeneous spaces of the Poincar\'{e} group is presented.
Different parametrizations of the field functions and harmonic analysis
on the homogeneous spaces are studied. It is shown that a direct product
of Minkowski spacetime and two-dimensional complex sphere is the most
suitable homogeneous space for the subsequent physical applications.
The Lagrangian formalism and field equations on the Poincar\'{e} group
are considered. A boundary value problem for the relativistically
invariant system is defined. General solutions of this problem are
expressed via an expansion in hyperspherical harmonics on the complex
two-sphere. A physical sense of the boundary conditions is discussed.
The boundary value problems of the same type are studied for the
Dirac and Maxwell fields. In turn, general solutions of these problems
are expressed via convergent Fourier type series. Field operators,
quantizations, causal commutators and vacuum expectation values of 
time ordered products of the field operators are defined for the Dirac
and Maxwell fields, respectively. Interacting fields and inclusion of
discrete symmetries into the framework of quantum electrodynamics on
the Poincar\'{e} group are discussed.
\end{abstract}

\section{Introduction}
An unification of spacetime and internal symmetries of elementary particles
is a long standing problem in physics. It is well known that the standard approach
in this area is a search of some unification group which includes the
Poincar\'{e} group $\cP$ and the group (or, groups) of internal symmetries
($SU(2)$, $SU(3)$ and so on) as a subgroup \cite{BR77}. However, in this approach
a physical sense of the unification group is unclear. The most natural way
for solving this problem
was proposed by Finkelstein \cite{Fin55}, he showed that elementary
particles models with internal degrees of freedom can be described
on manifolds larger then Minkowski spacetime (homogeneous spaces of the
Poincar\'{e} group). The quantum field theories on the Poincar\'{e} group
were discussed in papers 
\cite{Lur64,Kih70,BF74,Aro76,KLS95,Tol96,LSS96,Dre97,GS01,GL01}.

A consideration of the field models on the homogeneous spaces leads
naturally to a generalization of the concept of wave function
(fields on the Poincar\'{e} group).
The general form of these fields closely
relates with the structure of the Lorentz and Poincar\'{e} group
representations 
\cite{GMS,Nai58,BBTD88,GS01} 
and admits the following
factorization $f(x,\bz)=\phi^n(\bz)\psi_n(x)$, where $x\in T_4$ and
$\phi^n(\bz)$ form a basis in the representation space of the Lorentz
group. Since the functions $f(x,\bz)$ are considered as a starting point
of the present research, then it is very important to find a correct
parametrization for these functions. The parametrization of the field
functions is important also for the subsequent tasks of harmonic analysis
and solutions of relativistic wave equations. The most natural way to do
it is to express the functions $f(x,\bz)$ via the matrix elements of
the Poincar\'{e} group. The functions $\psi_n(x)$ can be expressed via
the exponentials. As is known, exponentials define unitary representations
of the translation subgroup $T_4$. In turn, the functions $\phi^n(\bz)$
can be parametrized in the form of matrix elements of the Lorentz group.
Matrix elements of irreducible representations of the Lorentz group
were studied in the works 
\cite{Dol56,DT59,DM59,Esk59,Esk61,Gol61,Str65,Str67,ST67,VD67,Kol70,HS70,SH70,Hus88},
where various expressions for these elements are found, however, in 
rather complicate and cumbersome form. The most simple form of the matrix
elements for spinor representations of the Lorentz group has been found in
the works \cite{Var022,Var042} in the basis of complex angular momentum,
that corresponds to a local isomorphism
$SL(2,C)\sim SU(2)\otimes SU(2)$. In essence, this form of the matrix
elements presents itself a four-dimensional analog of Legendre spherical
functions (hyperspherical functions). Due to this form a relationship
between matrix elements of the Lorentz group and special functions becomes
more clear.
Moreover, it allows us to investigate the functions $f(x,\bz)$ within the
framework of the powerful mathematical theory presented in the works of
Vilenkin, Klimyk, Miller and Talman 
\cite{Vil68,Mil68,Tal68,Kli79,VK90}.
In the sections 2--4 of the present work we study various forms of
matrix elements of the groups $SU(2)$, $SU(1,1)$ and $SL(2,\C)$ both
for finite- and infinite-dimensional representations. In parallel,
we consider basic facts concerning harmonic analysis on these groups.
As a rule, harmonic analysis on the noncompact groups is a very complicate
mathematical theory and by this reason physicists hardly used this
mathematics (obviously, it is one of the main obstacles that decelerate
development of the quantum field theory on the Poincar\'{e} group).
Therefore, it is very important to accommodate the abstract harmonic
analysis to physics, that is, to construct a physically meaningful
theory. With this end in view we restrict ourselves mainly by a consideration
of finite dimensional representations, since they lead to local physical
fields on the homogeneous spaces. Due to this fact, all the basic physical
quantities, such as solutions of relativistic wave equations, field
operators, causal commutators and so on, can be expressed via Fourier-type
series on the homogeneous spaces.

The following logical step consists in definition of the Lagrangian
formalism and field equations on the various homogeneous spaces of $\cP$.
The field equations for arbitrary spin are derived by the standard variation
procedure from a selected Lagrangian. The use of harmonic analysis on
the homogeneous spaces allows us to set up a boundary value problem for
relativistically invariant system. It is shown that solutions of this
problem are expressed via Fourier series on the two-dimensional complex
sphere. In the sections \ref{Sec:Dir} and \ref{Sec:Max} we study
boundary value problems of the same type for the Dirac and Maxwell
fields, respectively. The representation of the Dirac and Maxwell fields
as fields on the Poincar\'{e} group allows us to consider these fields
on an equal footing, from the one group theoretical viewpoint.
Moreover, a definition of the Maxwell field on the complex two-sphere
leads naturally to a Riemann-Silberstein representation for the
electromagnetic field \cite{Web01,Sil07,Bir96} (or Majorana-Oppenheimer
formulation of quantum electrodynamics). Such basic notions of quantum
field theory as field operators, quantization, causal commutators,
vacuum expectation values of time ordered products have been defined
for the Dirac and Maxwell fields in the sections \ref{Sec:Dir} and
\ref{Sec:Max}. Interacting fields are discussed in the section
\ref{Sec:Int}.

\section{The group $SU(2)$}
In this section we briefly consider some basic facts concerning the
group $SU(2)$ and its representations. The group $SU(2)$ is an universal
covering of the three-dimensional rotation group $SO(3)$. Any matrix
$u$ from $SU(2)$ has the form
\begin{equation}\label{SU1}
u=\begin{pmatrix}
\alpha & \beta\\
-\bar{\beta} & \bar{\alpha}
\end{pmatrix},
\end{equation}
where $\det u=1$ and, therefore, $|\alpha|^2+|\beta|^2=1$. An Euler
parameterization of (\ref{SU1}) has the following form:
\begin{equation}\label{SU2}
u=\begin{pmatrix}
\cos\frac{\theta}{2}e^{\frac{i(\varphi+\psi)}{2}} &
i\sin\frac{\theta}{2}e^{\frac{i(\varphi-\psi)}{2}} \\
i\sin\frac{\theta}{2}e^{\frac{i(\psi-\varphi)}{2}} &
\cos\frac{\theta}{2}e^{-\frac{i(\varphi+\psi)}{2}}
\end{pmatrix},
\end{equation}
where $0\leq\varphi\leq 2\pi$, $0<\theta<\pi$, $-2\pi\leq\varphi<2\pi$.
It is easy to verify that from (\ref{SU2}) it follows directly a
decomposition
\begin{equation}\label{SU3}
u(\varphi,\theta,\psi)=\begin{pmatrix}
e^{\frac{i\varphi}{2}} & 0\\
0 & e^{-\frac{i\varphi}{2}}
\end{pmatrix}\!\!
{\renewcommand{\arraystretch}{1.3}
\begin{pmatrix}
\cos\frac{\theta}{2} & i\sin\frac{\theta}{2}\\
i\sin\frac{\theta}{2} & \cos\frac{\theta}{2}
\end{pmatrix}}\!\!
\begin{pmatrix}
e^{\frac{i\psi}{2}} & 0\\
0 & e^{-\frac{i\psi}{2}}
\end{pmatrix}.
\end{equation}
The formula (\ref{SU3}) defines {\it a Cartan decomposition}
$G=KAK$ for the group $G=SU(2)$.

The group $SU(2)$ has three one-parameter subgroups
\[
\omega_1(t)=\begin{pmatrix}
\cos\frac{t}{2} & i\sin\frac{t}{2}\\
i\sin\frac{t}{2} & \cos\frac{t}{2}
\end{pmatrix},\quad
\omega_2(t)=\begin{pmatrix}
\cos\frac{t}{2} & -\sin\frac{t}{2}\\
\sin\frac{t}{2} & \cos\frac{t}{2}
\end{pmatrix},\quad
\omega_3(t)=\begin{pmatrix}
e^{\frac{it}{2}} & 0\\
0 & e^{-\frac{it}{2}}
\end{pmatrix}.
\]
The tangent matrices $A_i$ of these subgroups have the form
\begin{eqnarray}
A_1&=&\left.\frac{d\omega_1(t)}{dt}\right|_{t=0}=\frac{i}{2}\begin{pmatrix}
0 & 1\\
1 & 0
\end{pmatrix},\nonumber\\
A_2&=&\left.\frac{d\omega_2(t)}{dt}\right|_{t=0}=\frac{1}{2}\begin{pmatrix}
0 & -1\\
1 & 0
\end{pmatrix},\nonumber\\
A_3&=&\left.\frac{d\omega_3(t)}{dt}\right|_{t=0}=\frac{i}{2}\begin{pmatrix}
1 & 0\\
0 & -1
\end{pmatrix}.\nonumber
\end{eqnarray}
The elements $A_i$ form a basis of Lie algebra $\mathfrak{su}(2)$ and
satisfy the relations
\[
\ld A_1,A_2\rd=A_3,\quad\ld A_2,A_3\rd=A_1,\quad\ld A_3,A_1\rd=A_2.
\]
An Euler parameterization of infinitesimal operators $\sA_i$ has the form:
\begin{eqnarray}
\sA_1&=&\cos\psi\frac{\partial}{\partial\theta}+
\frac{\sin\psi}{\sin\theta}\frac{\partial}{\partial\varphi}-
\ctg\theta\sin\psi\frac{\partial}{\partial\psi},\label{Inf1}\\
\sA_2&=&-\sin\psi\frac{\partial}{\partial\theta}+
\frac{\cos\psi}{\sin\theta}\frac{\partial}{\partial\varphi}-
\ctg\theta\cos\psi\frac{\partial}{\partial\psi},\label{Inf2}\\
\sA_3&=&\frac{\partial}{\partial\psi}.\label{Inf3}
\end{eqnarray}
Since $SU(2)$ is a compact group, then all its representations are
finite-dimensional. They are realized in symmetric representation spaces
$\Sym_k$ ($k=2l$) of polynomials
\[
p(z_0,z_1)=\sum_{(\alpha_1,\ldots,\alpha_k)}\frac{1}{k!}
a^{\alpha_1\cdots\alpha_k}z_{\alpha_1}\cdots z_{\alpha_k},
\]
The dimension of $\Sym_k$ is equal to $2l+1$. Operators $T_l(u)$,
$u\in SU(2)$, act in $\Sym_k$ via the formula
\[
T_l(u)\varphi(\fz)=(\beta \fz+\bar{\alpha})^{2l}
\varphi\left(\frac{\alpha \fz+\bar{\beta}}{\beta \fz+\bar{\alpha}}\right),
\]
where $\fz=z_0/z_1$.
The matrix element\index{element!matrix}
$t^l_{mn}=e^{-i(m\varphi+n\psi)}\langle T_l(\theta)
\psi_n,\psi_m\rangle$ of the group $SU(2)$ in the polynomial basis
\begin{equation}\label{Basis}
\psi_n(\fz)=\frac{\fz^{l-n}}{\sqrt{\Gamma(l-n+1)\Gamma(l+n+1)}},
\quad -l\leq n\leq l,
\end{equation}
where
\[
T_l(\theta)\psi(\fz)=\left(i\sin\frac{\theta}{2}\fz+
\cos\frac{\theta}{2}\right)^{2l}\psi
\left(\frac{\cos\frac{\theta}{2}\fz+i\sin\frac{\theta}{2}}
{i\sin\frac{\theta}{2}\fz+\cos\frac{\theta}{2}}\right),
\]
has a form
\begin{multline}
t^l_{mn}(u)=e^{-i(m\varphi+n\psi)}\langle T_l(\theta)\psi_n,\psi_m\rangle
=\\[0.2cm]
\frac{e^{-i(m\varphi+n\psi)}\langle T_l(\theta)\fz^{l-n}\fz^{l-m}\rangle}
{\sqrt{\Gamma(l-m+1)\Gamma(l+m+1)\Gamma(l-n+1)\Gamma(l+n+1)}}=\\[0.2cm]
e^{-i(m\varphi+n\psi)}i^{m-n}\sqrt{\Gamma(l-m+1)\Gamma(l+m+1)\Gamma(l-n+1)
\Gamma(l+n+1)}\times\\
\cos^{2l}\frac{\theta}{2}\tg^{m-n}\frac{\theta}{2}\times\\
\sum^{\min(l-n,l+n)}_{j=\max(0,n-m)}
\frac{i^{2j}\tg^{2j}\dfrac{\theta}{2}}
{\Gamma(j+1)\Gamma(l-m-j+1)\Gamma(l+n-j+1)\Gamma(m-n+j+1)}.\label{Mat1}
\end{multline}
Further, using the formula
\begin{equation}\label{HG}
%{}_2F_1(\alpha,\beta;\gamma;z)
\hypergeom{2}{1}{\alpha,\beta}{\gamma}{z}
=\frac{\Gamma(\gamma)}{\Gamma(\alpha)
\Gamma(\beta)}\sum_{k\ge 0}\frac{\Gamma(\alpha+k)\Gamma(\beta+k)}
{\Gamma(\gamma+k)}\frac{z^k}{k!}
\end{equation}
we can express the matrix element (\ref{Mat1}) via the hypergeometric
function:\index{function!hypergeometric}
\begin{multline}
t^l_{mn}(u)=\frac{i^{m-n}e^{-i(m\varphi+n\psi)}}{\Gamma(m-n+1)}
\sqrt{\frac{\Gamma(l+m+1)\Gamma(l-n+1)}{\Gamma(l-m+1)\Gamma(l+n+1)}}\times\\
\cos^{2l}\frac{\theta}{2}\tg^{m-n}\frac{\theta}{2}
\hypergeom{2}{1}{m-l+1,1-l-n}{m-n+1}{i^2\tg^2\dfrac{\theta}{2}},\label{Mat2}
\end{multline}
where $m\ge n$. At $m<n$ in the right part of (\ref{Mat2}) it needs to
replace $m$ and $n$ by $-m$ and $-n$, respectively. Since $l,m$ and $n$
are finite numbers, then the hypergeometric series is interrupted.

The matrix elements (\ref{Mat1}) can be written in other form
\begin{multline}
t^l_{mn}(u)=i^{m-n}e^{-i(m\varphi+n\psi)}
\sqrt{\frac{\Gamma(l-m+1)\Gamma(l-n+1)}{\Gamma(l+m+1)\Gamma(l+n+1)}}\times\\
\ctg^{m+n}\frac{\theta}{2}\sum^l_{j=\max(m,n)}
\frac{\Gamma(l+j+1)i^{2j}}{\Gamma(l-j+1)\Gamma(j-m+1)\Gamma(j-n+1)}
\sin^{2j}\frac{\theta}{2}.\label{MSU}
\end{multline}
This form is derived from (\ref{Mat1}) by means of a factorization
$u=kz$, where 
$k=\begin{pmatrix}
\bar{\alpha}^{-1} & \beta\\
0 & \bar{\alpha}
\end{pmatrix}$ and
$z=\begin{pmatrix}
1 & 0\\
-\bar{\beta}/\bar{\alpha} & 1
\end{pmatrix}$. According to the Cartan decomposition (\ref{SU3}), the
matrix elements can be written as
\begin{eqnarray}
t^l_{mn}(u)&=&t^l_{mm}[u(\varphi,0,0)]t^l_{mn}(\theta)
t^l_{nn}[u(0,0,\psi)]\nonumber\\
&=&e^{-i(m\varphi+n\psi)}t^l_{mn}(\theta),\nonumber
\end{eqnarray}
where
\[
t^l_{mn}(\theta)=P^l_{mn}(\cos\theta),
\]
here $P^l_{mn}(\cos\theta)$ is a generalized spherical function
\cite{GMS,Vil68}. Therefore, the formula (\ref{MSU}) can be rewritten
as follows
\begin{equation}\label{Mat12}
t^l_{mn}(u)=e^{-i(m\varphi+n\psi)}P^l_{mn}(\cos\theta).
\end{equation}

The matrices $T_l(\theta)$ of irreducible representations of $SU(2)$ at
$l=0,\,\frac{1}{2},\,1$ have the following form:
\begin{gather}
T_0(\theta)=1,\nonumber\\
T_{\frac{1}{2}}(\theta)=\begin{pmatrix}
P^{\frac{1}{2}}_{-\frac{1}{2}-\frac{1}{2}} &
P^{\frac{1}{2}}_{\frac{1}{2}-\frac{1}{2}} \\
P^{\frac{1}{2}}_{-\frac{1}{2}\frac{1}{2}} &
P^{\frac{1}{2}}_{\frac{1}{2}\frac{1}{2}}
\end{pmatrix}=\begin{pmatrix}
\cos\frac{\theta}{2} & i\sin\frac{\theta}{2}\\
i\sin\frac{\theta}{2} & \cos\frac{\theta}{2}
\end{pmatrix},\nonumber\\
T_1(\theta)=\begin{pmatrix}
P^1_{-1-1} & P^1_{-10} & P^1_{-11}\\
P^1_{0-1} & P^1_{00} & P^1_{01}\\
P^1_{1-1} & P^1_{10} & P^1_{11}
\end{pmatrix}
=\begin{pmatrix}
\cos^2\frac{\theta}{2} & \frac{i\sin\theta}{\sqrt{2}} &-\sin^2\frac{\theta}{2}\\
\frac{i\sin\theta}{\sqrt{2}} & \cos\theta & \frac{i\sin\theta}{\sqrt{2}} \\
-\sin^2\frac{\theta}{2} & \frac{i\sin\theta}{\sqrt{2}} &\cos^2\frac{\theta}{2}
\end{pmatrix}.\nonumber
\end{gather}
Using (\ref{Inf1})--(\ref{Inf3}), it is easy to calculate the
Laplace-Beltrami operator $\Delta=\sA^2_1+\sA^2_2+\sA^2_3$ 
on the group $SU(2)$:
\begin{equation}\label{LB}
\Delta=\frac{\partial^2}{\partial\theta^2}+
\ctg\theta\frac{\partial}{\partial\theta}+
\frac{1}{\sin^2\theta}\left(\frac{\partial^2}{\partial\varphi^2}-
2\cos\theta\frac{\partial^2}{\partial\varphi\partial\psi}+
\frac{\partial^2}{\partial\psi^2}\right).
\end{equation}
Matrix elements (\ref{Mat12}) of irreducible representations of the group
$SU(2)$ are eigenfunctions of the operator (\ref{LB}):
\[
\Delta t^l_{mn}(u)=-l(l+1)t^l_{mn}(u).
\]
Substituting in this equality the expressions (\ref{LB}) and
(\ref{Mat12}) instead $\Delta$ and $t^l_{mn}(u)$ and supposing
$x=\cos\theta$, we come to the following differential equation
\begin{equation}\label{PE}
\left[(1-x^2)\frac{d^2}{dx^2}-2x\frac{d}{dx}-
\frac{m^2+n^2-2mnx}{1-x^2}\right] P^l_{mn}(x)=-l(l+1)P^l_{mn}(x).
\end{equation}
In particular, at $n=0$ we obtain differential equation for associated
Legendre functions:
\[
\left[(1-x^2)\frac{d^2}{dx^2}-2x\frac{d}{dx}-
\frac{m^2}{1-x^2}\right] P^m_l(x)=-l(l+1)P^m_l(x).
\]
Further, at $n=0$ and $m=0$ we come to the well known differential
equation for Legendre polynomials:
\[
(1-x^2)\frac{d^2P_l(x)}{dx^2}-2x\frac{d P_l(x)}{dx}+l(l+1)P_l(x)=0.
\]
\subsection{Harmonic analysis on the group $SU(2)$}
\label{HASU2}
It is well known \cite{Rud62,Vil68} that the classical Fourier analysis
is completely formulated in terms of an additive group of real numbers,
$\R$. $\R$ is an Abelian and non-compact group, and by this reason all
its unitary representations are one-dimensional and expressed via the
exponential $e^{ax}$, where $a=bi$. The regular representation of the
group $\R$ is constructed in the space $\mathfrak{S}$ of the square
integrable functions $f(x)$ defined on the group $\R$ (equally, on the
real axis $-\infty<x<\infty$) such that
\[
\int\limits^\infty_{-\infty}|f(x)|^2dx<+\infty.
\]
The operator $\R(x_0)$, transforming the function $f(x)$ into
$\R(x_0)f(x)=f(x+x_0)$, corresponds to the each element of the group
$\R$. $\R(x)$ is a regular representation of $\R$. At this point, any
function $f(x)\in\mathfrak{S}$ can be represented in the form of
continuous linear combination of the functions $e^{i\lambda x}$
(Fourier integral):
\[
f(x)=\int\limits^\infty_{-\infty}F(\lambda)e^{i\lambda x}dx,\quad
-\infty<x<\infty.
\]
The inverse transformation is defined by a formula
\[
F(\lambda)=\frac{1}{2\pi}\int\limits^\infty_{-\infty}f(x)e^{-i\lambda x} dx.
\]
Hence it follows a so called Plancherel formula
\[
\int\limits^\infty_{-\infty}|F(\lambda)|^2d\lambda=\frac{1}{2\pi}
\int\limits^\infty_{-\infty}|f(x)|^2dx.
\]

In the case of a quotient group $SO(2)=\R/\dZ_{2\pi}$
(rotation group of the euclidean plane), where the subgroup $\dZ_{2\pi}$
is generated by numbers of the form $2\pi k$, we come to the classical
Fourier series. Indeed, if $f(\varphi)$ is an irreducible unitary
representation of $SO(2)$, then the equality
\[
F(\varphi+2\pi k)=f(\varphi),\quad 0\leq\varphi<2\pi,
\]
defines the irreducible unitary representation $e^{iax}$ of $\R$. The
representations of $SO(2)$ are derived from $e^{iax}$ at the conditions
$F(2\pi)=f(0)=1$. They have the form $f(\varphi)=e^{in\varphi}$, where
$n$ is an integer number.

The integral on the group $SO(2)$ is defined as follows
\[
\int f(g)dg=\frac{1}{2\pi}\int\limits^{2\pi}_0f(\varphi)d\varphi,
\]
where $1/2\pi d\varphi$ is an invariant measure (Haar measure) on the
group $SO(2)$. This integral possesses the property
\[
\int f(gg_0)dg=\int f(g)dg.
\]

Any square integrable function $f(\varphi)$ on the group $SO(2)$ expands
into a convergent series
\[
f(\varphi)=\sum^\infty_{n=-\infty}c_ne^{in\varphi},
\]
where the coefficients are expressed via the formulae
\[
c_n=\frac{1}{2\pi}\int\limits^{2\pi}_0f(\varphi)e^{-in\varphi}d\varphi.
\]
Hence it immediately follows a Parseval equality
\[
\frac{1}{2\pi}\int\limits^{2\pi}_0|f(\varphi)|^2d\varphi=
\sum^\infty_{n=-\infty}|c_n|^2.
\]

The generalization of classical Fourier series for the case of non-Abelian
compact groups was first given by Peter and Weyl \cite{PW27}. The first
simplest case of such groups is $SU(2)$. Since $SU(2)$ is compact, then
there exists an invariant measure $du$ on this group such that
\begin{equation}\label{MES1}
\int f(u)du=\int f(u_0u)du=\int f(uu_0)du=\int f(u^{-1})du.
\end{equation}
This equality holds for all continuous functions $f(u)$ and all elements
$u_0$ from $SU(2)$. The invariant integral on the group $SU(2)$ is defined
in Euler parameterization by a formula
\begin{equation}\label{II1}
\int\limits_{SU(2)}f(u)du=\frac{1}{16\pi^2}\int\limits^{2\pi}_{-2\pi}
\int\limits^{2\pi}_0\int\limits^\pi_0f(\varphi,\theta,\psi)
\sin\theta d\theta d\varphi d\psi,
\end{equation}
where $1/16\pi^2\sin\theta d\theta d\varphi d\psi$ is a Haar measure on the
group $SU(2)$. It is easy to verify that the integral (\ref{II1}) satisfy
the equality (\ref{MES1}).

Since the dimensionality of the representation $T_l(u)$ of $SU(2)$ is
equal to $2l+1$, then the functions $\sqrt{2l+1}t^l_{mn}(u)$ form
a full orthogonal normalized system of functions with respect to the
invariant measure $du$ on this group. In other words, the functions
$t^l_{mn}(u)$ (matrix elements) satisfy the relations
\begin{equation}\label{ORT1}
\int\limits_{SU(2)}t^l_{mn}(u)\overline{t^s_{pq}(u)}du=
\frac{1}{2l+1}\delta_{ls}\delta_{mp}\delta_{nq}.
\end{equation}
Substituting 
into (\ref{ORT1}) instead the matrix elements $t^l_{mn}(u)$
their expressions via the Euler angles (see (\ref{Mat12})), we come to the
following formula
\begin{equation}\label{ORT2}
\int\limits^{2\pi}_{-2\pi}\int\limits^{2\pi}_0\int\limits^\pi_0
P^l_{mn}(\cos\theta)\overline{P^s_{pq}(\cos\theta)}\sin\theta
e^{i(p-m)\varphi}e^{i(q-n)\psi}d\theta d\varphi d\psi=
\frac{16\pi^2}{2l+1}\delta_{ls}\delta_{mp}\delta_{nq}.
\end{equation}
Due to the well known orthogonality relations for the functions
$P^l_{mn}(x)$ \cite{Vil68},
\[
\int\limits^1_{-1}P^l_{mn}(x)\overline{P^s_{mn}(x)}dx=
\frac{2}{2l+1}\delta_{ls},
\]
we see that the relation (\ref{ORT2}) holds for all $l$, $s$, $m$, $p$,
$n$, $q$.

Thus, any square integrable function $f(\varphi,\theta,\psi)$ on the
group $SU(2)$, $0\leq\varphi<2\pi$, $0\leq\theta<\pi$,
$-2\pi<\psi<2\pi$, such that
\[
\int\limits^{2\pi}_{-2\pi}\int\limits^{2\pi}_0\int\limits^\pi_0
|f(\varphi,\theta,\psi)|^2\sin\theta d\theta d\varphi d\psi\;<\;+\infty,
\]
is expanded in a convergent Fourier series on $SU(2)$,
\[
f(\varphi,\theta,\psi)=\sum_l\sum^l_{m=-l}\sum^l_{n=-l}
\alpha^l_{mn}e^{-i(m\varphi+n\psi)}P^l_{mn}(\cos\theta),
\]
where
\[
\alpha^l_{mn}=\frac{(-1)^{m-n}(2l+1)}{16\pi^2}
\int\limits^{2\pi}_{-2\pi}\int\limits^{2\pi}_0\int\limits^\pi_0
f(\varphi,\theta,\psi)e^{i(m\varphi+n\psi)}P^l_{mn}(\cos\theta)\sin\theta
d\theta d\varphi d\psi.
\]
The Parseval equality in this case has a form
\[
\sum_l\sum_{m=-l}^l\sum^l_{n=-l}|\alpha^l_{mn}|^2=
\frac{2l+1}{16\pi^2}
\int\limits^{2\pi}_{-2\pi}\int\limits^{2\pi}_0\int\limits^\pi_0
|f(\varphi,\theta,\psi)|^2\sin\theta d\theta d\varphi d\psi.
\]

\section{The group $SU(1,1)$}
The group $SU(1,1)$ (also known as a three-dimensional Lorentz group) is an
universal covering of the group $SH(3)$. The group $SH(3)$ is a group
of linear transformations of the space $E_3$ with a quadratic form
\[
\ld\bx,\bx\rd=x^2_1-x^2_2-x^2_3.
\]
Any matrix $g\in SU(1,1)$ has the form
\begin{equation}\label{FUNg}
g=\begin{pmatrix}
\gamma & \delta\\
\bar{\delta} & \bar{\gamma}
\end{pmatrix},
\end{equation}
where $\det g=1$ and, therefore, $|\gamma|^2-|\delta|^2=1$. An Euler
parametrization of (\ref{FUNg}) is
\[
g=\begin{pmatrix}
\ch\frac{\tau}{2}e^{\frac{i(\varphi+\psi)}{2}} &
\sh\frac{\tau}{2}e^{\frac{i(\varphi-\psi)}{2}} \\
\sh\frac{\tau}{2}e^{\frac{i(\psi-\varphi)}{2}} &
\ch\frac{\tau}{2}e^{\frac{-i(\varphi+\psi)}{2}} 
\end{pmatrix},
\]
where $0\leq\varphi\leq 2\pi$, $0<\tau<\infty$, $-2\pi\leq\psi<2\pi$.
Hence it immediately follows a Cartan decomposition for the group
$SU(1,1)$:
\begin{equation}\label{Car2}
g(\varphi,\tau,\psi)=\begin{pmatrix}
e^{\frac{i\varphi}{2}} & 0\\
0 & e^{-\frac{i\varphi}{2}}
\end{pmatrix}\!\!
{\renewcommand{\arraystretch}{1.3}
\begin{pmatrix}
\ch\frac{\tau}{2} & \sh\frac{\tau}{2}\\
\sh\frac{\tau}{2} & \ch\frac{\tau}{2}
\end{pmatrix}}\!\!
\begin{pmatrix}
e^{\frac{i\psi}{2}} & 0\\
0 & e^{-\frac{i\psi}{2}}
\end{pmatrix}.
\end{equation}
The group $SU(1,1)$ has three one-parameter subgroups
\[
\omega_1(t)=\begin{pmatrix}
\ch\frac{t}{2} & \sh\frac{t}{2}\\
\sh\frac{t}{2} & \ch\frac{t}{2}
\end{pmatrix},\quad
\omega_2(t)=\begin{pmatrix}
\ch\frac{t}{2} & i\sh\frac{t}{2}\\
-i\sh\frac{t}{2} & \ch\frac{t}{2}
\end{pmatrix},\quad
\omega_3(t)=\begin{pmatrix}
e^{\frac{it}{2}} & 0\\
0 & e^{-\frac{it}{2}}
\end{pmatrix}.
\]
The tangent matrices $A_i$ of these subgroups have the form
\begin{eqnarray}
A_1&=&\left.\frac{d\omega_1(t)}{dt}\right|_{t=0}=\frac{1}{2}
\begin{pmatrix}
0 & 1\\
1 & 0
\end{pmatrix},\nonumber\\
A_2&=&\left.\frac{d\omega_2(t)}{dt}\right|_{t=0}=\frac{i}{2}
\begin{pmatrix}
0 & 1\\
-1 & 0
\end{pmatrix},\nonumber\\
A_3&=&\left.\frac{d\omega_3(t)}{dt}\right|_{t=0}=\frac{i}{2}
\begin{pmatrix}
1 & 0\\
0 &-1
\end{pmatrix}.\nonumber
\end{eqnarray}
The elements $A_i$ form a basis of Lie algebra $\mathfrak{su}(1,1)$ and
satisfy the relations
\[
\ld A_1,A_2\rd=-A_3,\quad\ld A_2,A_3\rd=A_1,\quad\ld A_3,A_1\rd=A_2.
\]
In this case an Euler parametrization of infinitesimal operators
$\sA_i$ for the left regular representation has a form
\begin{eqnarray}
\sA_1&=&\cth\tau\sin\varphi\frac{\partial}{\partial\varphi}-
\cos\varphi\frac{\partial}{\partial\tau}-
\frac{\sin\varphi}{\sh\tau}\frac{\partial}{\partial\psi},\label{Inf4}\\
\sA_2&=&-\cth\tau\cos\varphi\frac{\partial}{\partial\varphi}-
\sin\varphi\frac{\partial}{\partial\tau}+
\frac{\cos\varphi}{\sh\tau}\frac{\partial}{\partial\psi},\label{Inf5}\\
\sA_3&=&-\frac{\partial}{\partial\varphi}.\label{Inf6}
\end{eqnarray}
Since $SU(1,1)\simeq SL(2,\R)$ is the noncompact group, then in this case
we have both finite- and infinite-dimensional representations.
Representations of the group $SL(2,\R)$ are defined by the two numbers
$\chi=(\tau,o)$, $\tau\in\C$, $o\in\{0,1/2\}$ \cite{Vil68}.
The representation $T_\chi\equiv T_{(\tau,o)}$ of the
principal nonunitary series of $SL(2,\R)$ is realized in the space of
functions $f(x)$ which depend on the real variable:
\[
T_\chi(g)f(x)=|\beta x+\delta|^{2\tau}\mbox{sign}^{2o}
(\beta x+\delta)f\left(\frac{\alpha x+\gamma}{\beta x+\delta}\right),
\]
where $g=\begin{pmatrix}
\alpha & \beta\\
\gamma & \delta
\end{pmatrix}\in SL(2,\R)$. At $\tau=i\rho-\frac{1}{2}$, $\rho\in\R$,
the representations $T_\chi$ are unitary with respect to a scalar product
in the Hilbert space $L^2(\R)$. They form the principal unitary 
representation series of $SL(2,\R)$. The group $SL(2,\R)$ has also
discrete series of unitary representations. Representations of the
negative discrete series $T^-_l$, $l=-1,-\frac{3}{2},-2,-\frac{5}{2},\ldots$,
act in the Hilbert space $H_l$ of the functions $F(w)$ which are analytic
in upper half-plane $\C_+$ and possess with a following scalar product
\[
(F_1,F_2)=\frac{i}{2\Gamma(-2l-1)}\int\limits_{\C_+}F_1(w)
\overline{F_2(w)}y^{-2l-2}dw\overline{dw},
\]
where $w=x+iy$, $dw\overline{dw}=-2idxdy$. The operators $T^-_l(g)$,
$g=\begin{pmatrix}
\alpha & \beta\\
\gamma & \delta
\end{pmatrix}$, are defined by
\[
T^-_l(g)F(w)=(\beta w+\delta)^{2l}F\left(\frac{\alpha w+\gamma}
{\beta w+\delta}\right).
\]
In like manner, representations of the positive discrete series
$T^+_l$, $l=-1,-\frac{3}{2},-2,\ldots$, are constructed in the Hilbert
space of functions which are analytic in lower half-plane $\C_-$.
Since, $SU(1,1)\simeq SL(2,\R$, then representations of the principal
nonunitary series of $SU(1,1)$ are defined by the same pair
$\chi=(l,o)$ as with the group $SL(2,\R)$. These representations
are realized on the subgroup $K=SO(2)$:
\begin{equation}\label{FinT}
T_\chi(g)f(e^{i\theta})=(\gamma e^{i\theta}+\bar{\delta})^{l+o}
(\bar{\gamma}e^{-i\theta}+\delta)^{l-o}
f\left(\frac{\delta e^{i\theta}+\bar{\gamma}}
{\gamma e^{i\theta}+\bar{\delta}}\right),
\end{equation}
where $g=\begin{pmatrix}
\gamma & \delta\\
\bar{\delta} & \bar{\gamma}
\end{pmatrix}\in SU(1,1)$.

As follows from (\ref{FinT}), matrix elements $t^\chi_{mn}(g)$ of the
representation $T_\chi(g)$ in the orthonormal basis
\[
\left\{\left.\frac{1}{\sqrt{2\pi}}e^{ip\varphi}\right|\;
p=0,\pm 1,\pm 2,\ldots\right\}
\]
have a following integral representation
\[
t^\chi_{mn}(g)=\frac{1}{2\pi}\int\limits^{2\pi}_0
(\gamma e^{i\theta}+\bar{\delta})^{l+n+o}
(\bar{\gamma}e^{-i\theta}+\delta)^{l-n-o}
e^{i(m-n)\varphi}d\varphi,
\]
where $m$ and $n$ are integer or half-integer numbers in accordance with
the values of $o$. This integral can be calculated by means
of the Newton binomial. In the case of finite-dimensional representations
we have
\begin{multline}
t^\chi_{mn}(g)=\Gamma(l+n+o+1)\Gamma(l-n-o+1)\delta^{l-n-o}
\bar{\delta}^{l+m+o}\gamma^{n-m}\times\\
\sum^{\min(l-n,l+m)}_{s=\max(0,m-n)}
\frac{|\gamma/\delta|^{2s}}{\Gamma(s+1)\Gamma(l-n-s-o+1)
\Gamma(n-m+s+1)\Gamma(l+m-s+o+1)}.\label{Mat1'}
\end{multline}
In accordance with the Cartan decomposition (\ref{Car2}), the matrix
elements can be written as follows
\begin{eqnarray}
t^\chi_{mn}(g)&=&t^\chi_{m^\prime m^\prime}[g(\varphi,0,0)]
t^\chi_{m^\prime n^\prime}(\tau)t^\chi_{n^\prime n^\prime}[g(0,0,\psi)]
\nonumber\\
&=&e^{-i(m^\prime\varphi+n^\prime\psi)}
\fP^l_{m^\prime n^\prime}(\ch\tau),\label{Mt1'}
\end{eqnarray}
where $\chi=(l,o)$, $m^\prime=m+o$, $n^\prime=n+o$. An explicit form
of the function $\fP^l_{mn}(\ch\tau)$ (Jacobi function) follows from
(\ref{Mat1'}) at $\gamma=\sh\frac{\tau}{2}$ and $\delta=\ch\frac{\tau}{2}$:
\begin{multline}
\fP^l_{mn}(\ch\tau)=\sqrt{\Gamma(l-n+1)\Gamma(l+n+1)\Gamma(l-m+1)
\Gamma(l+m+1)}\times\\
\ch^{2l}\frac{\tau}{2}\tnh^{n-m}\frac{\tau}{2}\times\\
\sum^{\min(l-n,l+m)}_{s=\max(0,m-n)}
\frac{\tnh^{2s}\frac{\tau}{2}}{\Gamma(s+1)\Gamma(l-n-s+1)\Gamma(n-m+s+1)
\Gamma(l+m-s+1)},\label{Jacobi}
\end{multline}
where $m,n=-l,-l+1,\ldots,l$.

Let us give explicit expressions for the matrices $T_l(\tau)$ at
$l=0,\,1/2,\,1$:
\begin{gather}
T_0(\tau)=1,\nonumber\\
T_{\frac{1}{2}}(\tau)=\begin{pmatrix}
\fP^{\frac{1}{2}}_{-\frac{1}{2}-\frac{1}{2}} &
\fP^{\frac{1}{2}}_{\frac{1}{2}-\frac{1}{2}} \\
\fP^{\frac{1}{2}}_{-\frac{1}{2}\frac{1}{2}} &
\fP^{\frac{1}{2}}_{\frac{1}{2}\frac{1}{2}}
\end{pmatrix}=\begin{pmatrix}
\ch\frac{\tau}{2} & \sh\frac{\tau}{2}\\
\sh\frac{\tau}{2} & \ch\frac{\tau}{2}
\end{pmatrix},\nonumber\\
T_1(\tau)=\begin{pmatrix}
\fP^1_{-1-1} & \fP^1_{-10} & \fP^1_{-11}\\
\fP^1_{0-1} & \fP^1_{00} & \fP^1_{01}\\
\fP^1_{1-1} & \fP^1_{10} & \fP^1_{11}
\end{pmatrix}
=\begin{pmatrix}
\ch^2\frac{\tau}{2} & \sh\tau & \sh^2\frac{\tau}{2}\\
\sh\tau & \ch\tau & \sh\tau\\
\sh^2\frac{\tau}{2} & \sh\tau & \ch^2\frac{\tau}{2}
\end{pmatrix}.\nonumber
\end{gather}
Using the formula (\ref{HG}), we can express the function 
$\fP^l_{mn}(\ch\tau)$ via the hypergeometric function:
\begin{multline}
\fP^l_{mn}(\ch\tau)=\frac{1}{\Gamma(n-m+1)}
\sqrt{\frac{\Gamma(l+n+1)\Gamma(l-m+1)}{\Gamma(l-n+1)\Gamma(l+m+1)}}\times\\
\ch^{2l}\frac{\tau}{2}\tnh^{n-m}\frac{\tau}{2}
\hypergeom{2}{1}{n-l+1,1-l-m}{n-m+1}{\tnh^2\frac{\tau}{2}},
\label{Mat2'}
\end{multline}
where $m\ge n$. At $m<n$ in the right part of (\ref{Mat2'}) it needs to
replace $m$ and $n$ by $-m$ and $-n$, respectively. Since $l$, $m$ and
$n$ are finite numbers, then the hypergeometric series in (\ref{Mat2'})
is finite also.

In the case of principal series of unitary representations, the function
(\ref{Jacobi}) takes a form
\begin{multline}
\fP^{-\frac{1}{2}+i\rho}_{mn}(\ch\tau)=
\sqrt{\Gamma(i\rho-n+\tfrac{1}{2})\Gamma(i\rho+n+\tfrac{1}{2})
\Gamma(i\rho-m+\tfrac{1}{2})
\Gamma(i\rho+m+\tfrac{1}{2})}\times\\
\ch^{2i\rho-1}\frac{\tau}{2}\tnh^{n-m}\frac{\tau}{2}\times\\
\sum^{\infty}_{s=\max(0,m-n)}
\frac{\tnh^{2s}\frac{\tau}{2}}{\Gamma(s+1)
\Gamma(i\rho-n-s+\tfrac{1}{2})\Gamma(n-m+s+1)
\Gamma(i\rho+m-s+\tfrac{1}{2})},\nonumber
\end{multline}
or
\begin{multline}
\fP^{-\frac{1}{2}+i\rho}_{mn}(\ch\tau)=\frac{1}{\Gamma(n-m+1)}
\sqrt{\frac{\Gamma(i\rho+n+\tfrac{1}{2})
\Gamma(i\rho-m+\tfrac{1}{2})}{\Gamma(i\rho-n+\tfrac{1}{2})
\Gamma(i\rho+m+\tfrac{1}{2})}}\times\\
\ch^{2i\rho-1}\frac{\tau}{2}\tnh^{n-m}\frac{\tau}{2}
\hypergeom{2}{1}{n-i\rho+\tfrac{3}{2},\tfrac{3}{2}-i\rho-m}
{n-m+1}{\tnh^2\frac{\tau}{2}}.
\label{Mat3}
\end{multline}
When $m=0$, $n=0$ (zonal functions), the function (\ref{Mat3}) transforms
into {\it a conical function} $\fP_{i\rho-\frac{1}{2}}(\ch\tau)$
(for more details about conical functions see \cite{Bat}).

Matrix elements $t^{l,-}_{mn}(g_\tau)$, $-\infty<m,n\leq l$, of the
discrete series $T^-_l$ of representations of the group $SU(1,1)$ are
expressed via the Jacobi polynomials $P^{(\alpha,\beta)}_n(\ch\tau)$
\cite{VK90}:
\[
\fP^{l,-}_{mn}(\ch\tau)=\left[\frac{\Gamma(l-m+1)\Gamma(-l-n)}
{\Gamma(l-n+1)\Gamma(-l-m)}\right]^{\frac{1}{2}}
\sh^{m-n}\frac{\tau}{2}\ch^{m+n}\frac{\tau}{2}
P^{(m-n,m+n)}_{l-m}(\ch\tau)
\]
if $n\leq m\leq l$ and
\[
\fP^{l,-}_{mn}(\ch\tau)=\left[\frac{\Gamma(l-n+1)\Gamma(-l-m)}
{\Gamma(l-m+1)\Gamma(-l-n)}\right]^{\frac{1}{2}}
\sh^{n-m}\frac{\tau}{2}\ch^{m+n}\frac{\tau}{2}
P^{(n-m,m+n)}_{l-n}(\ch\tau)
\]
if $m\leq n\leq l$.

On the group $SU(1,1)$ there exists the following Laplace-Beltrami
operator (or Casimir operator):
\[
\Delta=-\sA^2_1-\sA^2_2+\sA^2_3.
\]
Using (\ref{Inf4})--(\ref{Inf5}), we can express this operator via the
Euler angles:
\begin{equation}\label{LB2}
\Delta=-\frac{1}{\sh\tau}\frac{\partial}{\partial\tau}\sh\tau
\frac{\partial}{\partial\tau}-\frac{1}{\sh^2\tau}\left[
\frac{\partial^2}{\partial\varphi^2}-2\ch\tau\frac{\partial^2}
{\partial\varphi\partial\psi}+\frac{\partial^2}{\partial\psi^2}\right].
\end{equation}

Matrix elements $t^\chi_{mn}(g)$ of irreducible representations of the
group $SU(1,1)$ are eigenfunctions of the operator (\ref{LB2}):
\[
\Delta t^\chi_{mn}(g)=l(l+1)t^\chi_{mn}(g).
\]
Indeed, substituting in this equality the expression (\ref{LB2}) and
Euler parameterization of $t^\chi_{mn}(g)$ instead $\Delta$ and
$t^\chi_{mn}(g)$ and supposing $y=\ch\tau$, we come to a following
differential equation
\[
\left[(y^2-1)\frac{d^2}{dy^2}+2y\frac{d}{dy}-
\frac{m^2+n^2-2mny}{y^2-1}\right]\fP^l_{mn}(y)=l(l+1)\fP^l_{mn}(y).
\]
It is easy to see that this equation has the structure similar to
the equation
(\ref{PE}) for the functions $P^l_{mn}(x)$.

\subsection{Harmonic analysis on the group $SU(1,1)$}
Harmonic analysis on the group $SU(1,1)$ was studied by many authors
during long time (see, for example, \cite{KS60,Puk64,Vil68,Len85}).
Let $C^\infty_0(G)$, $G=SL(2,\R)$, be a space of infinitely differentiable
functions on the group $SL(2,\R)$, then for the representation $T_\chi$
of the principal nonunitary series of $SL(2,\R)$ there exists on
$C^\infty_0(G)$ the following transformation (Fourier transform)
\begin{equation}\label{Har1}
T^f_\chi=\int\limits_{SL(2,\R)}f(g)T^\ast_\chi(g)dg,\quad
f\in C^\infty_0(G).
\end{equation}
The same transformation can be defined for the representations $T^-_l$ and
$T^+_l$ of the discrete series of $SL(2,\R)$:
\begin{equation}\label{Har2}
T^{f,\pm}_l=\int\limits_{SL(2,\R)}f(g)\ld T^\pm_l(g)\rd^\ast dg,
\end{equation}
where $T^f_\chi$, $T^{f,\pm}_\chi$ are operators with the trace.

The inverse Fourier transform is defined by a formula
\begin{multline}
f(g)=\sum_{o=0,\frac{1}{2}}\frac{1}{4\pi^2}\int\limits^\infty_0
\Tr\ld T^f_{(i\rho-1/2,o)}T_{(i\rho-1/2,o)}(g)\rd\rho\tnh\pi(\rho+io)dp+\\
+\frac{1}{4\pi^2}\sum_l\left(l+\frac{1}{2}\right)
\Tr\ld T^{f,+}_{-l-1}T^+_{-l-1}(g)+T^{f,-}_{-l-1}T^-_{-l-1}(g)\rd,\label{Har3}
\end{multline}
where the index $l$ in the second sum runs the values $0,\,\frac{1}{2},\,1,\,
\frac{3}{2},\,\ldots$.

In this case the Plancherel formula can be written as follows
\begin{multline}
\int\limits_{SL(2,\R)}|f(g)|^2dg=\frac{1}{4\pi^2}\left\{
\sum_{o=0,\frac{1}{2}}\int\limits^\infty_0\Tr\ld T^f_{(i\rho-1/2,o)}
(T^f_{(i\rho-1/2,o)})^\ast\rd\right.\times\\
\times\rho\tnh\pi(\rho+io)d\rho+\sum_l\left(l+\frac{1}{2}\right)\Tr\left[
T^{f,+}_{-l-1}(T^{f,+}_{-l-1})^\ast+\right.\\
\left.\left.+T^{f,-}_{-l-1}(T^{f,-}_{-l-1})^\ast\right]\right\}.\label{Har4}
\end{multline}

The invariant measure on the group $SU(1,1)$ has a form
\[
dg=\frac{1}{4\pi^2}\sh\tau d\varphi d\tau d\psi.
\]
Taking into account this expression, we can rewrite the formulae
(\ref{Har1})--(\ref{Har4}) via the matrix elements (\ref{Mat1'}). Indeed,
any square integrable function $f(g)$ on the group $SU(1,1)$, such that
\[
\int\limits_{SU(1,1)}|f(g)|^2dg<+\infty,
\]
is expanded in matrix elements as follows
\begin{multline}
f(\varphi,\tau,\psi)=\frac{1}{4\pi^2}\sum_{m,n,o}\left[
\int\limits^\infty_0a^o_{mn}(\rho)e^{-i(m\varphi+n\psi)}
\fP^{-\frac{1}{2}+i\rho,o}_{mn}(\ch\tau)\rho\tnh\pi(\rho+oi)d\rho+\right]\\
+\left.\sum^\infty_{l=1-o}\left(l-\frac{1}{2}\right)b^o_{mn}(l)
e^{-i(m\varphi+n\psi)}\fP^{l,o}_{mn}(\ch\tau)\right],
\label{Har5}
\end{multline}
where the values of the coefficients $a^o_{mn}(\rho)$ and $b^o_{mn}(l)$
are expressed via the formulae
\[
a^o_{mn}(\rho)=\int\limits^{2\pi}_{-2\pi}\int\limits^{2\pi}_0
\int\limits^\infty_0f(\varphi,\tau,\psi)e^{i(m\varphi+n\psi)}
\overline{\fP^{-\frac{1}{2}+i\rho,o}_{mn}(\ch\tau)}\sh\tau d\tau d\varphi
d\psi,
\]
\begin{multline}
b^o_{mn}(l)=\frac{(-1)^{m-n}\Gamma(l+m+o+1)\Gamma(l-m-o+1)}
{\Gamma(l+n+o+1)\Gamma(l-n-o+1)}\times\\
\int\limits^{2\pi}_{-2\pi}\int\limits^{2\pi}_0\int\limits^\infty_0
f(\varphi,\tau,\psi)e^{i(m\varphi+n\psi)}
\overline{\fP^{l,o}_{mn}(\ch\tau)}\sh\tau d\tau d\varphi d\psi.\nonumber
\end{multline}
Further, the Plancherel formula takes a form
\begin{multline}
\int\limits_{SU(1,1)}|f(g)|^2dg=\frac{1}{4\pi^2}\sum_{m,n,o}
\left[\int\limits^\infty_0|a^o_{mn}(\rho)|^2\rho\tnh\pi(\rho+oi)d\rho+\right.\\
+\left.(-1)^{m-n}\sum^\infty_{l=1-o}
\frac{\Gamma(l+n+o+1)\Gamma(l-n-o+1)}{\Gamma(l+m+o+1)\Gamma(l-m-o+1)}
\left(l-\frac{1}{2}\right)|b^o_{mn}(l)|^2\right].
\nonumber
\end{multline}
\section{The group $SL(2,\C)$}\label{SL2C}
As is known, the group $SL(2,\C)$ is an universal covering of the proper
orthochronous Lorentz group $L^\uparrow_+$.
The group $SL(2,\C)$ of all complex matrices
\[\ar
\fg=
\begin{pmatrix}
\alpha & \beta\\
\gamma & \delta
\end{pmatrix}
\]
of 2-nd order with the determinant $\alpha\delta-\gamma\beta=1$, is
a {\it complexification} of the group $SU(2)$. The group $SU(2)$ is one of
the real forms of $SL(2,\C)$. The transition from $SU(2)$ to $SL(2,\C)$
is realized via the complexification of three real parameters
$\varphi,\,\theta,\,\psi$ (Euler angles). Let $\theta^c=\theta-i\tau$,
$\varphi^c=\varphi-i\epsilon$, $\psi^c=\psi-i\varepsilon$ be complex
Euler angles,\index{Euler angles!complex} where
\begin{equation}\label{CEA}
{\renewcommand{\arraystretch}{1.05}
\begin{array}{ccccc}
0 &\leq&\re\theta^c=\theta& \leq& \pi,\\
0 &\leq&\re\varphi^c=\varphi& <&2\pi,\\
-2\pi&\leq&\re\psi^c=\psi&<&2\pi,
\end{array}\quad\quad
\begin{array}{ccccc}
-\infty &<&\im\theta^c=\tau&<&+\infty,\\
-\infty&<&\im\varphi^c=\epsilon&<&+\infty,\\
-\infty&<&\im\psi^c=\varepsilon&<&+\infty.
\end{array}}
\end{equation}
Replacing in (\ref{SU2}) the angles $\varphi,\,\theta,\,\psi$ by the
complex angles $\varphi^c,\theta^c,\psi^c$, we come to the following matrix
%\[
\begin{gather}
{\renewcommand{\arraystretch}{1.3}
\mathfrak{g}=
\begin{pmatrix}
\cos\frac{\theta^c}{2}e^{\frac{i(\varphi^c+\psi^c)}{2}} &
i\sin\frac{\theta^c}{2}e^{\frac{i(\varphi^c-\psi^c)}{2}}\\
i\sin\frac{\theta^c}{2}e^{\frac{i(\psi^c-\varphi^c)}{2}} &
\cos\frac{\theta^c}{2}e^{-\frac{i(\varphi^c+\psi^c)}{2}}
\end{pmatrix}}=
\nonumber\\
%\]
%\begin{multline}
{\renewcommand{\arraystretch}{1.3}
\begin{pmatrix}
\left[\cos\frac{\theta}{2}\ch\frac{\tau}{2}+
i\sin\frac{\theta}{2}\sh\frac{\tau}{2}\right]
e^{\frac{\epsilon+\varepsilon+i(\varphi+\psi)}{2}} &
\left[\cos\frac{\theta}{2}\sh\frac{\tau}{2}+
i\sin\frac{\theta}{2}\ch\frac{\tau}{2}\right]
e^{\frac{\epsilon-\varepsilon+i(\varphi-\psi)}{2}} \\
%\end{array}
%\right.}\\
%{\renewcommand{\arraystretch}{1.1}
%\left.\begin{array}{c}
\left[\cos\frac{\theta}{2}\sh\frac{\tau}{2}+
i\sin\frac{\theta}{2}\ch\frac{\tau}{2}\right]
e^{\frac{\varepsilon-\epsilon+i(\psi-\varphi)}{2}} &
\left[\cos\frac{\theta}{2}\ch\frac{\tau}{2}+
i\sin\frac{\theta}{2}\sh\frac{\tau}{2}\right]
e^{\frac{-\epsilon-\varepsilon-i(\varphi+\psi)}{2}}
\end{pmatrix}},\label{SL1}
\end{gather}
since $\cos\frac{1}{2}(\theta-i\tau)=\cos\frac{\theta}{2}\ch\frac{\tau}{2}+
i\sin\frac{\theta}{2}\sh\frac{\tau}{2}$, and 
$\sin\frac{1}{2}(\theta-i\tau)=\sin\frac{\theta}{2}\ch\frac{\tau}{2}-
i\cos\frac{\theta}{2}\sh\frac{\tau}{2}$. It is easy to see that in this case
a Cartan decomposition for
$SL(2,\C)$  has the form
\begin{multline}
\mathfrak{g}(\varphi,\,\epsilon,\,\theta,\,\tau,\,\psi,\,\varepsilon)=\\[0.2cm]
%{\renewcommand{\arraystretch}{1.05}
\begin{pmatrix}
e^{i\frac{\varphi}{2}} & 0\\
0 & e^{-i\frac{\varphi}{2}}
\end{pmatrix}{\renewcommand{\arraystretch}{1.1}\!\!\!\!\begin{pmatrix}
e^{\frac{\epsilon}{2}} & 0\\
0 & e^{-\frac{\epsilon}{2}}
\end{pmatrix}}\!\!\!\!{\renewcommand{\arraystretch}{1.3}\begin{pmatrix}
\cos\frac{\theta}{2} & i\sin\frac{\theta}{2}\\
i\sin\frac{\theta}{2} & \cos\frac{\theta}{2}
\end{pmatrix}\!\!\!\!
%{\renewcommand{\arraystretch}{1.05}
\begin{pmatrix}
\ch\frac{\tau}{2} & \sh\frac{\tau}{2}\\
\sh\frac{\tau}{2} & \ch\frac{\tau}{2}
\end{pmatrix}}\!\!\!\!{\renewcommand{\arraystretch}{1.1}\begin{pmatrix}
e^{i\frac{\psi}{2}} & 0\\
0 & e^{-i\frac{\psi}{2}}
\end{pmatrix}}\!\!\!\!
%{\renewcommand{\arraystretch}{1.05}
\begin{pmatrix}
e^{\frac{\varepsilon}{2}} & 0\\
0 & e^{-\frac{\varepsilon}{2}}
\end{pmatrix}.\label{FUN}
\end{multline}
If we restrict the parameters $\im\varphi^c=\epsilon$, 
$\im\psi^c=\varepsilon$ within the limits $0\leq\epsilon\leq 2\pi$,
$-2\pi\leq\varepsilon<2\pi$, then we come to the following Cartan
decomposition
\begin{equation}\label{Car3}
\fg=\begin{pmatrix}
\alpha & \beta\\
-\bar{\beta} & \bar{\alpha}
\end{pmatrix}\!\begin{pmatrix}
\gamma & \delta\\
\bar{\delta} & \bar{\gamma}
\end{pmatrix},
\end{equation}
where 
$\begin{pmatrix}
\alpha & \beta\\
-\bar{\beta} & \bar{\alpha}
\end{pmatrix}\in SU(2)$ and
$\begin{pmatrix}
\gamma & \delta\\
\bar{\delta} & \bar{\gamma}
\end{pmatrix}\in SU(1,1)$.

The group $SL(2,\C)$ has six one-parameter subgroups
\begin{gather}
a_1(t)=\begin{pmatrix}
\cos\frac{t}{2} & i\sin\frac{t}{2}\\
i\sin\frac{t}{2} & \cos\frac{t}{2}
\end{pmatrix},\quad a_2(t)=\begin{pmatrix}
\cos\frac{t}{2} & -\sin\frac{t}{2}\\
\sin\frac{t}{2} & \cos\frac{t}{2}
\end{pmatrix},\quad a_3(t)=\begin{pmatrix}
e^{\frac{it}{2}} & 0\\
0 & e^{-\frac{it}{2}}
\end{pmatrix},\nonumber\\
b_1(t)=\begin{pmatrix}
\ch\frac{t}{2} & \sh\frac{t}{2}\\
\sh\frac{t}{2} & \ch\frac{t}{2}
\end{pmatrix},\quad b_2(t)=\begin{pmatrix}
\ch\frac{t}{2} & i\sh\frac{t}{2}\\
-i\sh\frac{t}{2} & \ch\frac{t}{2}
\end{pmatrix},\quad b_3(t)=\begin{pmatrix}
e^{\frac{t}{2}} & 0\\
0 & e^{-\frac{t}{2}}
\end{pmatrix}.\nonumber
\end{gather}
The tangent matrices $A_i$ and $B_i$ of these subgroups are defined
as follows
\begin{eqnarray}
A_1&=&\left.\frac{d a_1(t)}{dt}\right|_{t=0}=\frac{i}{2}\begin{pmatrix}
0 & 1\\
1 & 0
\end{pmatrix},\nonumber\\
A_2&=&\left.\frac{d a_2(t)}{dt}\right|_{t=0}=\frac{1}{2}\begin{pmatrix}
0 & -1\\
1 & 0
\end{pmatrix},\nonumber\\
A_3&=&\left.\frac{d a_3(t)}{dt}\right|_{t=0}=\frac{i}{2}\begin{pmatrix}
1 & 0\\
0 & -1
\end{pmatrix},\nonumber\\
B_1&=&\left.\frac{d b_1(t)}{dt}\right|_{t=0}=\frac{1}{2}\begin{pmatrix}
0 & 1\\
1 & 0
\end{pmatrix},\nonumber\\
B_2&=&\left.\frac{d b_2(t)}{dt}\right|_{t=0}=\frac{i}{2}\begin{pmatrix}
0 & 1\\
-1 & 0
\end{pmatrix},\nonumber\\
B_3&=&\left.\frac{d b_3(t)}{dt}\right|_{t=0}=\frac{1}{2}\begin{pmatrix}
1 & 0\\
0 &-1
\end{pmatrix}.\nonumber
\end{eqnarray}
The elements $\sA_i$ and $\sB_i$ form a basis of Lie algebra 
$\mathfrak{sl}(2,\C)$ and satisfy the relations
\begin{equation}\label{Com1}
\left.\begin{array}{lll}
\ld\sA_1,\sA_2\rd=\sA_3, & \ld\sA_2,\sA_3\rd=\sA_1, &
\ld\sA_3,\sA_1\rd=\sA_2,\\[0.1cm]
\ld\sB_1,\sB_2\rd=-\sA_3, & \ld\sB_2,\sB_3\rd=-\sA_1, &
\ld\sB_3,\sB_1\rd=-\sA_2,\\[0.1cm]
\ld\sA_1,\sB_1\rd=0, & \ld\sA_2,\sB_2\rd=0, &
\ld\sA_3,\sB_3\rd=0,\\[0.1cm]
\ld\sA_1,\sB_2\rd=\sB_3, & \ld\sA_1,\sB_3\rd=-\sB_2, & \\[0.1cm]
\ld\sA_2,\sB_3\rd=\sB_1, & \ld\sA_2,\sB_1\rd=-\sB_3, & \\[0.1cm]
\ld\sA_3,\sB_1\rd=\sB_2, & \ld\sA_3,\sB_2\rd=-\sB_1. &
\end{array}\right\}
\end{equation}
Denoting $\sI^{23}=\sA_1$, $\sI^{31}=\sA_2$,
$\sI^{12}=\sA_3$, and $\sI^{01}=\sB_1$, $\sI^{02}=\sB_2$, $\sI^{03}=\sB_3$
we can write the relations (\ref{Com1}) in a more compact form:
\[
\ld\sI^{\mu\nu},\sI^{\lambda\rho}\rd=\delta_{\mu\rho}\sI^{\lambda\nu}+
\delta_{\nu\lambda}\sI^{\mu\rho}-\delta_{\nu\rho}\sI^{\mu\lambda}-
\delta_{\mu\lambda}\sI^{\nu\rho}.
\]
Let us consider the operators
\begin{gather}
\sX_l=\frac{1}{2}i(\sA_l+i\sB_l),\quad\sY_l=\frac{1}{2}i(\sA_l-i\sB_l),
\label{SL25}\\
(l=1,2,3).\nonumber
\end{gather}
Using the relations (\ref{Com1}), we find that
\begin{equation}\label{Com2}
\ld\sX_k,\sX_l\rd=i\varepsilon_{klm}\sX_m,\quad
\ld\sY_l,\sY_m\rd=i\varepsilon_{lmn}\sY_n,\quad
\ld\sX_l,\sY_m\rd=0.
\end{equation}
Further, introducing generators of the form
\begin{equation}\label{SL26}
\left.\begin{array}{cc}
\sX_+=\sX_1+i\sX_2, & \sX_-=\sX_1-i\sX_2,\\[0.1cm]
\sY_+=\sY_1+i\sY_2, & \sY_-=\sY_1-i\sY_2,
\end{array}\right\}
\end{equation}
we see that in virtue of commutativity of the relations (\ref{Com2}) a
space of an irreducible finite--dimensional representation of the group
$SL(2,\C)$ can be spanned on the totality of $(2l+1)(2\dot{l}+1)$ basis
vectors $\mid l,m;\dot{l},\dot{m}\rangle$, where $l,m,\dot{l},\dot{m}$
are integer or half--integer numbers, $-l\leq m\leq l$,
$-\dot{l}\leq \dot{m}\leq \dot{l}$. Therefore,
\begin{eqnarray}
&&\sX_-\mid l,m;\dot{l},\dot{m}\rangle=
\sqrt{(l+m)(l-m+1)}\mid l,m-1,\dot{l},\dot{m}\rangle
\;\;(m>-l),\nonumber\\
&&\sX_+\mid l,m;\dot{l},\dot{m}\rangle=
\sqrt{(l-m)(l+m+1)}\mid l,m+1;\dot{l},\dot{m}\rangle
\;\;(m<l),\nonumber\\
&&\sX_3\mid l,m;\dot{l},\dot{m}\rangle=
m\mid l,m;\dot{l},\dot{m}\rangle,\nonumber\\
&&\sY_-\mid l,m;\dot{l},\dot{m}\rangle=
\sqrt{(\dot{l}+\dot{m})(\dot{l}-\dot{m}+1)}\mid l,m;\dot{l},\dot{m}-1
\rangle\;\;(\dot{m}>-\dot{l}),\nonumber\\
&&\sY_+\mid l,m;\dot{l},\dot{m}\rangle=
\sqrt{(\dot{l}-\dot{m})(\dot{l}+\dot{m}+1)}\mid l,m;\dot{l},\dot{m}+1
\rangle\;\;(\dot{m}<\dot{l}),\nonumber\\
&&\sY_3\mid l,m;\dot{l},\dot{m}\rangle=
\dot{m}\mid l,m;\dot{l},\dot{m}\rangle.\label{Waerden}
\end{eqnarray}
From the relations (\ref{Com2}) it follows that each of the sets of 
infinitesimal operators $\sX$ and $\sY$ generates the group $SU(2)$ and these
two groups commute with each other. Thus, from the relations (\ref{Com2})
and (\ref{Waerden}) it follows that the group $SL(2,\C)$, in essence,
is equivalent locally to the group $SU(2)\otimes SU(2)$. In contrast to the
Gel'fand--Naimark representation\index{representation!Gel'fand-Naimark}
for the Lorentz group \cite{GMS,Nai58},
which does not find a broad application in physics,
a representation (\ref{Waerden}) is a most useful in theoretical physics
(see, for example, \cite{AB,Sch61,RF,Ryd85}). This representation for the
Lorentz group was first given by Van der Waerden in his brilliant book
\cite{Wa32}.\index{representation!Van der Waerden}
It should be noted here that the representation basis, defined by the
formulae (\ref{SL25})--(\ref{Waerden}), has an evident physical meaning.
For example, in the case of $(1,0)\oplus(0,1)$--representation space
there is an analogy with the photon spin states. Namely, the operators
$\sX$ and $\sY$ correspond to the right and left polarization states of the
photon. For that reason we will call the canonical basis consisting of the
vectors $\mid lm;\dot{l}\dot{m}\rangle$ as
{\it a helicity basis}.
Infinitesimal operators of $SL(2,\C)$ in the helicity basis have a very simple
form
\begin{eqnarray}
\sA_1\mid l,m;\do{l},\dot{m}\rangle&=&-\frac{i}{2}\boldsymbol{\alpha}^l_m
\mid l,m-1;\dot{l},\dot{m}\rangle-
\frac{i}{2}\boldsymbol{\alpha}^l_{m+1}\mid l,m+1;\dot{l}\dot{m}\rangle,
\nonumber\\
\sA_2\mid l,m;\dot{l},\dot{m}\rangle&=&\frac{1}{2}\boldsymbol{\alpha}^l_m
\mid l,m-1;\dot{l},\dot{m}\rangle-
\frac{1}{2}\boldsymbol{\alpha}^l_{m+1}
\mid l,m+1;\dot{l},\dot{m}\rangle,\label{OpA}\\
\sA_3\mid l,m;\dot{l},\dot{m}\rangle&=&-im
\mid l,m;\dot{l},\dot{m}\rangle,\nonumber
\end{eqnarray}
\begin{eqnarray}
\sB_1\mid l,m;\dot{l},\dot{m}\rangle&=&-\frac{1}{2}\boldsymbol{\alpha}^l_m
\mid l,m-1;\dot{l},\dot{m}\rangle-
\frac{1}{2}\boldsymbol{\alpha}^l_{m+1}
\mid l,m+1;\dot{l},\dot{m}\rangle,\nonumber\\
\sB_2\mid l,m;\dot{l},\dot{m}\rangle&=&-\frac{i}{2}\boldsymbol{\alpha}^l_m
\mid l,m-1;\dot{l},\dot{m}\rangle+
\frac{i}{2}\boldsymbol{\alpha}^l_{m+1}
\mid l,m+1;\dot{l},\dot{m}\rangle,\label{OpB}\\
\sB_3\mid l,m;\dot{l},\dot{m}\rangle&=&-m\mid l,m;\dot{l},\dot{m}\rangle,
\nonumber
\end{eqnarray}
\begin{eqnarray}
\widetilde{\sA}_1\mid l,m;\dot{l},\dot{m}\rangle&=&
-\frac{i}{2}\boldsymbol{\alpha}^{\dot{l}}_{\dot{m}}
\mid l,m;\dot{l},\dot{m}-1\rangle-
\frac{i}{2}\boldsymbol{\alpha}^{\dot{l}}_{\dot{m}+1}
\mid l,m;\dot{l},\dot{m}+1\rangle,\nonumber\\
\widetilde{\sA}_2\mid l,m;\dot{l},\dot{m}\rangle&=&
\frac{1}{2}\boldsymbol{\alpha}^{\dot{l}}_{\dot{m}}
\mid l,m;\dot{l},\dot{m}-1\rangle-
\frac{1}{2}\boldsymbol{\alpha}^{\dot{l}}_{\dot{m}+1}
\mid l,m;\dot{l},\dot{m}+1\rangle,\label{DopA}\\
\widetilde{\sA}_3\mid l,m;\dot{l},\dot{m}\rangle&=&
-i\dot{m}\mid l,m;\dot{l},\dot{m}\rangle,\nonumber
\end{eqnarray}
\begin{eqnarray}
\widetilde{\sB}_1\mid l,m;\dot{l},\dot{m}\rangle&=&
\frac{1}{2}\boldsymbol{\alpha}^{\dot{l}}_{\dot{m}}
\mid l,m;\dot{l},\dot{m}-1\rangle+
\frac{1}{2}\boldsymbol{\alpha}^{\dot{l}}_{\dot{m}+1}
\mid l,m;\dot{l},\dot{m}+1\rangle,\nonumber\\
\widetilde{\sB}_2\mid l,m;\dot{l},\dot{m}\rangle&=&
\frac{i}{2}\boldsymbol{\alpha}^{\dot{l}}_{\dot{m}}
\mid l,m;\dot{l},\dot{m}-1\rangle-
\frac{i}{2}\boldsymbol{\alpha}^{\dot{l}}_{\dot{m}+1}
\mid l,m;\dot{l},\dot{m}+1\rangle,\label{DopB}\\
\widetilde{\sB}_3\mid l,m;\dot{l},\dot{m}\rangle&=&
-\dot{m}\mid l,m;\dot{l},\dot{m}\rangle,\nonumber
\end{eqnarray}
where
\[
\boldsymbol{\alpha}^l_m=\sqrt{(l+m)(l-m+1)}.
\]
The representation of the group $SL(2,\C)$ in the space $\Sym(k,r)$ has a form
\begin{eqnarray}
T_gq(\fz,\bar{\fz})&=&\frac{1}{z^k_1\,\bar{z}^r_1}T_g\left[
z^k_1\bar{z}^r_1q\left(\frac{z_0}{z_1},\frac{\bar{z}_0}{\bar{z}_1}\right)
\right]= \nonumber\\
&=&(\gamma\fz+\delta)^k(\overset{\ast}{\gamma}\bar{\fz}+
\overset{\ast}{\delta})^rq\left(\frac{\alpha\fz+\beta}{\gamma\fz+\delta},
\frac{\overset{\ast}{\alpha}\bar{\fz}+\overset{\ast}{\beta}}
{\overset{\ast}{\gamma}\bar{\fz}+\overset{\ast}{\delta}}\right),
\label{Rep}
\end{eqnarray}
where
\[
\fz=\frac{z_0}{z_1},\quad\bar{\fz}=\frac{\bar{z}_0}{\bar{z}_1}.
\]
In turn, every space $\Sym_{(k,r)}$
can be represented by 
a space of polynomials\index{space!of polynomials}
\begin{gather}
p(z_0,z_1,\bar{z}_0,\bar{z}_1)=\sum_{\substack{(\alpha_1,\ldots,\alpha_k)\\
(\dot{\alpha}_1,\ldots,\dot{\alpha}_r)}}\frac{1}{k!\,r!}
a^{\alpha_1\cdots\alpha_k\dot{\alpha}_1\cdots\dot{\alpha}_r}
z_{\alpha_1}\cdots z_{\alpha_k}\bar{z}_{\dot{\alpha}_1}\cdots
\bar{z}_{\dot{\alpha}_r}.\label{SF}\\
(\alpha_i,\dot{\alpha}_i=0,1)\nonumber
\end{gather}
where the numbers 
$a^{\alpha_1\cdots\alpha_k\dot{\alpha}_1\cdots\dot{\alpha}_r}$
are unaffected at the permutations of indices. 
The expressions (\ref{SF}) can be understood as {\it functions on the
Lorentz group}.

The infinitesimal operators $\sA_i$ and $\sB_i$ can be written via the
complex Euler angles (\ref{CEA}) as follows
(for more details see
\cite{Var022})
\begin{eqnarray}
\sA_1&=&\cos\psi^c\frac{\partial}{\partial\theta}+
\frac{\sin\psi^c}{\sin\theta^c}\frac{\partial}{\partial\varphi}-
\ctg\theta^c\sin\psi^c\frac{\partial}{\partial\psi},\label{SL12}\\
\sA_2&=&-\sin\psi^c\frac{\partial}{\partial\theta}+
\frac{\cos\psi^c}{\sin\theta^c}\frac{\partial}{\partial\varphi}-
\ctg\theta^c\cos\psi^c\frac{\partial}{\partial\psi},\label{SL24'}\\
\sA_3&=&\frac{\partial}{\partial\psi},\label{SL8}\\
\sB_1&=&\cos\psi^c\frac{\partial}{\partial\tau}+
\frac{\sin\psi^c}{\sin\theta^c}\frac{\partial}{\partial\epsilon}-
\ctg\theta^c\sin\psi^c\frac{\partial}{\partial\varepsilon},\label{SL13}\\
\sB_2&=&-\sin\psi^c\frac{\partial}{\partial\tau}+
\frac{\cos\psi^c}{\sin\theta^c}\frac{\partial}{\partial\epsilon}-
\ctg\theta^c\cos\psi^c\frac{\partial}{\partial\varepsilon},\label{SL24''}\\
\sB_3&=&\frac{\partial}{\partial\varepsilon}.\label{SL8'}
\end{eqnarray}
It is easy to verify that operators $\sA_i$, $\sB_i$,
defined by the formulae (\ref{SL12})--(\ref{SL8'}),
are satisfy the commutation relations (\ref{Com1}).

Further, taking into account the expressions (\ref{SL12})--(\ref{SL8'}),
we can write
the operators (\ref{SL25}) in the form
\begin{eqnarray}
\sX_1&=&\cos\psi^c\frac{\partial}{\partial\theta^c}+
\frac{\sin\psi^c}{\sin\theta^c}\frac{\partial}{\partial\varphi^c}-
\ctg\theta^c\sin\psi^c\frac{\partial}{\partial\psi^c},\label{X1}\\
\sX_2&=&-\sin\psi^c\frac{\partial}{\partial\theta^c}+
\frac{\cos\psi^c}{\sin\theta^c}\frac{\partial}{\partial\varphi^c}-
\ctg\theta^c\cos\psi^c\frac{\partial}{\partial\psi^c},\label{X2}\\
\sX_3&=&\frac{\partial}{\partial\psi^c},\label{X3}\\
\sY_1&=&\cos\dot{\psi}^c\frac{\partial}{\partial\dot{\theta}^c}+
\frac{\sin\dot{\psi}^c}{\sin\dot{\theta}^c}
\frac{\partial}{\partial\dot{\varphi}^c}-
\ctg\dot{\theta}^c\sin\dot{\psi}^c\frac{\partial}{\partial\dot{\psi}^c},
\label{Y1}\\
\sY_2&=&-\sin\dot{\psi}^c\frac{\partial}{\partial\dot{\theta}^c}+
\frac{\cos\dot{\psi}^c}{\sin\dot{\theta}^c}
\frac{\partial}{\partial\dot{\varphi}^c}-
\ctg\dot{\theta}^c\cos\dot{\psi}^c\frac{\partial}{\partial\dot{\psi}^c},
\label{Y2}\\
\sY_3&=&\frac{\partial}{\partial\dot{\psi}^c},\label{Y3}
\end{eqnarray}
where
\begin{gather}
\frac{\partial}{\partial\theta^c}=\frac{1}{2}\left(
\frac{\partial}{\partial\theta}+i\frac{\partial}{\partial\tau}\right),\;\;
\frac{\partial}{\partial\varphi^c}=\frac{1}{2}\left(
\frac{\partial}{\partial\varphi}+i\frac{\partial}{\partial\epsilon}\right),\;\;
\frac{\partial}{\partial\psi^c}=\frac{1}{2}\left(
\frac{\partial}{\partial\psi}+i\frac{\partial}{\partial\varepsilon}\right),
\nonumber\\
\frac{\partial}{\partial\dot{\theta}^c}=\frac{1}{2}\left(
\frac{\partial}{\partial\theta}-i\frac{\partial}{\partial\tau}\right),\;\;
\frac{\partial}{\partial\dot{\varphi}^c}=\frac{1}{2}\left(
\frac{\partial}{\partial\varphi}-i\frac{\partial}{\partial\epsilon}\right),\;\;
\frac{\partial}{\partial\dot{\psi}^c}=\frac{1}{2}\left(
\frac{\partial}{\partial\psi}-i\frac{\partial}{\partial\varepsilon}\right),
\nonumber
\end{gather}
On the group $SL(2,\C)$ there exist the following Laplace-Beltrami operators:
\begin{eqnarray}
\sX^2&=&\sX^2_1+\sX^2_2+\sX^2_3=\frac{1}{4}(\sA^2-\sB^2+2i\sA\sB),\nonumber\\
\sY^2&=&\sY^2_1+\sY^2_2+\sY^2_3=
\frac{1}{4}(\widetilde{\sA}^2-\widetilde{\sB}^2-
2i\widetilde{\sA}\widetilde{\sB}).\label{KO}
\end{eqnarray}
At this point, we see that operators (\ref{KO}) contain the well known
Casimir operators $\sA^2-\sB^2$, $\sA\sB$ of the Lorentz group.
Substituting (\ref{X1})-(\ref{Y3}) into (\ref{KO}), we obtain 
an Euler parametrization of the Laplace-Beltrami operators:
\begin{eqnarray}
\sX^2&=&\frac{\partial^2}{\partial\theta^c{}^2}+
\ctg\theta^c\frac{\partial}{\partial\theta^c}+\frac{1}{\sin^2\theta^c}\left[
\frac{\partial^2}{\partial\varphi^c{}^2}-
2\cos\theta^c\frac{\partial}{\partial\varphi^c}
\frac{\partial}{\partial\psi^c}+
\frac{\partial^2}{\partial\psi^c{}^2}\right],\nonumber\\
\sY^2&=&\frac{\partial^2}{\partial\dot{\theta}^c{}^2}+
\ctg\dot{\theta}^c\frac{\partial}{\partial\dot{\theta}^c}+
\frac{1}{\sin^2\dot{\theta}^c}\left[
\frac{\partial^2}{\partial\dot{\varphi}^c{}^2}-
2\cos\dot{\theta}^c\frac{\partial}{\partial\dot{\varphi}^c}
\frac{\partial}{\partial\dot{\psi}^c}+
\frac{\partial^2}{\partial\dot{\psi}^c{}^2}\right].\label{KO2}
\end{eqnarray}
Matrix elements $t^l_{mn}(\fg)=\fM^l_{mn}(\varphi^c,\theta^c,\psi^c)$ of 
irreducible representations of $SL(2,\C)$ are
eigenfunctions of the operators (\ref{KO2}):
\begin{eqnarray}
\left[\sX^2+l(l+1)\right]\fM^l_{mn}(\varphi^c,\theta^c,\psi^c)&=&0,\nonumber\\
\left[\sY^2+\dot{l}(\dot{l}+1)\right]\fM^{\dot{l}}_{\dot{m}\dot{n}}
(\dot{\varphi}^c,\dot{\theta}^c,\dot{\psi}^c)&=&0,\label{EQ}
\end{eqnarray}
where
\begin{eqnarray}
\fM^l_{mn}(\varphi^c,\theta^c,\psi^c)&=&e^{-i(m\varphi^c+n\psi^c)}
Z^l_{mn}(\cos\theta^c),\nonumber\\
\fM^{\dot{l}}_{\dot{m}\dot{n}}(\dot{\varphi}^c,\dot{\theta}^c,\dot{\psi}^c)&=&
e^{-i(\dot{m}\dot{\varphi}^c+\dot{n}\dot{\varphi}^c)}
Z^{\dot{l}}_{\dot{m}\dot{n}}(\cos\dot{\theta}^c).\label{HF3'}
\end{eqnarray}
Substituting the hyperspherical functions (\ref{HF3'}) into (\ref{EQ}) and
taking into account the operators (\ref{KO2}), we obtain 
\begin{eqnarray}
\left[\frac{d^2}{d\theta^c{}^2}+\ctg\theta^c\frac{d}{d\theta^c}-
\frac{m^2+n^2-2mn\cos\theta^c}{\sin^2\theta^c}+l(l+1)\right]
Z^l_{mn}(\cos\theta^c)&=&0,\nonumber\\
\left[\frac{d^2}{d\dot{\theta}^c{}^2}+\ctg\dot{\theta}^c
\frac{d}{d\dot{\theta}^c}-\frac{\dot{m}^2+\dot{n}^2-
2\dot{m}\dot{n}\cos\dot{\theta}^c}{\sin^2\dot{\theta}^c}+
\dot{l}(\dot{l}+1)\right]Z^{\dot{l}}_{\dot{m}\dot{n}}(\cos\dot{\theta}^c)
&=&0.\nonumber
\end{eqnarray}
Finally, after substitutions $z=\cos\theta^c$ and 
$\overset{\ast}{z}=\cos\dot{\theta}^c$, we come to the following
differential equations (a complex analog of the Legendre equations)
\begin{eqnarray}
\left[(1-z^2)\frac{d^2}{dz^2}-2z\frac{d}{dz}-
\frac{m^2+n^2-2mnz}{1-z^2}+l(l+1)\right]Z^l_{mn}(z)&=&0,\label{Leg1}\\
\left[(1-\overset{\ast}{z}{}^2)\frac{d^2}{d\overset{\ast}{z}{}^2}-
2\overset{\ast}{z}\frac{d}{d\overset{\ast}{z}}-
\frac{\dot{m}^2+\dot{n}^2-2\dot{m}\dot{n}\overset{\ast}{z}}
{1-\overset{\ast}{z}{}^2}+\dot{l}(\dot{l}+1)\right]
Z^{\dot{l}}_{\dot{m}\dot{n}}(\overset{\ast}{z})&=&0.\label{Leg2}
\end{eqnarray}
The latter equations have three singular points $-1$, $+1$, $\infty$.
Solutions of (\ref{Leg1}) have the form
\begin{multline}
Z^l_{mn}=
\sum^l_{k=-l}i^{m-k}
\sqrt{\Gamma(l-m+1)\Gamma(l+m+1)\Gamma(l-k+1)\Gamma(l+k+1)}\times\\
\cos^{2l}\frac{\theta}{2}\tg^{m-k}\frac{\theta}{2}\times\\[0.2cm]
\sum^{\min(l-m,l+k)}_{j=\max(0,k-m)}
\frac{i^{2j}\tg^{2j}\dfrac{\theta}{2}}
{\Gamma(j+1)\Gamma(l-m-j+1)\Gamma(l+k-j+1)\Gamma(m-k+j+1)}\times\\[0.2cm]
\sqrt{\Gamma(l-n+1)\Gamma(l+n+1)\Gamma(l-k+1)\Gamma(l+k+1)}
\ch^{2l}\frac{\tau}{2}\tnh^{n-k}\frac{\tau}{2}\times\\[0.2cm]
\sum^{\min(l-n,l+k)}_{s=\max(0,k-n)}
\frac{\tnh^{2s}\dfrac{\tau}{2}}
{\Gamma(s+1)\Gamma(l-n-s+1)\Gamma(l+k-s+1)\Gamma(n-k+s+1)}.\label{HS}
\end{multline}
We will call the functions $Z^l_{mn}$ in (\ref{HS}) as
{\it hyperspherical functions}\footnote{The hyperspherical functions (or hyperspherical
harmonics) are known in mathematics for a long time 
(see, for example, \cite{Bat2}). These functions are generalizations
of the three-dimensional spherical functions on the case of $n$-dimensional
euclidean spaces. For that reason we retain this name (hyperspherical
functions) for the case of pseudo-euclidean spaces.}. 
This form of the functions $Z^l_{mn}$ immediately follows from the
Cartan decompositions (\ref{FUN}) and (\ref{Car3}). Thus, matrix elements
$t^l_{mn}(\fg)$ are expressed by means of the function
({\it a generalized 
hyperspherical function})
\begin{equation}\label{HS2}
\fT^l_{mn}(\mathfrak{g})=e^{-m(\epsilon+i\varphi)}Z^l_{mn}(\cos\theta^c)
e^{-n(\varepsilon+i\psi)},
\end{equation}
where
\begin{equation}\label{HS3}
Z^l_{mn}(\cos\theta^c)
=\sum^l_{k=-l}P^l_{mk}(\cos\theta)\mathfrak{P}^l_{kn}(\ch\tau),
\end{equation}
here $P^l_{mn}(\cos\theta)$ is a 
generalized spherical 
function\index{function!spherical!generalized} on the
group $SU(2)$ (see (\ref{Mat1}), (\ref{Mat12})), 
and $\mathfrak{P}^l_{mn}(\ch\tau)$ is an analog of
the generalized spherical function for the group $SU(1,1)$ 
(see (\ref{Mat1'}), (\ref{Jacobi})).

Further, using (\ref{Mat2}) and (\ref{Mat2'}), we can write the
hyperspherical functions (\ref{HS3}) via the hypergeometric series:
\begin{multline}
Z^l_{mn}(\cos\theta^c)=\cos^{2l}\frac{\theta}{2}\ch^{2l}\frac{\tau}{2}
\sum^l_{k=-l}i^{m-k}\tg^{m-k}\frac{\theta}{2}
\tnh^{n-k}\frac{\tau}{2}\times\\[0.2cm]
\hypergeom{2}{1}{m-l+1,1-l-k}{m-k+1}{i^2\tg^2\dfrac{\theta}{2}}
\hypergeom{2}{1}{n-l+1,1-l-k}{n-k+1}{\tnh^2\dfrac{\tau}{2}}.\label{HS1}
\end{multline}
It is obvious that solutions $Z^{\dot{l}}_{\dot{m}\dot{n}}$ of the
equation (\ref{Leg2}) have the same structure.

Further, from (\ref{HS1}) we see that the function $Z^l_{mn}$ depends on
two variables $\theta$ and $\tau$. Therefore, using Bateman factorization we can
express the hyperspherical functions $Z^l_{mn}$ via Appell functions
$F_1$--$F_4$ (hypergeometric series of two variables \cite{AK26,Bat}).
For more details about this relationship see \cite{Var042}.

Using the formula (\ref{HS}), let us find explicit expressions for the
matrices $T_l(\theta,\tau)$ of the finite--dimensional representations of
$SL(2,\C)$ at $l=0,\frac{1}{2},1$:
\begin{gather}
T_0(\theta,\tau)=1,\nonumber\\[0.3cm]
T_{\frac{1}{2}}(\theta,\tau)=\begin{pmatrix}
Z^{\frac{1}{2}}_{-\frac{1}{2}-\frac{1}{2}} &
Z^{\frac{1}{2}}_{\frac{1}{2}-\frac{1}{2}}\\
Z^{\frac{1}{2}}_{-\frac{1}{2}\frac{1}{2}} &
Z^{\frac{1}{2}}_{\frac{1}{2}\frac{1}{2}}
\end{pmatrix}=\nonumber\\[0.3cm]
{\renewcommand{\arraystretch}{1.3}
\begin{pmatrix}
\cos\frac{\theta}{2}\ch\frac{\tau}{2}+
i\sin\frac{\theta}{2}\sh\frac{\tau}{2} &
\cos\frac{\theta}{2}\sh\frac{\tau}{2}+
i\sin\frac{\theta}{2}\ch\frac{\tau}{2} \\
\cos\frac{\theta}{2}\sh\frac{\tau}{2}+
i\sin\frac{\theta}{2}\ch\frac{\tau}{2} &
\cos\frac{\theta}{2}\ch\frac{\tau}{2}+
i\sin\frac{\theta}{2}\sh\frac{\tau}{2} 
\end{pmatrix}},\label{T1}\\[0.4cm]
T_1(\theta,\tau)=\begin{pmatrix}
Z^1_{-1-1} & Z^1_{-10} & Z^1_{-11}\\
Z^1_{0-1} & Z^1_{00} & Z^1_{01}\\
Z^1_{1-1} & Z^1_{10} & Z^1_{11}
\end{pmatrix}=\nonumber
\end{gather}
\begin{multline}
{\renewcommand{\arraystretch}{1.8}\left(\begin{array}{cc}\s
\cos^2\tfrac{\theta}{2}\ch^2\tfrac{\tau}{2}+\tfrac{i\sin\theta\sh\tau}{2}-
\sin^2\tfrac{\theta}{2}\sh^2\tfrac{\tau}{2} &\s
\tfrac{1}{\sqrt{2}}(\cos\theta\sh\tau+i\sin\theta\ch\tau) \\
\s\tfrac{1}{\sqrt{2}}(\cos\theta\sh\tau+i\sin\theta\ch\tau) &
\s\cos\theta\ch\tau+i\sin\theta\sh\tau \\
\s\cos^2\tfrac{\theta}{2}\sh^2\tfrac{\tau}{2}+\tfrac{i\sin\theta\sh\tau}{2}-
\sin^2\tfrac{\theta}{2}\ch^2\tfrac{\tau}{2} &
\s
\tfrac{1}{\sqrt{2}}(\cos\theta\sh\tau+i\sin\theta\ch\tau) 
\end{array}\right.}\\
{\renewcommand{\arraystretch}{1.8}\left.\begin{array}{c}\s
\cos^2\tfrac{\theta}{2}\sh^2\tfrac{\tau}{2}+\tfrac{i\sin\theta\sh\tau}{2}-
\sin^2\tfrac{\theta}{2}\ch^2\tfrac{\tau}{2} \\
\s\tfrac{1}{\sqrt{2}}(\cos\theta\sh\tau+i\sin\theta\ch\tau) \\
\s\cos^2\tfrac{\theta}{2}\ch^2\tfrac{\tau}{2}+\tfrac{i\sin\theta\sh\tau}{2}-
\sin^2\tfrac{\theta}{2}\sh^2\tfrac{\tau}{2} 
\end{array}\right).}\label{T2}
\end{multline}

\subsection{Associated hyperspherical functions and a two-dimensional complex
sphere}
Let us consider now associated hyperspherical functions of the
representation $T_\chi(\fg)$, $\chi=(l,0)$, that is, the matrix elements
$t^l_{m0}(\fg)$ standing in one column with the function $t^l_{00}(\fg)$.
In this case from (\ref{HS2}) we have
\begin{equation}\label{Associated}
t^l_{m0}(\fg)=\fM^l_{m0}(\fg)=e^{-m(\epsilon+i\varphi)}Z^l_{m0}(\theta,\tau).
\end{equation}
Hence it follows that matrix elements $t^l_{m0}(\fg)$ do not depend on
the Euler angles $\varepsilon$ and $\psi$, that is, $t^l_{m0}(\fg)$
are constant on the left adjacency classes formed by the subgroup
$\Omega^c_\psi$ of the diagonal matrices
$\begin{pmatrix} e^{\frac{i\psi^c}{2}} & 0\\ 0 & e^{-\frac{i\psi^c}{2}}
\end{pmatrix}$. Therefore,
\[
t^l_{m0}(\fg h)=t^l_{m0}(\fg),\quad h\in\Omega^c_\psi.
\]
We will denote the functions $Z^{m0}(\theta,\tau)$ via $Z^m_l(\theta,\tau)$.
From (\ref{HS}) we obtain an explicit expression for the
{\it associated hyperspherical function} $Z^m_l(\theta,\tau)$:
\begin{multline}
Z^m_l(\theta,\tau)=
\sum^l_{k=-l}i^{m-k}
\sqrt{\Gamma(l-m+1)\Gamma(l+m+1)\Gamma(l-k+1)\Gamma(l+k+1)}\times\\
\cos^{2l}\frac{\theta}{2}\tg^{m-k}\frac{\theta}{2}\times\\[0.2cm]
\sum^{\min(l-m,l+k)}_{j=\max(0,k-m)}
\frac{i^{2j}\tg^{2j}\dfrac{\theta}{2}}
{\Gamma(j+1)\Gamma(l-m-j+1)\Gamma(l+k-j+1)\Gamma(m-k+j+1)}\times\\[0.2cm]
\Gamma(l+1)\sqrt{\Gamma(l-k+1)\Gamma(l+k+1)}
\ch^{2l}\frac{\tau}{2}\tnh^{-k}\frac{\tau}{2}\times\\[0.2cm]
\sum^{\min(l,l+k)}_{s=\max(0,k)}
\frac{\tnh^{2s}\dfrac{\tau}{2}}
{\Gamma(s+1)\Gamma(l-s+1)\Gamma(l+k-s+1)\Gamma(s-k+1)}.\label{AHS}
\end{multline}
Further, from (\ref{HS1}) it follows that
\begin{multline}
Z^l_{mn}=\cos^{2l}\frac{\theta}{2}\ch^{2l}\frac{\tau}{2}
\sum^l_{k=-l}i^{m-k}\tg^{m-k}\frac{\theta}{2}
\tnh^{-k}\frac{\tau}{2}\times\\[0.2cm]
\hypergeom{2}{1}{m-l+1,1-l-k}{m-k+1}{i^2\tg^2\dfrac{\theta}{2}}
\hypergeom{2}{1}{-l+1,1-l-k}{-k+1}{\tnh^2\dfrac{\tau}{2}}.\label{AHS1}
\end{multline}
Associated hyperspherical functions admit a very elegant geometric
interpretation, namely, they are the functions on the surface of the
two-dimensional complex sphere. 
Indeed, let us construct in $\C^3$ the two--dimensional complex sphere from the
quantities $z_k=x_k+iy_k$, $\overset{\ast}{z}_k=x_k-iy_k$
as follows (see Figure 1)
\begin{equation}\label{CS}
\bz^2=z^2_1+z^2_2+z^2_3=\bx^2-\by^2+2i\bx\by=r^2
\end{equation}
and its complex conjugate (dual) sphere
\begin{equation}\label{DS}
\overset{\ast}{\bz}{}^2=\overset{\ast}{z}_1{}^2+\overset{\ast}{z}_2{}^2+
\overset{\ast}{z}_3{}^2=\bx^2-\by^2-2i\bx\by=\overset{\ast}{r}{}^2.
\end{equation}
For more details about the two-dimensional complex sphere see
\cite{Hus70,HS70,SH70}.
It is well-known that both quantities $\bx^2-\by^2$, $\bx\by$ are
invariant with respect to the Lorentz transformations, since a surface of
the complex sphere is invariant 
(Casimir operators\index{operator!Casimir} of the Lorentz group are
constructed from such quantities, see also (\ref{KO})).
Moreover, since the real and imaginary parts of the complex two-sphere
transform like the electric and magnetic fields, respectively,
the invariance of $\bz^2\sim(\bE+i\bB)^2$ under proper Lorentz
transformations is evident. At this point, the quantities
$\bx^2-\by^2$, $\bx\by$ are similar to the well known electromagnetic
invariants $E^2-B^2$, $\bE\bB$. This intriguing relationship between
the Laplace-Beltrami operators (\ref{KO}), Casimir operators of the
Lorentz group and electromagnetic invariants
$E^2-B^2\sim\bx^2-\by^2$, $\bE\bB\sim\bx\by$ leads naturally to a
Riemann-Silberstein representation of the electromagnetic field
(see the section \ref{Sec:Max}). In other words, the two-dimensional sphere,
considered as a homogeneous space of the Poincar\'{e} group, is the most
suitable arena for the subsequent investigations in quantum electrodynamics.
It is easy to see that three--dimensional complex space $\C^3$ is
isometric to a real space $\R^{3,3}$ with a basis
$\left\{i\e_1,i\e_2,i\e_3,\e_4,\e_5,\e_6\right\}$. 
\begin{figure}
\[
\unitlength=0.3mm
%\linethickness{0.4pt}
\begin{picture}(100.00,175.00)(0,25)
\put(50,50){\vector(0,1){51}}
\put(50,150){\vector(0,-1){48}}
\put(47,49.5){$\bullet$}
\put(47,150){$\bullet$}
\put(54,75){$r^\ast$}
\put(54,125){$r$}
\put(58,13){$\bullet$}
\put(58,183){$\bullet$}
\put(60,15){\vector(1,1){23}}
\put(60,15){\vector(1,-1){23}}
\put(60,185){\vector(1,1){23}}
\put(60,185){\vector(1,-1){23}}
\put(46,10){$P^\prime$}
\put(46,180){$P$}
\put(-95,50){$\mathfrak{M}_{\dot{l}}^{\dot{m}}
(\dot{\varphi}^c,\dot{\theta}^c,0)$}
\put(-95,150){$\mathfrak{M}_{l}^{m}(\varphi^c,\theta^c,0)$}
\put(85,-12){$\boldsymbol{e}_{\dot{\varphi}^c}$}
\put(80,210){$\boldsymbol{e}_{\varphi^c}$}
\put(80,30){$\boldsymbol{e}_{\dot{\theta}^c}$}
\put(80,170){$\boldsymbol{e}_{\theta^c}$}
\put(105,18){$\dot{\varphi}^c=\varphi+i\epsilon$}
\put(105,2){$\dot{\theta}^c=\theta+i\tau$}
\put(105,198){$\varphi^c=\varphi-i\epsilon$}
\put(105,182){$\theta^c=\theta-i\tau$}
%ellips
\put(50,34){$\cdot$}
\put(49.5,34){$\cdot$}
\put(49,34){$\cdot$}
\put(48.5,34){$\cdot$}
\put(48,34.01){$\cdot$}
\put(47.5,34.02){$\cdot$}
\put(47,34.03){$\cdot$}
\put(46.5,34.04){$\cdot$}
\put(46,34.05){$\cdot$}
\put(45.5,34.06){$\cdot$}
\put(45,34.08){$\cdot$}
\put(44.5,34.10){$\cdot$}
\put(44,34.11){$\cdot$}
\put(43.5,34.13){$\cdot$}
\put(43,34.16){$\cdot$}
\put(42.5,34.18){$\cdot$}
\put(42,34.20){$\cdot$}
\put(41.5,34.23){$\cdot$}
\put(41,34.26){$\cdot$}
\put(40.5,34.29){$\cdot$}
\put(40,34.32){$\cdot$}
\put(39.5,34.36){$\cdot$}
\put(39,34.39){$\cdot$}
\put(38.5,34.43){$\cdot$}
\put(38,34.47){$\cdot$}
\put(37.5,34.51){$\cdot$}
\put(37,34.55){$\cdot$}
\put(36.5,34.59){$\cdot$}
\put(36,34.64){$\cdot$}
\put(35.5,34.69){$\cdot$}
\put(35,34.74){$\cdot$}
\put(34.5,34.79){$\cdot$}
\put(34,34.84){$\cdot$}
\put(33.5,34.90){$\cdot$}
\put(33,34.95){$\cdot$}
\put(32.5,35.01){$\cdot$}
\put(32,35.07){$\cdot$}
\put(31.5,35.13){$\cdot$}
\put(31,35.20){$\cdot$}
\put(30.5,35.27){$\cdot$}
\put(30,35.33){$\cdot$}
\put(29.5,35.40){$\cdot$}
\put(29,35.48){$\cdot$}
\put(28.5,35.55){$\cdot$}
\put(28,35.63){$\cdot$}
\put(27.5,35.71){$\cdot$}
\put(27,35.79){$\cdot$}
\put(26.5,35.88){$\cdot$}
\put(26,35.96){$\cdot$}
\put(25.5,36.05){$\cdot$}
\put(25,36.14){$\cdot$}
\put(24.5,36.24){$\cdot$}
\put(24,36.33){$\cdot$}
\put(23.5,36.43){$\cdot$}
\put(23,36.53){$\cdot$}
\put(22.5,36.64){$\cdot$}
\put(22,36.74){$\cdot$}
\put(21.5,36.85){$\cdot$}
\put(21,36.97){$\cdot$}
\put(20.5,37.08){$\cdot$}
\put(20,37.20){$\cdot$}
\put(19.5,37.32){$\cdot$}
\put(19,37.45){$\cdot$}
\put(18.5,37.57){$\cdot$}
\put(18,37.70){$\cdot$}
\put(17.5,37.84){$\cdot$}
\put(17,37.98){$\cdot$}
\put(16.5,38.12){$\cdot$}
\put(16,38.27){$\cdot$}
\put(15.5,38.42){$\cdot$}
\put(15,38.57){$\cdot$}
\put(14.5,38.73){$\cdot$}
\put(14,38.90){$\cdot$}
\put(13.5,39.06){$\cdot$}
\put(13,39.24){$\cdot$}
\put(12.5,39.42){$\cdot$}
\put(12,39.60){$\cdot$}
\put(11.5,39.79){$\cdot$}
\put(11,39.99){$\cdot$}
\put(10.5,40.19){$\cdot$}
\put(10,40.40){$\cdot$}
\put(9.5,40.62){$\cdot$}
\put(9,40.84){$\cdot$}
\put(8.5,41.07){$\cdot$}
\put(8,41.32){$\cdot$}
\put(7.5,41.57){$\cdot$}
\put(7,41.83){$\cdot$}
\put(6.5,42.11){$\cdot$}
\put(6,42.40){$\cdot$}
\put(5.5,42.70){$\cdot$}
\put(5,43.02){$\cdot$}
\put(4.5,43.36){$\cdot$}
\put(4,43.73){$\cdot$}
\put(3.5,44.12){$\cdot$}
\put(3,44.54){$\cdot$}
\put(2.5,45.00){$\cdot$}
\put(2,45.52){$\cdot$}
\put(1.5,46.11){$\cdot$}
\put(1,46.82){$\cdot$}
\put(0.5,47.74){$\cdot$}
\put(0,50){$\cdot$}
\put(50,34){$\cdot$}
\put(50.5,34){$\cdot$}
\put(51,34){$\cdot$}
\put(51.5,34){$\cdot$}
\put(52,34.01){$\cdot$}
\put(52.5,34.02){$\cdot$}
\put(53,34.03){$\cdot$}
\put(53.5,34.04){$\cdot$}
\put(54,34.05){$\cdot$}
\put(54.5,34.06){$\cdot$}
\put(55,34.08){$\cdot$}
\put(55.5,34.10){$\cdot$}
\put(56,34.11){$\cdot$}
\put(56.5,34.13){$\cdot$}
\put(57,34.16){$\cdot$}
\put(57.5,34.18){$\cdot$}
\put(58,34.20){$\cdot$}
\put(58.5,34.23){$\cdot$}
\put(59,34.26){$\cdot$}
\put(59.5,34.29){$\cdot$}
\put(60,34.32){$\cdot$}
\put(60.5,34.36){$\cdot$}
\put(61,34.39){$\cdot$}
\put(61.5,34.43){$\cdot$}
\put(62,34.47){$\cdot$}
\put(62.5,34.51){$\cdot$}
\put(63,34.55){$\cdot$}
\put(63.5,34.59){$\cdot$}
\put(64,34.64){$\cdot$}
\put(64.5,34.69){$\cdot$}
\put(65,34.74){$\cdot$}
\put(65.5,34.79){$\cdot$}
\put(66,34.84){$\cdot$}
\put(66.5,34.90){$\cdot$}
\put(67,34.95){$\cdot$}
\put(67.5,35.01){$\cdot$}
\put(68,35.07){$\cdot$}
\put(68.5,35.13){$\cdot$}
\put(69,35.20){$\cdot$}
\put(69.5,35.27){$\cdot$}
\put(70,35.33){$\cdot$}
\put(70.5,35.40){$\cdot$}
\put(71,35.48){$\cdot$}
\put(71.5,35.55){$\cdot$}
\put(72,35.63){$\cdot$}
\put(72.5,35.71){$\cdot$}
\put(73,35.79){$\cdot$}
\put(73.5,35.88){$\cdot$}
\put(74,35.96){$\cdot$}
\put(74.5,36.05){$\cdot$}
\put(75,36.14){$\cdot$}
\put(75.5,36.24){$\cdot$}
\put(76,36.33){$\cdot$}
\put(76.5,36.43){$\cdot$}
\put(77,36.53){$\cdot$}
\put(77.5,36.64){$\cdot$}
\put(78,36.74){$\cdot$}
\put(78.5,36.85){$\cdot$}
\put(79,36.97){$\cdot$}
\put(79.5,37.08){$\cdot$}
\put(80,37.20){$\cdot$}
\put(80.5,37.32){$\cdot$}
\put(81,37.45){$\cdot$}
\put(81.5,37.57){$\cdot$}
\put(82,37.70){$\cdot$}
\put(82.5,37.84){$\cdot$}
\put(83,37.98){$\cdot$}
\put(83.5,38.12){$\cdot$}
\put(84,38.27){$\cdot$}
\put(84.5,38.42){$\cdot$}
\put(85,38.57){$\cdot$}
\put(85.5,38.73){$\cdot$}
\put(86,38.90){$\cdot$}
\put(86.5,39.06){$\cdot$}
\put(87,39.24){$\cdot$}
\put(87.5,39.42){$\cdot$}
\put(88,39.60){$\cdot$}
\put(88.5,39.79){$\cdot$}
\put(89,39.99){$\cdot$}
\put(89.5,40.19){$\cdot$}
\put(90,40.40){$\cdot$}
\put(90.5,40.62){$\cdot$}
\put(91,40.84){$\cdot$}
\put(91.5,41.07){$\cdot$}
\put(92,41.32){$\cdot$}
\put(92.5,41.57){$\cdot$}
\put(93,41.83){$\cdot$}
\put(93.5,42.11){$\cdot$}
\put(94,42.40){$\cdot$}
\put(94.5,42.70){$\cdot$}
\put(95,43.02){$\cdot$}
\put(95.5,43.36){$\cdot$}
\put(96,43.73){$\cdot$}
\put(96.5,44.12){$\cdot$}
\put(97,44.54){$\cdot$}
\put(97.5,45.00){$\cdot$}
\put(98,45.52){$\cdot$}
\put(98.5,46.11){$\cdot$}
\put(99,46.82){$\cdot$}
\put(99.5,47.74){$\cdot$}
\put(100,50){$\cdot$}

\put(50,134){$\cdot$}
\put(49.5,134){$\cdot$}
\put(49,134){$\cdot$}
\put(48.5,134){$\cdot$}
\put(48,134.01){$\cdot$}
\put(47.5,134.02){$\cdot$}
\put(47,134.03){$\cdot$}
\put(46.5,134.04){$\cdot$}
\put(46,134.05){$\cdot$}
\put(45.5,134.06){$\cdot$}
\put(45,134.08){$\cdot$}
\put(44.5,134.10){$\cdot$}
\put(44,134.11){$\cdot$}
\put(43.5,134.13){$\cdot$}
\put(43,134.16){$\cdot$}
\put(42.5,134.18){$\cdot$}
\put(42,134.20){$\cdot$}
\put(41.5,134.23){$\cdot$}
\put(41,134.26){$\cdot$}
\put(40.5,134.29){$\cdot$}
\put(40,134.32){$\cdot$}
\put(39.5,134.36){$\cdot$}
\put(39,134.39){$\cdot$}
\put(38.5,134.43){$\cdot$}
\put(38,134.47){$\cdot$}
\put(37.5,134.51){$\cdot$}
\put(37,134.55){$\cdot$}
\put(36.5,134.59){$\cdot$}
\put(36,134.64){$\cdot$}
\put(35.5,134.69){$\cdot$}
\put(35,134.74){$\cdot$}
\put(34.5,134.79){$\cdot$}
\put(34,134.84){$\cdot$}
\put(33.5,134.90){$\cdot$}
\put(33,134.95){$\cdot$}
\put(32.5,135.01){$\cdot$}
\put(32,135.07){$\cdot$}
\put(31.5,135.13){$\cdot$}
\put(31,135.20){$\cdot$}
\put(30.5,135.27){$\cdot$}
\put(30,135.33){$\cdot$}
\put(29.5,135.40){$\cdot$}
\put(29,135.48){$\cdot$}
\put(28.5,135.55){$\cdot$}
\put(28,135.63){$\cdot$}
\put(27.5,135.71){$\cdot$}
\put(27,135.79){$\cdot$}
\put(26.5,135.88){$\cdot$}
\put(26,135.96){$\cdot$}
\put(25.5,136.05){$\cdot$}
\put(25,136.14){$\cdot$}
\put(24.5,136.24){$\cdot$}
\put(24,136.33){$\cdot$}
\put(23.5,136.43){$\cdot$}
\put(23,136.53){$\cdot$}
\put(22.5,136.64){$\cdot$}
\put(22,136.74){$\cdot$}
\put(21.5,136.85){$\cdot$}
\put(21,136.97){$\cdot$}
\put(20.5,137.08){$\cdot$}
\put(20,137.20){$\cdot$}
\put(19.5,137.32){$\cdot$}
\put(19,137.45){$\cdot$}
\put(18.5,137.57){$\cdot$}
\put(18,137.70){$\cdot$}
\put(17.5,137.84){$\cdot$}
\put(17,137.98){$\cdot$}
\put(16.5,138.12){$\cdot$}
\put(16,138.27){$\cdot$}
\put(15.5,138.42){$\cdot$}
\put(15,138.57){$\cdot$}
\put(14.5,138.73){$\cdot$}
\put(14,138.90){$\cdot$}
\put(13.5,139.06){$\cdot$}
\put(13,139.24){$\cdot$}
\put(12.5,139.42){$\cdot$}
\put(12,139.60){$\cdot$}
\put(11.5,139.79){$\cdot$}
\put(11,139.99){$\cdot$}
\put(10.5,140.19){$\cdot$}
\put(10,140.40){$\cdot$}
\put(9.5,140.62){$\cdot$}
\put(9,140.84){$\cdot$}
\put(8.5,141.07){$\cdot$}
\put(8,141.32){$\cdot$}
\put(7.5,141.57){$\cdot$}
\put(7,141.83){$\cdot$}
\put(6.5,142.11){$\cdot$}
\put(6,142.40){$\cdot$}
\put(5.5,142.70){$\cdot$}
\put(5,143.02){$\cdot$}
\put(4.5,143.36){$\cdot$}
\put(4,143.73){$\cdot$}
\put(3.5,144.12){$\cdot$}
\put(3,144.54){$\cdot$}
\put(2.5,145.00){$\cdot$}
\put(2,145.52){$\cdot$}
\put(1.5,146.11){$\cdot$}
\put(1,146.82){$\cdot$}
\put(0.5,147.74){$\cdot$}
\put(0,150){$\cdot$}
\put(50,134){$\cdot$}
\put(50.5,134){$\cdot$}
\put(51,134){$\cdot$}
\put(51.5,134){$\cdot$}
\put(52,134.01){$\cdot$}
\put(52.5,134.02){$\cdot$}
\put(53,134.03){$\cdot$}
\put(53.5,134.04){$\cdot$}
\put(54,134.05){$\cdot$}
\put(54.5,134.06){$\cdot$}
\put(55,134.08){$\cdot$}
\put(55.5,134.10){$\cdot$}
\put(56,134.11){$\cdot$}
\put(56.5,134.13){$\cdot$}
\put(57,134.16){$\cdot$}
\put(57.5,134.18){$\cdot$}
\put(58,134.20){$\cdot$}
\put(58.5,134.23){$\cdot$}
\put(59,134.26){$\cdot$}
\put(59.5,134.29){$\cdot$}
\put(60,134.32){$\cdot$}
\put(60.5,134.36){$\cdot$}
\put(61,134.39){$\cdot$}
\put(61.5,134.43){$\cdot$}
\put(62,134.47){$\cdot$}
\put(62.5,134.51){$\cdot$}
\put(63,134.55){$\cdot$}
\put(63.5,134.59){$\cdot$}
\put(64,134.64){$\cdot$}
\put(64.5,134.69){$\cdot$}
\put(65,134.74){$\cdot$}
\put(65.5,134.79){$\cdot$}
\put(66,134.84){$\cdot$}
\put(66.5,134.90){$\cdot$}
\put(67,134.95){$\cdot$}
\put(67.5,135.01){$\cdot$}
\put(68,135.07){$\cdot$}
\put(68.5,135.13){$\cdot$}
\put(69,135.20){$\cdot$}
\put(69.5,135.27){$\cdot$}
\put(70,135.33){$\cdot$}
\put(70.5,135.40){$\cdot$}
\put(71,135.48){$\cdot$}
\put(71.5,135.55){$\cdot$}
\put(72,135.63){$\cdot$}
\put(72.5,135.71){$\cdot$}
\put(73,135.79){$\cdot$}
\put(73.5,135.88){$\cdot$}
\put(74,135.96){$\cdot$}
\put(74.5,136.05){$\cdot$}
\put(75,136.14){$\cdot$}
\put(75.5,136.24){$\cdot$}
\put(76,136.33){$\cdot$}
\put(76.5,136.43){$\cdot$}
\put(77,136.53){$\cdot$}
\put(77.5,136.64){$\cdot$}
\put(78,136.74){$\cdot$}
\put(78.5,136.85){$\cdot$}
\put(79,136.97){$\cdot$}
\put(79.5,137.08){$\cdot$}
\put(80,137.20){$\cdot$}
\put(80.5,137.32){$\cdot$}
\put(81,137.45){$\cdot$}
\put(81.5,137.57){$\cdot$}
\put(82,137.70){$\cdot$}
\put(82.5,137.84){$\cdot$}
\put(83,137.98){$\cdot$}
\put(83.5,138.12){$\cdot$}
\put(84,138.27){$\cdot$}
\put(84.5,138.42){$\cdot$}
\put(85,138.57){$\cdot$}
\put(85.5,138.73){$\cdot$}
\put(86,138.90){$\cdot$}
\put(86.5,139.06){$\cdot$}
\put(87,139.24){$\cdot$}
\put(87.5,139.42){$\cdot$}
\put(88,139.60){$\cdot$}
\put(88.5,139.79){$\cdot$}
\put(89,139.99){$\cdot$}
\put(89.5,140.19){$\cdot$}
\put(90,140.40){$\cdot$}
\put(90.5,140.62){$\cdot$}
\put(91,140.84){$\cdot$}
\put(91.5,141.07){$\cdot$}
\put(92,141.32){$\cdot$}
\put(92.5,141.57){$\cdot$}
\put(93,141.83){$\cdot$}
\put(93.5,142.11){$\cdot$}
\put(94,142.40){$\cdot$}
\put(94.5,142.70){$\cdot$}
\put(95,143.02){$\cdot$}
\put(95.5,143.36){$\cdot$}
\put(96,143.73){$\cdot$}
\put(96.5,144.12){$\cdot$}
\put(97,144.54){$\cdot$}
\put(97.5,145.00){$\cdot$}
\put(98,145.52){$\cdot$}
\put(98.5,146.11){$\cdot$}
\put(99,146.82){$\cdot$}
\put(99.5,147.74){$\cdot$}
\put(100,150){$\cdot$}
\put(50,66){$\cdot$}
\put(49.5,65.99){$\cdot$}
\put(49,65.99){$\cdot$}
\put(48.5,65.99){$\cdot$}
\put(48,65.98){$\cdot$}
\put(45,65.91){$\cdot$}
\put(44.5,65.90){$\cdot$}
\put(44,65.88){$\cdot$}
\put(43.5,65.86){$\cdot$}
\put(43,65.84){$\cdot$}
\put(40,65.68){$\cdot$}
\put(39.5,65.64){$\cdot$}
\put(39,65.60){$\cdot$}
\put(38.5,65.57){$\cdot$}
\put(38,65.53){$\cdot$}
\put(35,65.26){$\cdot$}
\put(34.5,65.21){$\cdot$}
\put(34,65.16){$\cdot$}
\put(33.5,65.10){$\cdot$}
\put(33,65.04){$\cdot$}
\put(30,64.66){$\cdot$}
\put(29.5,64.59){$\cdot$}
\put(29,64.52){$\cdot$}
\put(28.5,64.44){$\cdot$}
\put(28,64.37){$\cdot$}
\put(25,63.86){$\cdot$}
\put(24.5,63.76){$\cdot$}
\put(24,63.66){$\cdot$}
\put(23.5,63.57){$\cdot$}
\put(23,63.47){$\cdot$}
\put(20,62.80){$\cdot$}
\put(19.5,62.68){$\cdot$}
\put(19,62.55){$\cdot$}
\put(18.5,62.42){$\cdot$}
\put(18,62.29){$\cdot$}
\put(15,61.43){$\cdot$}
\put(14.5,61.27){$\cdot$}
\put(14,61.10){$\cdot$}
\put(13.5,60.93){$\cdot$}
\put(13,60.76){$\cdot$}
\put(10,59.60){$\cdot$}
\put(9.5,59.38){$\cdot$}
\put(9,59.16){$\cdot$}
\put(8.5,58.92){$\cdot$}
\put(8,58.68){$\cdot$}
\put(5,56.97){$\cdot$}
\put(4.5,56.63){$\cdot$}
\put(4,56.27){$\cdot$}
\put(3.5,55.88){$\cdot$}
\put(3,55.46){$\cdot$}
\put(53,65.98){$\cdot$}
\put(53.5,65.97){$\cdot$}
\put(54,65.96){$\cdot$}
\put(54.5,65.95){$\cdot$}
\put(55,65.93){$\cdot$}
\put(58,65.82){$\cdot$}
\put(58.5,65.79){$\cdot$}
\put(59,65.77){$\cdot$}
\put(59.5,65.74){$\cdot$}
\put(60,65.70){$\cdot$}
\put(63,65.49){$\cdot$}
\put(63.5,65.45){$\cdot$}
\put(64,65.40){$\cdot$}
\put(64.5,65.36){$\cdot$}
\put(65,65.31){$\cdot$}
\put(68,64.99){$\cdot$}
\put(68.5,64.93){$\cdot$}
\put(69,64.86){$\cdot$}
\put(69.5,64.79){$\cdot$}
\put(70,64.73){$\cdot$}
\put(73,64.29){$\cdot$}
\put(73.5,64.20){$\cdot$}
\put(74,64.12){$\cdot$}
\put(74.5,64.04){$\cdot$}
\put(75,63.95){$\cdot$}
\put(78,63.36){$\cdot$}
\put(78.5,63.25){$\cdot$}
\put(79,63.15){$\cdot$}
\put(79.5,63.03){$\cdot$}
\put(80,62.92){$\cdot$}
\put(83,62.16){$\cdot$}
\put(83.5,62.02){$\cdot$}
\put(84,61.88){$\cdot$}
\put(84.5,61.73){$\cdot$}
\put(85,61.58){$\cdot$}
\put(88,60.58){$\cdot$}
\put(88.5,60.40){$\cdot$}
\put(89,60.20){$\cdot$}
\put(89.5,60.01){$\cdot$}
\put(90,59.81){$\cdot$}
\put(93,58.43){$\cdot$}
\put(93.5,58.16){$\cdot$}
\put(94,57.88){$\cdot$}
\put(94.5,57.59){$\cdot$}
\put(95,57.29){$\cdot$}
\put(98,54.99){$\cdot$}
\put(98.5,54.48){$\cdot$}
\put(99,53.89){$\cdot$}
\put(99.5,53.18){$\cdot$}
\put(100,52.26){$\cdot$}
\put(50,166){$\cdot$}
\put(49.5,165.99){$\cdot$}
\put(49,165.99){$\cdot$}
\put(48.5,165.99){$\cdot$}
\put(48,165.98){$\cdot$}
\put(45,165.91){$\cdot$}
\put(44.5,165.90){$\cdot$}
\put(44,165.88){$\cdot$}
\put(43.5,165.86){$\cdot$}
\put(43,165.84){$\cdot$}
\put(40,165.68){$\cdot$}
\put(39.5,165.64){$\cdot$}
\put(39,165.60){$\cdot$}
\put(38.5,165.57){$\cdot$}
\put(38,165.53){$\cdot$}
\put(35,165.26){$\cdot$}
\put(34.5,165.21){$\cdot$}
\put(34,165.16){$\cdot$}
\put(33.5,165.10){$\cdot$}
\put(33,165.04){$\cdot$}
\put(30,164.66){$\cdot$}
\put(29.5,164.59){$\cdot$}
\put(29,164.52){$\cdot$}
\put(28.5,164.44){$\cdot$}
\put(28,164.37){$\cdot$}
\put(25,163.86){$\cdot$}
\put(24.5,163.76){$\cdot$}
\put(24,163.66){$\cdot$}
\put(23.5,163.57){$\cdot$}
\put(23,163.47){$\cdot$}
\put(20,162.80){$\cdot$}
\put(19.5,162.68){$\cdot$}
\put(19,162.55){$\cdot$}
\put(18.5,162.42){$\cdot$}
\put(18,162.29){$\cdot$}
\put(15,161.43){$\cdot$}
\put(14.5,161.27){$\cdot$}
\put(14,161.10){$\cdot$}
\put(13.5,160.93){$\cdot$}
\put(13,160.76){$\cdot$}
\put(10,159.60){$\cdot$}
\put(9.5,159.38){$\cdot$}
\put(9,159.16){$\cdot$}
\put(8.5,158.92){$\cdot$}
\put(8,158.68){$\cdot$}
\put(5,156.97){$\cdot$}
\put(4.5,156.63){$\cdot$}
\put(4,156.27){$\cdot$}
\put(3.5,155.88){$\cdot$}
\put(3,155.46){$\cdot$}
\put(53,165.98){$\cdot$}
\put(53.5,165.97){$\cdot$}
\put(54,165.96){$\cdot$}
\put(54.5,165.95){$\cdot$}
\put(55,165.93){$\cdot$}
\put(58,165.82){$\cdot$}
\put(58.5,165.79){$\cdot$}
\put(59,165.77){$\cdot$}
\put(59.5,165.74){$\cdot$}
\put(60,165.70){$\cdot$}
\put(63,165.49){$\cdot$}
\put(63.5,165.45){$\cdot$}
\put(64,165.40){$\cdot$}
\put(64.5,165.36){$\cdot$}
\put(65,165.31){$\cdot$}
\put(68,164.99){$\cdot$}
\put(68.5,164.93){$\cdot$}
\put(69,164.86){$\cdot$}
\put(69.5,164.79){$\cdot$}
\put(70,164.73){$\cdot$}
\put(73,164.29){$\cdot$}
\put(73.5,164.20){$\cdot$}
\put(74,164.12){$\cdot$}
\put(74.5,164.04){$\cdot$}
\put(75,163.95){$\cdot$}
\put(78,163.36){$\cdot$}
\put(78.5,163.25){$\cdot$}
\put(79,163.15){$\cdot$}
\put(79.5,163.03){$\cdot$}
\put(80,162.92){$\cdot$}
\put(83,162.16){$\cdot$}
\put(83.5,162.02){$\cdot$}
\put(84,161.88){$\cdot$}
\put(84.5,161.73){$\cdot$}
\put(85,161.58){$\cdot$}
\put(88,160.58){$\cdot$}
\put(88.5,160.40){$\cdot$}
\put(89,160.20){$\cdot$}
\put(89.5,160.01){$\cdot$}
\put(90,159.81){$\cdot$}
\put(93,158.43){$\cdot$}
\put(93.5,158.16){$\cdot$}
\put(94,157.88){$\cdot$}
\put(94.5,157.59){$\cdot$}
\put(95,157.29){$\cdot$}
\put(98,154.99){$\cdot$}
\put(98.5,154.48){$\cdot$}
\put(99,153.89){$\cdot$}
\put(99.5,153.18){$\cdot$}
\put(100,152.26){$\cdot$}
%shere I quadrant
\put(50,100){$\cdot$}
\put(49.5,99.99){$\cdot$}
\put(49,99.98){$\cdot$}
\put(48.5,99.97){$\cdot$}
\put(48,99.96){$\cdot$}
\put(47.5,99.94){$\cdot$}
\put(47,99.90){$\cdot$}
\put(46.5,99.88){$\cdot$}
\put(46,99.84){$\cdot$}
\put(45.5,99.80){$\cdot$}
\put(45,99.75){$\cdot$}
\put(44.5,99.70){$\cdot$}
\put(44,99.64){$\cdot$}
\put(43.5,99.58){$\cdot$}
\put(43,99.51){$\cdot$}
\put(42.5,99.43){$\cdot$}
\put(42,99.35){$\cdot$}
\put(41.5,99.27){$\cdot$}
\put(41,99.18){$\cdot$}
\put(40.5,99.09){$\cdot$}
\put(40,98.99){$\cdot$}
\put(39.5,98.88){$\cdot$}
\put(39,98.77){$\cdot$}
\put(38.5,98.66){$\cdot$}
\put(38,98.54){$\cdot$}
\put(37.5,98.41){$\cdot$}
\put(37,98.28){$\cdot$}
\put(36.5,98.14){$\cdot$}
\put(36,98.00){$\cdot$}
\put(35.5,97.85){$\cdot$}
\put(35,97.70){$\cdot$}
\put(34.5,97.54){$\cdot$}
\put(34,97.37){$\cdot$}
\put(33.5,97.20){$\cdot$}
\put(33,97.02){$\cdot$}
\put(32.5,96.84){$\cdot$}
\put(32,96.65){$\cdot$}
\put(31.5,96.45){$\cdot$}
\put(31,96.25){$\cdot$}
\put(30.5,96.04){$\cdot$}
\put(30,95.82){$\cdot$}
\put(29.5,95.60){$\cdot$}
\put(29,95.38){$\cdot$}
\put(28.5,95.14){$\cdot$}
\put(28,94.90){$\cdot$}
\put(27.5,94.65){$\cdot$}
\put(27,94.39){$\cdot$}
\put(26.5,94.13){$\cdot$}
\put(26,93.86){$\cdot$}
\put(25.5,93.59){$\cdot$}
\put(25,93.30){$\cdot$}
\put(24.5,93.01){$\cdot$}
\put(24,92.71){$\cdot$}
\put(23.5,92.40){$\cdot$}
\put(23,92.08){$\cdot$}
\put(22.5,91.76){$\cdot$}
\put(22,91.42){$\cdot$}
\put(21.5,91.08){$\cdot$}
\put(21,90.73){$\cdot$}
\put(20.5,90.37){$\cdot$}
\put(20,90.00){$\cdot$}
\put(19.5,89.62){$\cdot$}
\put(19,89.23){$\cdot$}
\put(18.5,88.83){$\cdot$}
\put(18,88.42){$\cdot$}
\put(17.5,87.99){$\cdot$}
\put(17,87.56){$\cdot$}
\put(16.5,87.12){$\cdot$}
\put(16,86.66){$\cdot$}
\put(15.5,86.19){$\cdot$}
\put(15,85.70){$\cdot$}
\put(14.5,85.21){$\cdot$}
\put(14,84.70){$\cdot$}
\put(13.5,84.17){$\cdot$}
\put(13,83.63){$\cdot$}
\put(12.5,83.07){$\cdot$}
\put(12,82.49){$\cdot$}
\put(11.5,81.90){$\cdot$}
\put(11,81.29){$\cdot$}
\put(10.5,80.65){$\cdot$}
\put(10,80.00){$\cdot$}
\put(9.5,79.32){$\cdot$}
\put(9,78.62){$\cdot$}
\put(8.5,77.88){$\cdot$}
\put(8,77.13){$\cdot$}
\put(7.5,76.34){$\cdot$}
\put(7,75.51){$\cdot$}
\put(6.5,74.65){$\cdot$}
\put(6,73.75){$\cdot$}
\put(5.5,72.80){$\cdot$}
\put(5,71.79){$\cdot$}
\put(4.5,70.73){$\cdot$}
\put(4,69.59){$\cdot$}
\put(3.5,68.38){$\cdot$}
\put(3,67.06){$\cdot$}
\put(2.5,65.61){$\cdot$}
\put(2,64.00){$\cdot$}
\put(1.5,62.15){$\cdot$}
\put(1,59.95){$\cdot$}
\put(0.5,57.05){$\cdot$}
\put(0,50){$\cdot$}
%  II quadrant
\put(50,100){$\cdot$}
\put(50.5,99.99){$\cdot$}
\put(51,99.98){$\cdot$}
\put(51.5,99.97){$\cdot$}
\put(52,99.96){$\cdot$}
\put(52.5,99.94){$\cdot$}
\put(53,99.90){$\cdot$}
\put(53.5,99.88){$\cdot$}
\put(54,99.84){$\cdot$}
\put(54.5,99.80){$\cdot$}
\put(55,99.75){$\cdot$}
\put(55.5,99.70){$\cdot$}
\put(56,99.64){$\cdot$}
\put(56.5,99.58){$\cdot$}
\put(57,99.51){$\cdot$}
\put(57.5,99.43){$\cdot$}
\put(58,99.35){$\cdot$}
\put(58.5,99.27){$\cdot$}
\put(59,99.18){$\cdot$}
\put(59.5,99.09){$\cdot$}
\put(60,98.99){$\cdot$}
\put(60.5,98.88){$\cdot$}
\put(61,98.77){$\cdot$}
\put(61.5,98.66){$\cdot$}
\put(62,98.54){$\cdot$}
\put(62.5,98.41){$\cdot$}
\put(63,98.28){$\cdot$}
\put(63.5,98.14){$\cdot$}
\put(64,98.00){$\cdot$}
\put(64.5,97.85){$\cdot$}
\put(65,97.70){$\cdot$}
\put(65.5,97.54){$\cdot$}
\put(66,97.37){$\cdot$}
\put(66.5,97.20){$\cdot$}
\put(67,97.02){$\cdot$}
\put(67.5,96.84){$\cdot$}
\put(68,96.65){$\cdot$}
\put(68.5,96.45){$\cdot$}
\put(69,96.25){$\cdot$}
\put(69.5,96.04){$\cdot$}
\put(70,95.82){$\cdot$}
\put(70.5,95.60){$\cdot$}
\put(71,95.38){$\cdot$}
\put(71.5,95.14){$\cdot$}
\put(72,94.90){$\cdot$}
\put(72.5,94.65){$\cdot$}
\put(73,94.39){$\cdot$}
\put(73.5,94.13){$\cdot$}
\put(74,93.86){$\cdot$}
\put(74.5,93.59){$\cdot$}
\put(75,93.30){$\cdot$}
\put(75.5,93.01){$\cdot$}
\put(76,92.71){$\cdot$}
\put(76.5,92.40){$\cdot$}
\put(77,92.08){$\cdot$}
\put(77.5,91.76){$\cdot$}
\put(78,91.42){$\cdot$}
\put(78.5,91.08){$\cdot$}
\put(79,90.73){$\cdot$}
\put(79.5,90.37){$\cdot$}
\put(80,90.00){$\cdot$}
\put(80.5,89.62){$\cdot$}
\put(81,89.23){$\cdot$}
\put(81.5,88.83){$\cdot$}
\put(82,88.42){$\cdot$}
\put(82.5,87.99){$\cdot$}
\put(83,87.56){$\cdot$}
\put(83.5,87.12){$\cdot$}
\put(84,86.66){$\cdot$}
\put(84.5,86.19){$\cdot$}
\put(85,85.70){$\cdot$}
\put(85.5,85.21){$\cdot$}
\put(86,84.70){$\cdot$}
\put(86.5,84.17){$\cdot$}
\put(87,83.63){$\cdot$}
\put(87.5,83.07){$\cdot$}
\put(88,82.49){$\cdot$}
\put(88.5,81.90){$\cdot$}
\put(89,81.29){$\cdot$}
\put(89.5,80.65){$\cdot$}
\put(90,80.00){$\cdot$}
\put(90.5,79.32){$\cdot$}
\put(91,78.62){$\cdot$}
\put(91.5,77.88){$\cdot$}
\put(92,77.13){$\cdot$}
\put(92.5,76.34){$\cdot$}
\put(93,75.51){$\cdot$}
\put(93.5,74.65){$\cdot$}
\put(94,73.75){$\cdot$}
\put(94.5,72.80){$\cdot$}
\put(95,71.79){$\cdot$}
\put(95.5,70.73){$\cdot$}
\put(96,69.59){$\cdot$}
\put(96.5,68.38){$\cdot$}
\put(97,67.06){$\cdot$}
\put(97.5,65.61){$\cdot$}
\put(98,64.00){$\cdot$}
\put(98.5,62.15){$\cdot$}
\put(99,59.95){$\cdot$}
\put(99.5,57.05){$\cdot$}
\put(100,50){$\cdot$}
%shere I quadrant 2
\put(50,200){$\cdot$}
\put(49.5,199.99){$\cdot$}
\put(49,199.98){$\cdot$}
\put(48.5,199.97){$\cdot$}
\put(48,199.96){$\cdot$}
\put(47.5,199.94){$\cdot$}
\put(47,199.90){$\cdot$}
\put(46.5,199.88){$\cdot$}
\put(46,199.84){$\cdot$}
\put(45.5,199.80){$\cdot$}
\put(45,199.75){$\cdot$}
\put(44.5,199.70){$\cdot$}
\put(44,199.64){$\cdot$}
\put(43.5,199.58){$\cdot$}
\put(43,199.51){$\cdot$}
\put(42.5,199.43){$\cdot$}
\put(42,199.35){$\cdot$}
\put(41.5,199.27){$\cdot$}
\put(41,199.18){$\cdot$}
\put(40.5,199.09){$\cdot$}
\put(40,198.99){$\cdot$}
\put(39.5,198.88){$\cdot$}
\put(39,198.77){$\cdot$}
\put(38.5,198.66){$\cdot$}
\put(38,198.54){$\cdot$}
\put(37.5,198.41){$\cdot$}
\put(37,198.28){$\cdot$}
\put(36.5,198.14){$\cdot$}
\put(36,198.00){$\cdot$}
\put(35.5,197.85){$\cdot$}
\put(35,197.70){$\cdot$}
\put(34.5,197.54){$\cdot$}
\put(34,197.37){$\cdot$}
\put(33.5,197.20){$\cdot$}
\put(33,197.02){$\cdot$}
\put(32.5,196.84){$\cdot$}
\put(32,196.65){$\cdot$}
\put(31.5,196.45){$\cdot$}
\put(31,196.25){$\cdot$}
\put(30.5,196.04){$\cdot$}
\put(30,195.82){$\cdot$}
\put(29.5,195.60){$\cdot$}
\put(29,195.38){$\cdot$}
\put(28.5,195.14){$\cdot$}
\put(28,194.90){$\cdot$}
\put(27.5,194.65){$\cdot$}
\put(27,194.39){$\cdot$}
\put(26.5,194.13){$\cdot$}
\put(26,193.86){$\cdot$}
\put(25.5,193.59){$\cdot$}
\put(25,193.30){$\cdot$}
\put(24.5,193.01){$\cdot$}
\put(24,192.71){$\cdot$}
\put(23.5,192.40){$\cdot$}
\put(23,192.08){$\cdot$}
\put(22.5,191.76){$\cdot$}
\put(22,191.42){$\cdot$}
\put(21.5,191.08){$\cdot$}
\put(21,190.73){$\cdot$}
\put(20.5,190.37){$\cdot$}
\put(20,190.00){$\cdot$}
\put(19.5,189.62){$\cdot$}
\put(19,189.23){$\cdot$}
\put(18.5,188.83){$\cdot$}
\put(18,188.42){$\cdot$}
\put(17.5,187.99){$\cdot$}
\put(17,187.56){$\cdot$}
\put(16.5,187.12){$\cdot$}
\put(16,186.66){$\cdot$}
\put(15.5,186.19){$\cdot$}
\put(15,185.70){$\cdot$}
\put(14.5,185.21){$\cdot$}
\put(14,184.70){$\cdot$}
\put(13.5,184.17){$\cdot$}
\put(13,183.63){$\cdot$}
\put(12.5,183.07){$\cdot$}
\put(12,182.49){$\cdot$}
\put(11.5,181.90){$\cdot$}
\put(11,181.29){$\cdot$}
\put(10.5,180.65){$\cdot$}
\put(10,180.00){$\cdot$}
\put(9.5,179.32){$\cdot$}
\put(9,178.62){$\cdot$}
\put(8.5,177.88){$\cdot$}
\put(8,177.13){$\cdot$}
\put(7.5,176.34){$\cdot$}
\put(7,175.51){$\cdot$}
\put(6.5,174.65){$\cdot$}
\put(6,173.75){$\cdot$}
\put(5.5,172.80){$\cdot$}
\put(5,171.79){$\cdot$}
\put(4.5,170.73){$\cdot$}
\put(4,169.59){$\cdot$}
\put(3.5,168.38){$\cdot$}
\put(3,167.06){$\cdot$}
\put(2.5,165.61){$\cdot$}
\put(2,164.00){$\cdot$}
\put(1.5,162.15){$\cdot$}
\put(1,159.95){$\cdot$}
\put(0.5,157.05){$\cdot$}
\put(0,150){$\cdot$}
%  II quadrant 2
\put(50,200){$\cdot$}
\put(50.5,199.99){$\cdot$}
\put(51,199.98){$\cdot$}
\put(51.5,199.97){$\cdot$}
\put(52,199.96){$\cdot$}
\put(52.5,199.94){$\cdot$}
\put(53,199.90){$\cdot$}
\put(53.5,199.88){$\cdot$}
\put(54,199.84){$\cdot$}
\put(54.5,199.80){$\cdot$}
\put(55,199.75){$\cdot$}
\put(55.5,199.70){$\cdot$}
\put(56,199.64){$\cdot$}
\put(56.5,199.58){$\cdot$}
\put(57,199.51){$\cdot$}
\put(57.5,199.43){$\cdot$}
\put(58,199.35){$\cdot$}
\put(58.5,199.27){$\cdot$}
\put(59,199.18){$\cdot$}
\put(59.5,199.09){$\cdot$}
\put(60,198.99){$\cdot$}
\put(60.5,198.88){$\cdot$}
\put(61,198.77){$\cdot$}
\put(61.5,198.66){$\cdot$}
\put(62,198.54){$\cdot$}
\put(62.5,198.41){$\cdot$}
\put(63,198.28){$\cdot$}
\put(63.5,198.14){$\cdot$}
\put(64,198.00){$\cdot$}
\put(64.5,197.85){$\cdot$}
\put(65,197.70){$\cdot$}
\put(65.5,197.54){$\cdot$}
\put(66,197.37){$\cdot$}
\put(66.5,197.20){$\cdot$}
\put(67,197.02){$\cdot$}
\put(67.5,196.84){$\cdot$}
\put(68,196.65){$\cdot$}
\put(68.5,196.45){$\cdot$}
\put(69,196.25){$\cdot$}
\put(69.5,196.04){$\cdot$}
\put(70,195.82){$\cdot$}
\put(70.5,195.60){$\cdot$}
\put(71,195.38){$\cdot$}
\put(71.5,195.14){$\cdot$}
\put(72,194.90){$\cdot$}
\put(72.5,194.65){$\cdot$}
\put(73,194.39){$\cdot$}
\put(73.5,194.13){$\cdot$}
\put(74,193.86){$\cdot$}
\put(74.5,193.59){$\cdot$}
\put(75,193.30){$\cdot$}
\put(75.5,193.01){$\cdot$}
\put(76,192.71){$\cdot$}
\put(76.5,192.40){$\cdot$}
\put(77,192.08){$\cdot$}
\put(77.5,191.76){$\cdot$}
\put(78,191.42){$\cdot$}
\put(78.5,191.08){$\cdot$}
\put(79,190.73){$\cdot$}
\put(79.5,190.37){$\cdot$}
\put(80,190.00){$\cdot$}
\put(80.5,189.62){$\cdot$}
\put(81,189.23){$\cdot$}
\put(81.5,188.83){$\cdot$}
\put(82,188.42){$\cdot$}
\put(82.5,187.99){$\cdot$}
\put(83,187.56){$\cdot$}
\put(83.5,187.12){$\cdot$}
\put(84,186.66){$\cdot$}
\put(84.5,186.19){$\cdot$}
\put(85,185.70){$\cdot$}
\put(85.5,185.21){$\cdot$}
\put(86,184.70){$\cdot$}
\put(86.5,184.17){$\cdot$}
\put(87,183.63){$\cdot$}
\put(87.5,183.07){$\cdot$}
\put(88,182.49){$\cdot$}
\put(88.5,181.90){$\cdot$}
\put(89,181.29){$\cdot$}
\put(89.5,180.65){$\cdot$}
\put(90,180.00){$\cdot$}
\put(90.5,179.32){$\cdot$}
\put(91,178.62){$\cdot$}
\put(91.5,177.88){$\cdot$}
\put(92,177.13){$\cdot$}
\put(92.5,176.34){$\cdot$}
\put(93,175.51){$\cdot$}
\put(93.5,174.65){$\cdot$}
\put(94,173.75){$\cdot$}
\put(94.5,172.80){$\cdot$}
\put(95,171.79){$\cdot$}
\put(95.5,170.73){$\cdot$}
\put(96,169.59){$\cdot$}
\put(96.5,168.38){$\cdot$}
\put(97,167.06){$\cdot$}
\put(97.5,165.61){$\cdot$}
\put(98,164.00){$\cdot$}
\put(98.5,162.15){$\cdot$}
\put(99,159.95){$\cdot$}
\put(99.5,157.05){$\cdot$}
\put(100,150){$\cdot$}
% shere III quadrant
\put(50,0){$\cdot$}
\put(49.5,0){$\cdot$}
\put(49,0.01){$\cdot$}
\put(48.5,0.02){$\cdot$}
\put(48,0.04){$\cdot$}
\put(47.5,0.06){$\cdot$}
\put(47,0.09){$\cdot$}
\put(46.5,0.12){$\cdot$}
\put(46,0.16){$\cdot$}
\put(45.5,0.2){$\cdot$}
\put(45,0.25){$\cdot$}
\put(44.5,0.3){$\cdot$}
\put(44,0.36){$\cdot$}
\put(43.5,0.42){$\cdot$}
\put(43,0.49){$\cdot$}
\put(42.5,0.56){$\cdot$}
\put(42,0.64){$\cdot$}
\put(41.5,0.73){$\cdot$}
\put(41,0.82){$\cdot$}
\put(40.5,0.91){$\cdot$}
\put(40,1.01){$\cdot$}
\put(39.5,1.11){$\cdot$}
\put(39,1.22){$\cdot$}
\put(38.5,1.34){$\cdot$}
\put(38,1.46){$\cdot$}
\put(37.5,1.59){$\cdot$}
\put(37,1.72){$\cdot$}
\put(36.5,1.86){$\cdot$}
\put(36,2.0){$\cdot$}
\put(35.5,2.15){$\cdot$}
\put(35,2.3){$\cdot$}
\put(34.5,2.46){$\cdot$}
\put(34,2.63){$\cdot$}
\put(33.5,2.8){$\cdot$}
\put(33,2.98){$\cdot$}
\put(32.5,3.16){$\cdot$}
\put(32,3.35){$\cdot$}
\put(31.5,3.55){$\cdot$}
\put(31,3.75){$\cdot$}
\put(30.5,3.96){$\cdot$}
\put(30,4.17){$\cdot$}
\put(29.5,4.39){$\cdot$}
\put(29,4.62){$\cdot$}
\put(28.5,4.86){$\cdot$}
\put(28,5.1){$\cdot$}
\put(27.5,5.35){$\cdot$}
\put(27,5.6){$\cdot$}
\put(26.5,5.87){$\cdot$}
\put(26,6.14){$\cdot$}
\put(25.5,6.41){$\cdot$}
\put(25,6.7){$\cdot$}
\put(24.5,6.99){$\cdot$}
\put(24,7.29){$\cdot$}
\put(23.5,7.6){$\cdot$}
\put(23,7.9){$\cdot$}
\put(22.5,8.24){$\cdot$}
\put(22,8.57){$\cdot$}
\put(21.5,8.92){$\cdot$}
\put(21,9.27){$\cdot$}
\put(20.5,9.63){$\cdot$}
\put(20,10.0){$\cdot$}
\put(19.5,10.38){$\cdot$}
\put(19,10.77){$\cdot$}
\put(18.5,11.17){$\cdot$}
\put(18,11.58){$\cdot$}
\put(17.5,12.0){$\cdot$}
\put(17,12.44){$\cdot$}
\put(16.5,12.88){$\cdot$}
\put(16,13.34){$\cdot$}
\put(15.5,13.81){$\cdot$}
\put(15,14.29){$\cdot$}
\put(14.5,14.79){$\cdot$}
\put(14,15.3){$\cdot$}
\put(13.5,15.83){$\cdot$}
\put(13,16.37){$\cdot$}
\put(12.5,16.93){$\cdot$}
\put(12,17.5){$\cdot$}
\put(11.5,18.1){$\cdot$}
\put(11,18.71){$\cdot$}
\put(10.5,19.34){$\cdot$}
\put(10,20.00){$\cdot$}
\put(9.5,20.68){$\cdot$}
\put(9,21.38){$\cdot$}
\put(8.5,22.11){$\cdot$}
\put(8,22.87){$\cdot$}
\put(7.5,23.66){$\cdot$}
\put(7,24.48){$\cdot$}
\put(6.5,25.35){$\cdot$}
\put(6,26.25){$\cdot$}
\put(5.5,27.2){$\cdot$}
\put(5,28.2){$\cdot$}
\put(4.5,29.27){$\cdot$}
\put(4,30.4){$\cdot$}
\put(3.5,31.62){$\cdot$}
\put(3,32.94){$\cdot$}
\put(2.5,34.39){$\cdot$}
\put(2,36.00){$\cdot$}
\put(1.5,37.84){$\cdot$}
\put(1,40.05){$\cdot$}
\put(0.5,42.95){$\cdot$}
\put(0,50){$\cdot$}
% shere IV quadrant
\put(50,0){$\cdot$}
\put(50.5,0){$\cdot$}
\put(51,0.01){$\cdot$}
\put(51.5,0.02){$\cdot$}
\put(52,0.04){$\cdot$}
\put(52.5,0.06){$\cdot$}
\put(53,0.09){$\cdot$}
\put(53.5,0.12){$\cdot$}
\put(54,0.16){$\cdot$}
\put(54.5,0.2){$\cdot$}
\put(55,0.25){$\cdot$}
\put(55.5,0.3){$\cdot$}
\put(56,0.36){$\cdot$}
\put(56.5,0.42){$\cdot$}
\put(57,0.49){$\cdot$}
\put(57.5,0.56){$\cdot$}
\put(58,0.64){$\cdot$}
\put(58.5,0.73){$\cdot$}
\put(59,0.82){$\cdot$}
\put(59.5,0.91){$\cdot$}
\put(60,1.01){$\cdot$}
\put(60.5,1.11){$\cdot$}
\put(61,1.22){$\cdot$}
\put(61.5,1.34){$\cdot$}
\put(62,1.46){$\cdot$}
\put(62.5,1.59){$\cdot$}
\put(63,1.72){$\cdot$}
\put(63.5,1.86){$\cdot$}
\put(64,2.0){$\cdot$}
\put(64.5,2.15){$\cdot$}
\put(65,2.3){$\cdot$}
\put(65.5,2.46){$\cdot$}
\put(66,2.63){$\cdot$}
\put(66.5,2.8){$\cdot$}
\put(67,2.98){$\cdot$}
\put(67.5,3.16){$\cdot$}
\put(68,3.35){$\cdot$}
\put(68.5,3.55){$\cdot$}
\put(69,3.75){$\cdot$}
\put(69.5,3.96){$\cdot$}
\put(70,4.17){$\cdot$}
\put(70.5,4.39){$\cdot$}
\put(71,4.62){$\cdot$}
\put(71.5,4.86){$\cdot$}
\put(72,5.1){$\cdot$}
\put(72.5,5.35){$\cdot$}
\put(73,5.6){$\cdot$}
\put(73.5,5.87){$\cdot$}
\put(74,6.14){$\cdot$}
\put(74.5,6.41){$\cdot$}
\put(75,6.7){$\cdot$}
\put(75.5,6.99){$\cdot$}
\put(76,7.29){$\cdot$}
\put(76.5,7.6){$\cdot$}
\put(77,7.9){$\cdot$}
\put(77.5,8.24){$\cdot$}
\put(78,8.57){$\cdot$}
\put(78.5,8.92){$\cdot$}
\put(79,9.27){$\cdot$}
\put(79.5,9.63){$\cdot$}
\put(80,10.0){$\cdot$}
\put(80.5,10.38){$\cdot$}
\put(81,10.77){$\cdot$}
\put(81.5,11.17){$\cdot$}
\put(82,11.58){$\cdot$}
\put(82.5,12.0){$\cdot$}
\put(83,12.44){$\cdot$}
\put(83.5,12.88){$\cdot$}
\put(84,13.34){$\cdot$}
\put(84.5,13.81){$\cdot$}
\put(85,14.29){$\cdot$}
\put(85.5,14.79){$\cdot$}
\put(86,15.3){$\cdot$}
\put(86.5,15.83){$\cdot$}
\put(87,16.37){$\cdot$}
\put(87.5,16.93){$\cdot$}
\put(88,17.5){$\cdot$}
\put(88.5,18.1){$\cdot$}
\put(89,18.71){$\cdot$}
\put(89.5,19.34){$\cdot$}
\put(90,20.00){$\cdot$}
\put(90.5,20.68){$\cdot$}
\put(91,21.38){$\cdot$}
\put(91.5,22.11){$\cdot$}
\put(92,22.87){$\cdot$}
\put(92.5,23.66){$\cdot$}
\put(93,24.48){$\cdot$}
\put(93.5,25.35){$\cdot$}
\put(94,26.25){$\cdot$}
\put(94.5,27.2){$\cdot$}
\put(95,28.2){$\cdot$}
\put(95.5,29.27){$\cdot$}
\put(96,30.4){$\cdot$}
\put(96.5,31.62){$\cdot$}
\put(97,32.94){$\cdot$}
\put(97.5,34.39){$\cdot$}
\put(98,36.00){$\cdot$}
\put(98.5,37.84){$\cdot$}
\put(99,40.05){$\cdot$}
\put(99.5,42.95){$\cdot$}
\put(100,50){$\cdot$}
% shere III quadrant 2
\put(50,100){$\cdot$}
\put(49.5,100){$\cdot$}
\put(49,100.01){$\cdot$}
\put(48.5,100.02){$\cdot$}
\put(48,100.04){$\cdot$}
\put(47.5,100.06){$\cdot$}
\put(47,100.09){$\cdot$}
\put(46.5,100.12){$\cdot$}
\put(46,100.16){$\cdot$}
\put(45.5,100.2){$\cdot$}
\put(45,100.25){$\cdot$}
\put(44.5,100.3){$\cdot$}
\put(44,100.36){$\cdot$}
\put(43.5,100.42){$\cdot$}
\put(43,100.49){$\cdot$}
\put(42.5,100.56){$\cdot$}
\put(42,100.64){$\cdot$}
\put(41.5,100.73){$\cdot$}
\put(41,100.82){$\cdot$}
\put(40.5,100.91){$\cdot$}
\put(40,101.01){$\cdot$}
\put(39.5,101.11){$\cdot$}
\put(39,101.22){$\cdot$}
\put(38.5,101.34){$\cdot$}
\put(38,101.46){$\cdot$}
\put(37.5,101.59){$\cdot$}
\put(37,101.72){$\cdot$}
\put(36.5,101.86){$\cdot$}
\put(36,102.0){$\cdot$}
\put(35.5,102.15){$\cdot$}
\put(35,102.3){$\cdot$}
\put(34.5,102.46){$\cdot$}
\put(34,102.63){$\cdot$}
\put(33.5,102.8){$\cdot$}
\put(33,102.98){$\cdot$}
\put(32.5,103.16){$\cdot$}
\put(32,103.35){$\cdot$}
\put(31.5,103.55){$\cdot$}
\put(31,103.75){$\cdot$}
\put(30.5,103.96){$\cdot$}
\put(30,104.17){$\cdot$}
\put(29.5,104.39){$\cdot$}
\put(29,104.62){$\cdot$}
\put(28.5,104.86){$\cdot$}
\put(28,105.1){$\cdot$}
\put(27.5,105.35){$\cdot$}
\put(27,105.6){$\cdot$}
\put(26.5,105.87){$\cdot$}
\put(26,106.14){$\cdot$}
\put(25.5,106.41){$\cdot$}
\put(25,106.7){$\cdot$}
\put(24.5,106.99){$\cdot$}
\put(24,107.29){$\cdot$}
\put(23.5,107.6){$\cdot$}
\put(23,107.9){$\cdot$}
\put(22.5,108.24){$\cdot$}
\put(22,108.57){$\cdot$}
\put(21.5,108.92){$\cdot$}
\put(21,109.27){$\cdot$}
\put(20.5,109.63){$\cdot$}
\put(20,110.0){$\cdot$}
\put(19.5,110.38){$\cdot$}
\put(19,110.77){$\cdot$}
\put(18.5,111.17){$\cdot$}
\put(18,111.58){$\cdot$}
\put(17.5,112.0){$\cdot$}
\put(17,112.44){$\cdot$}
\put(16.5,112.88){$\cdot$}
\put(16,113.34){$\cdot$}
\put(15.5,113.81){$\cdot$}
\put(15,114.29){$\cdot$}
\put(14.5,114.79){$\cdot$}
\put(14,115.3){$\cdot$}
\put(13.5,115.83){$\cdot$}
\put(13,116.37){$\cdot$}
\put(12.5,116.93){$\cdot$}
\put(12,117.5){$\cdot$}
\put(11.5,118.1){$\cdot$}
\put(11,118.71){$\cdot$}
\put(10.5,119.34){$\cdot$}
\put(10,120.00){$\cdot$}
\put(9.5,120.68){$\cdot$}
\put(9,121.38){$\cdot$}
\put(8.5,122.11){$\cdot$}
\put(8,122.87){$\cdot$}
\put(7.5,123.66){$\cdot$}
\put(7,124.48){$\cdot$}
\put(6.5,125.35){$\cdot$}
\put(6,126.25){$\cdot$}
\put(5.5,127.2){$\cdot$}
\put(5,128.2){$\cdot$}
\put(4.5,129.27){$\cdot$}
\put(4,130.4){$\cdot$}
\put(3.5,131.62){$\cdot$}
\put(3,132.94){$\cdot$}
\put(2.5,134.39){$\cdot$}
\put(2,136.00){$\cdot$}
\put(1.5,137.84){$\cdot$}
\put(1,140.05){$\cdot$}
\put(0.5,142.95){$\cdot$}
\put(0,150){$\cdot$}
% shere IV quadrant 2
\put(50,100){$\cdot$}
\put(50.5,100){$\cdot$}
\put(51,100.01){$\cdot$}
\put(51.5,100.02){$\cdot$}
\put(52,100.04){$\cdot$}
\put(52.5,100.06){$\cdot$}
\put(53,100.09){$\cdot$}
\put(53.5,100.12){$\cdot$}
\put(54,100.16){$\cdot$}
\put(54.5,100.2){$\cdot$}
\put(55,100.25){$\cdot$}
\put(55.5,100.3){$\cdot$}
\put(56,100.36){$\cdot$}
\put(56.5,100.42){$\cdot$}
\put(57,100.49){$\cdot$}
\put(57.5,100.56){$\cdot$}
\put(58,100.64){$\cdot$}
\put(58.5,100.73){$\cdot$}
\put(59,100.82){$\cdot$}
\put(59.5,100.91){$\cdot$}
\put(60,101.01){$\cdot$}
\put(60.5,101.11){$\cdot$}
\put(61,101.22){$\cdot$}
\put(61.5,101.34){$\cdot$}
\put(62,101.46){$\cdot$}
\put(62.5,101.59){$\cdot$}
\put(63,101.72){$\cdot$}
\put(63.5,101.86){$\cdot$}
\put(64,102.0){$\cdot$}
\put(64.5,102.15){$\cdot$}
\put(65,102.3){$\cdot$}
\put(65.5,102.46){$\cdot$}
\put(66,102.63){$\cdot$}
\put(66.5,102.8){$\cdot$}
\put(67,102.98){$\cdot$}
\put(67.5,103.16){$\cdot$}
\put(68,103.35){$\cdot$}
\put(68.5,103.55){$\cdot$}
\put(69,103.75){$\cdot$}
\put(69.5,103.96){$\cdot$}
\put(70,104.17){$\cdot$}
\put(70.5,104.39){$\cdot$}
\put(71,104.62){$\cdot$}
\put(71.5,104.86){$\cdot$}
\put(72,105.1){$\cdot$}
\put(72.5,105.35){$\cdot$}
\put(73,105.6){$\cdot$}
\put(73.5,105.87){$\cdot$}
\put(74,106.14){$\cdot$}
\put(74.5,106.41){$\cdot$}
\put(75,106.7){$\cdot$}
\put(75.5,106.99){$\cdot$}
\put(76,107.29){$\cdot$}
\put(76.5,107.6){$\cdot$}
\put(77,107.9){$\cdot$}
\put(77.5,108.24){$\cdot$}
\put(78,108.57){$\cdot$}
\put(78.5,108.92){$\cdot$}
\put(79,109.27){$\cdot$}
\put(79.5,109.63){$\cdot$}
\put(80,110.0){$\cdot$}
\put(80.5,110.38){$\cdot$}
\put(81,110.77){$\cdot$}
\put(81.5,111.17){$\cdot$}
\put(82,111.58){$\cdot$}
\put(82.5,112.0){$\cdot$}
\put(83,112.44){$\cdot$}
\put(83.5,112.88){$\cdot$}
\put(84,113.34){$\cdot$}
\put(84.5,113.81){$\cdot$}
\put(85,114.29){$\cdot$}
\put(85.5,114.79){$\cdot$}
\put(86,115.3){$\cdot$}
\put(86.5,115.83){$\cdot$}
\put(87,116.37){$\cdot$}
\put(87.5,116.93){$\cdot$}
\put(88,117.5){$\cdot$}
\put(88.5,118.1){$\cdot$}
\put(89,118.71){$\cdot$}
\put(89.5,119.34){$\cdot$}
\put(90,120.00){$\cdot$}
\put(90.5,120.68){$\cdot$}
\put(91,121.38){$\cdot$}
\put(91.5,122.11){$\cdot$}
\put(92,122.87){$\cdot$}
\put(92.5,123.66){$\cdot$}
\put(93,124.48){$\cdot$}
\put(93.5,125.35){$\cdot$}
\put(94,126.25){$\cdot$}
\put(94.5,127.2){$\cdot$}
\put(95,128.2){$\cdot$}
\put(95.5,129.27){$\cdot$}
\put(96,130.4){$\cdot$}
\put(96.5,131.62){$\cdot$}
\put(97,132.94){$\cdot$}
\put(97.5,134.39){$\cdot$}
\put(98,136.00){$\cdot$}
\put(98.5,137.84){$\cdot$}
\put(99,140.05){$\cdot$}
\put(99.5,142.95){$\cdot$}
\put(100,150){$\cdot$}
%bondary
\put(0,50){$\cdot$}
\put(0.01,50.5){$\cdot$}
\put(0.01,51){$\cdot$}
\put(0.02,51.5){$\cdot$}
\put(0.04,52){$\cdot$}
\put(0.06,52.5){$\cdot$}
\put(0.09,53){$\cdot$}
\put(0.12,53.5){$\cdot$}
\put(0.16,54){$\cdot$}
\put(0.2,54.5){$\cdot$}
\put(0.25,55){$\cdot$}
\put(0.3,55.5){$\cdot$}
\put(0.36,56){$\cdot$}
\put(0.42,56.5){$\cdot$}
\put(0.49,57){$\cdot$}
\put(0.56,57.5){$\cdot$}
\put(0.64,58){$\cdot$}
\put(0.78,58.5){$\cdot$}
\put(0.82,59){$\cdot$}
\put(0.91,59.5){$\cdot$}
\put(1.01,60){$\cdot$}
\put(1.11,60.5){$\cdot$}
\put(1.22,61){$\cdot$}
\put(1.34,61.5){$\cdot$}
\put(1.46,62){$\cdot$}
\put(1.59,62.5){$\cdot$}
\put(1.72,63){$\cdot$}
\put(1.86,63.5){$\cdot$}
\put(2,64){$\cdot$}
\put(2.15,64.5){$\cdot$}
\put(2.3,65){$\cdot$}
\put(2.46,65.5){$\cdot$}
\put(2.63,66){$\cdot$}
\put(2.6,66.5){$\cdot$}
\put(2.98,67){$\cdot$}
\put(3.16,67.5){$\cdot$}
\put(3.35,68){$\cdot$}
\put(3.54,68.5){$\cdot$}
\put(3.75,69){$\cdot$}
\put(3.96,69.5){$\cdot$}
\put(4.17,70){$\cdot$}
\put(4.39,70.5){$\cdot$}
\put(4.62,71){$\cdot$}
\put(4.86,71.5){$\cdot$}
\put(5.1,72){$\cdot$}
\put(5.35,72.5){$\cdot$}
\put(5.6,73){$\cdot$}
\put(5.87,73.5){$\cdot$}
\put(6.14,74){$\cdot$}
\put(6.41,74.5){$\cdot$}
\put(6.7,75){$\cdot$}

\put(0,50){$\cdot$}
\put(0.01,49.5){$\cdot$}
\put(0.01,49){$\cdot$}
\put(0.02,48.5){$\cdot$}
\put(0.04,48){$\cdot$}
\put(0.06,47.5){$\cdot$}
\put(0.09,47){$\cdot$}
\put(0.12,46.5){$\cdot$}
\put(0.16,46){$\cdot$}
\put(0.2,45.5){$\cdot$}
\put(0.25,45){$\cdot$}
\put(0.3,44.5){$\cdot$}
\put(0.36,44){$\cdot$}
\put(0.42,43.5){$\cdot$}
\put(0.49,43){$\cdot$}
\put(0.56,42.5){$\cdot$}
\put(0.64,42){$\cdot$}
\put(0.78,41.5){$\cdot$}
\put(0.82,41){$\cdot$}
\put(0.91,40.5){$\cdot$}
\put(1.01,40){$\cdot$}
\put(1.11,39.5){$\cdot$}
\put(1.22,39){$\cdot$}
\put(1.34,38.5){$\cdot$}
\put(1.46,38){$\cdot$}
\put(1.59,37.5){$\cdot$}
\put(1.72,37){$\cdot$}
\put(1.86,36.5){$\cdot$}
\put(2,36){$\cdot$}
\put(2.15,35.5){$\cdot$}
\put(2.3,35){$\cdot$}
\put(2.46,34.5){$\cdot$}
\put(2.63,34){$\cdot$}
\put(2.6,33.5){$\cdot$}
\put(2.98,33){$\cdot$}
\put(3.16,32.5){$\cdot$}
\put(3.35,32){$\cdot$}
\put(3.54,31.5){$\cdot$}
\put(3.75,31){$\cdot$}
\put(3.96,30.5){$\cdot$}
\put(4.17,30){$\cdot$}
\put(4.39,29.5){$\cdot$}
\put(4.62,29){$\cdot$}
\put(4.86,28.5){$\cdot$}
\put(5.1,28){$\cdot$}
\put(5.35,27.5){$\cdot$}
\put(5.6,27){$\cdot$}
\put(5.87,26.5){$\cdot$}
\put(6.14,26){$\cdot$}
\put(6.41,25.5){$\cdot$}
\put(6.7,25){$\cdot$}
%bondary up
\put(0,150){$\cdot$}
\put(0.01,150.5){$\cdot$}
\put(0.01,151){$\cdot$}
\put(0.02,151.5){$\cdot$}
\put(0.04,152){$\cdot$}
\put(0.06,152.5){$\cdot$}
\put(0.09,153){$\cdot$}
\put(0.12,153.5){$\cdot$}
\put(0.16,154){$\cdot$}
\put(0.2,154.5){$\cdot$}
\put(0.25,155){$\cdot$}
\put(0.3,155.5){$\cdot$}
\put(0.36,156){$\cdot$}
\put(0.42,156.5){$\cdot$}
\put(0.49,157){$\cdot$}
\put(0.56,157.5){$\cdot$}
\put(0.64,158){$\cdot$}
\put(0.78,158.5){$\cdot$}
\put(0.82,159){$\cdot$}
\put(0.91,159.5){$\cdot$}
\put(1.01,160){$\cdot$}
\put(1.11,160.5){$\cdot$}
\put(1.22,161){$\cdot$}
\put(1.34,161.5){$\cdot$}
\put(1.46,162){$\cdot$}
\put(1.59,162.5){$\cdot$}
\put(1.72,163){$\cdot$}
\put(1.86,163.5){$\cdot$}
\put(2,164){$\cdot$}
\put(2.15,164.5){$\cdot$}
\put(2.3,165){$\cdot$}
\put(2.46,165.5){$\cdot$}
\put(2.63,166){$\cdot$}
\put(2.6,166.5){$\cdot$}
\put(2.98,167){$\cdot$}
\put(3.16,167.5){$\cdot$}
\put(3.35,168){$\cdot$}
\put(3.54,168.5){$\cdot$}
\put(3.75,169){$\cdot$}
\put(3.96,169.5){$\cdot$}
\put(4.17,170){$\cdot$}
\put(4.39,170.5){$\cdot$}
\put(4.62,171){$\cdot$}
\put(4.86,171.5){$\cdot$}
\put(5.1,172){$\cdot$}
\put(5.35,172.5){$\cdot$}
\put(5.6,173){$\cdot$}
\put(5.87,173.5){$\cdot$}
\put(6.14,174){$\cdot$}
\put(6.41,174.5){$\cdot$}
\put(6.7,175){$\cdot$}

\put(0,150){$\cdot$}
\put(0.01,149.5){$\cdot$}
\put(0.01,149){$\cdot$}
\put(0.02,148.5){$\cdot$}
\put(0.04,148){$\cdot$}
\put(0.06,147.5){$\cdot$}
\put(0.09,147){$\cdot$}
\put(0.12,146.5){$\cdot$}
\put(0.16,146){$\cdot$}
\put(0.2,145.5){$\cdot$}
\put(0.25,145){$\cdot$}
\put(0.3,144.5){$\cdot$}
\put(0.36,144){$\cdot$}
\put(0.42,143.5){$\cdot$}
\put(0.49,143){$\cdot$}
\put(0.56,142.5){$\cdot$}
\put(0.64,142){$\cdot$}
\put(0.78,141.5){$\cdot$}
\put(0.82,141){$\cdot$}
\put(0.91,140.5){$\cdot$}
\put(1.01,140){$\cdot$}
\put(1.11,139.5){$\cdot$}
\put(1.22,139){$\cdot$}
\put(1.34,138.5){$\cdot$}
\put(1.46,138){$\cdot$}
\put(1.59,137.5){$\cdot$}
\put(1.72,137){$\cdot$}
\put(1.86,136.5){$\cdot$}
\put(2,136){$\cdot$}
\put(2.15,135.5){$\cdot$}
\put(2.3,135){$\cdot$}
\put(2.46,134.5){$\cdot$}
\put(2.63,134){$\cdot$}
\put(2.6,133.5){$\cdot$}
\put(2.98,133){$\cdot$}
\put(3.16,132.5){$\cdot$}
\put(3.35,132){$\cdot$}
\put(3.54,131.5){$\cdot$}
\put(3.75,131){$\cdot$}
\put(3.96,130.5){$\cdot$}
\put(4.17,130){$\cdot$}
\put(4.39,129.5){$\cdot$}
\put(4.62,129){$\cdot$}
\put(4.86,128.5){$\cdot$}
\put(5.1,128){$\cdot$}
\put(5.35,127.5){$\cdot$}
\put(5.6,127){$\cdot$}
\put(5.87,126.5){$\cdot$}
\put(6.14,126){$\cdot$}
\put(6.41,125.5){$\cdot$}
\put(6.7,125){$\cdot$}

%bondary down right
\put(100,50){$\cdot$}
\put(99.99,50.5){$\cdot$}
\put(99.99,51){$\cdot$}
\put(99.98,51.5){$\cdot$}
\put(99.96,52){$\cdot$}
\put(99.94,52.5){$\cdot$}
\put(99.91,53){$\cdot$}
\put(99.88,53.5){$\cdot$}
\put(99.84,54){$\cdot$}
\put(99.8,54.5){$\cdot$}
\put(99.75,55){$\cdot$}
\put(99.7,55.5){$\cdot$}
\put(99.64,56){$\cdot$}
\put(99.58,56.5){$\cdot$}
\put(99.51,57){$\cdot$}
\put(99.44,57.5){$\cdot$}
\put(99.36,58){$\cdot$}
\put(99.22,58.5){$\cdot$}
\put(99.18,59){$\cdot$}
\put(99.09,59.5){$\cdot$}
\put(98.99,60){$\cdot$}
\put(98.89,60.5){$\cdot$}
\put(98.78,61){$\cdot$}
\put(98.66,61.5){$\cdot$}
\put(98.54,62){$\cdot$}
\put(98.41,62.5){$\cdot$}
\put(98.28,63){$\cdot$}
\put(98.14,63.5){$\cdot$}
\put(98,64){$\cdot$}
\put(97.85,64.5){$\cdot$}
\put(97.7,65){$\cdot$}
\put(97.54,65.5){$\cdot$}
\put(97.37,66){$\cdot$}
\put(97.4,66.5){$\cdot$}
\put(97.02,67){$\cdot$}
\put(96.84,67.5){$\cdot$}
\put(96.65,68){$\cdot$}
\put(96.46,68.5){$\cdot$}
\put(96.25,69){$\cdot$}
\put(96.04,69.5){$\cdot$}
\put(95.83,70){$\cdot$}
\put(95.61,70.5){$\cdot$}
\put(95.38,71){$\cdot$}
\put(95.14,71.5){$\cdot$}
\put(94.9,72){$\cdot$}
\put(94.65,72.5){$\cdot$}
\put(94.4,73){$\cdot$}
\put(94.13,73.5){$\cdot$}
\put(93.86,74){$\cdot$}
\put(93.59,74.5){$\cdot$}
\put(93.3,75){$\cdot$}

\put(100,50){$\cdot$}
\put(99.99,49.5){$\cdot$}
\put(99.99,49){$\cdot$}
\put(99.98,48.5){$\cdot$}
\put(99.96,48){$\cdot$}
\put(99.94,47.5){$\cdot$}
\put(99.91,47){$\cdot$}
\put(99.88,46.5){$\cdot$}
\put(99.84,46){$\cdot$}
\put(99.8,45.5){$\cdot$}
\put(99.75,45){$\cdot$}
\put(99.7,44.5){$\cdot$}
\put(99.64,44){$\cdot$}
\put(99.58,43.5){$\cdot$}
\put(99.51,43){$\cdot$}
\put(99.44,42.5){$\cdot$}
\put(99.36,42){$\cdot$}
\put(99.22,41.5){$\cdot$}
\put(99.18,41){$\cdot$}
\put(99.09,40.5){$\cdot$}
\put(98.99,40){$\cdot$}
\put(98.89,39.5){$\cdot$}
\put(98.78,39){$\cdot$}
\put(98.66,38.5){$\cdot$}
\put(98.54,38){$\cdot$}
\put(98.41,37.5){$\cdot$}
\put(98.28,37){$\cdot$}
\put(98.14,36.5){$\cdot$}
\put(98,36){$\cdot$}
\put(97.85,35.5){$\cdot$}
\put(97.7,35){$\cdot$}
\put(97.54,34.5){$\cdot$}
\put(97.37,34){$\cdot$}
\put(97.4,33.5){$\cdot$}
\put(97.02,33){$\cdot$}
\put(96.84,32.5){$\cdot$}
\put(96.65,32){$\cdot$}
\put(96.46,31.5){$\cdot$}
\put(96.25,31){$\cdot$}
\put(96.04,30.5){$\cdot$}
\put(95.83,30){$\cdot$}
\put(95.61,29.5){$\cdot$}
\put(95.38,29){$\cdot$}
\put(95.14,28.5){$\cdot$}
\put(94.9,28){$\cdot$}
\put(94.65,27.5){$\cdot$}
\put(94.4,27){$\cdot$}
\put(94.13,26.5){$\cdot$}
\put(93.86,26){$\cdot$}
\put(93.59,25.5){$\cdot$}
\put(93.3,25){$\cdot$}

%bondary up right
\put(100,150){$\cdot$}
\put(99.99,150.5){$\cdot$}
\put(99.99,151){$\cdot$}
\put(99.98,151.5){$\cdot$}
\put(99.96,152){$\cdot$}
\put(99.94,152.5){$\cdot$}
\put(99.91,153){$\cdot$}
\put(99.88,153.5){$\cdot$}
\put(99.84,154){$\cdot$}
\put(99.8,154.5){$\cdot$}
\put(99.75,155){$\cdot$}
\put(99.7,155.5){$\cdot$}
\put(99.64,156){$\cdot$}
\put(99.58,156.5){$\cdot$}
\put(99.51,157){$\cdot$}
\put(99.44,157.5){$\cdot$}
\put(99.36,158){$\cdot$}
\put(99.22,158.5){$\cdot$}
\put(99.18,159){$\cdot$}
\put(99.09,159.5){$\cdot$}
\put(98.99,160){$\cdot$}
\put(98.89,160.5){$\cdot$}
\put(98.78,161){$\cdot$}
\put(98.66,161.5){$\cdot$}
\put(98.54,162){$\cdot$}
\put(98.41,162.5){$\cdot$}
\put(98.28,163){$\cdot$}
\put(98.14,163.5){$\cdot$}
\put(98,164){$\cdot$}
\put(97.85,164.5){$\cdot$}
\put(97.7,165){$\cdot$}
\put(97.54,165.5){$\cdot$}
\put(97.37,166){$\cdot$}
\put(97.4,166.5){$\cdot$}
\put(97.02,167){$\cdot$}
\put(96.84,167.5){$\cdot$}
\put(96.65,168){$\cdot$}
\put(96.46,168.5){$\cdot$}
\put(96.25,169){$\cdot$}
\put(96.04,169.5){$\cdot$}
\put(95.83,170){$\cdot$}
\put(95.61,170.5){$\cdot$}
\put(95.38,171){$\cdot$}
\put(95.14,171.5){$\cdot$}
\put(94.9,172){$\cdot$}
\put(94.65,172.5){$\cdot$}
\put(94.4,173){$\cdot$}
\put(94.13,173.5){$\cdot$}
\put(93.86,174){$\cdot$}
\put(93.59,174.5){$\cdot$}
\put(93.3,175){$\cdot$}

\put(100,150){$\cdot$}
\put(99.99,149.5){$\cdot$}
\put(99.99,149){$\cdot$}
\put(99.98,148.5){$\cdot$}
\put(99.96,148){$\cdot$}
\put(99.94,147.5){$\cdot$}
\put(99.91,147){$\cdot$}
\put(99.88,146.5){$\cdot$}
\put(99.84,146){$\cdot$}
\put(99.8,145.5){$\cdot$}
\put(99.75,145){$\cdot$}
\put(99.7,144.5){$\cdot$}
\put(99.64,144){$\cdot$}
\put(99.58,143.5){$\cdot$}
\put(99.51,143){$\cdot$}
\put(99.44,142.5){$\cdot$}
\put(99.36,142){$\cdot$}
\put(99.22,141.5){$\cdot$}
\put(99.18,141){$\cdot$}
\put(99.09,140.5){$\cdot$}
\put(98.99,140){$\cdot$}
\put(98.89,139.5){$\cdot$}
\put(98.78,139){$\cdot$}
\put(98.66,138.5){$\cdot$}
\put(98.54,138){$\cdot$}
\put(98.41,137.5){$\cdot$}
\put(98.28,137){$\cdot$}
\put(98.14,136.5){$\cdot$}
\put(98,136){$\cdot$}
\put(97.85,135.5){$\cdot$}
\put(97.7,135){$\cdot$}
\put(97.54,134.5){$\cdot$}
\put(97.37,134){$\cdot$}
\put(97.4,133.5){$\cdot$}
\put(97.02,133){$\cdot$}
\put(96.84,132.5){$\cdot$}
\put(96.65,132){$\cdot$}
\put(96.46,131.5){$\cdot$}
\put(96.25,131){$\cdot$}
\put(96.04,130.5){$\cdot$}
\put(95.83,130){$\cdot$}
\put(95.61,129.5){$\cdot$}
\put(95.38,129){$\cdot$}
\put(95.14,128.5){$\cdot$}
\put(94.9,128){$\cdot$}
\put(94.65,127.5){$\cdot$}
\put(94.4,127){$\cdot$}
\put(94.13,126.5){$\cdot$}
\put(93.86,126){$\cdot$}
\put(93.59,125.5){$\cdot$}
\put(93.3,125){$\cdot$}

\end{picture}
\]
\vspace{4ex}
%\caption{
\begin{center}
\begin{minipage}{25pc}{{\renewcommand{\baselinestretch}{1.2}\small
{\rm Figure 1}\;
Two--dimensional complex sphere $z^2_1+z^2_2+z^2_3=r^2$ in
three--dimensional complex space $\C^3$. The space $\C^3$ is isometric
to the bivector space $\R^6$. The dual (complex conjugate) sphere
$\overset{\ast}{z}_1{}^2+\overset{\ast}{z}_2{}^2+\overset{\ast}{z}_3{}^2=
\overset{\ast}{r}{}^2$ is a mirror image of the complex sphere with
respect to the hyperplane. The associated hyperspherical functions
$\fM_l^{m}(\varphi^c,\theta^c,0)$ 
($\fM_{\dot{l}}^{\dot{m}}(\dot{\varphi}^c,\dot{\theta}^c,0)$)
are defined on the surface of the complex (dual) sphere.}}
\end{minipage}
\end{center}
\label{Sphere}
%\medskip
\end{figure}
\subsection{Matrix elements of principal and supplementary series of
representations}
As it has been shown in \cite{Nai58}, for the case of principal unitary series
representations of $SL(2,\C)$
there exists an analog of the spinor representation
formula (\ref{Rep}):
\begin{equation}\label{Principal}
T^\alpha f(z)=(a_{12}z+a_{22})^{\frac{\lambda}{2}+i\frac{\rho}{2}-1}
\overline{(a_{12}z+a_{22})}^{-\frac{\lambda}{2}+i\frac{\rho}{2}-1}
f\left(\frac{a_{11}z+a_{21}}{a_{12}z+a_{22}}\right),
\end{equation}
where $f(z)$ is a measurable functions of the Hilbert space $L_2(Z)$,
satisfying the condition $\int|f(z)|^2dz<\infty$, $z=x+iy$. At this point,
the numbers
$l_0$, $l_1$ and $\lambda$, $\rho$ are related by the formulae
\begin{gather}
l_0=\left|\frac{\lambda}{2}\right|,\quad
l_1=-i(\sign\lambda)\frac{\rho}{2}\quad\text{if $m\neq 0$},\nonumber\\
l_0=0,\quad
l_1=\pm i\frac{\rho}{2}\quad\text{if $m=0$}.\nonumber
\end{gather}
A totality of all representations
$a\rightarrow T^\alpha$, corresponding to all possible pairs
$\lambda$, $\rho$, is called a principal series of representations of the
group
$SL(2,\C)$. At this point, a comparison of (\ref{Principal}) 
with the formula (\ref{Rep}) for the spinor representation
$\fS_l$ shows that the both formulas have the same structure;
only the exponents at the factors
$(a_{12}z+a_{22})$, $\overline{(a_{12}z+a_{22})}$ and the functions $f(z)$
are different. In the case of spinor representations the functions
$f(z)$ are polynomials
$p(z,\bar{z})$ in the spaces
$\Sym_{(k,r)}$, and in the case of a representation
$\fS_{\lambda,\rho}$ of the principal series $f(z)$ are functions from
the Hilbert space $L_2(Z)$.

%As known, a representation $S_l$ of the group $SU(2)$ is realized in terms
%of the functions $P^l_{mn}(\cos\theta)$. 
%\begin{theorem}[{\rm Naimark \cite{Nai58}}]
%The representation $S_l$ is contained in
%$\fS_{\lambda,\rho}$ no more then one time. At this point, $S_l$ is
%contained in
%$\fS_{\lambda,\rho}$, when $\frac{\lambda}{2}$ is one from the numbers
%$-l,-l+1,\ldots,l$.
%\end{theorem}
Therefore, matrix elements of the principal series representations of the
Lorentz group, making infinite-dimensional matrix, have the form
\begin{multline}
t^{-\frac{1}{2}+i\rho}_{mn}(\mathfrak{g})=
e^{-m(\epsilon+i\varphi)-n(\varepsilon+i\psi)}
Z^{-\frac{1}{2}+i\rho}_{mn}=
e^{-m(\epsilon+i\varphi)-n(\varepsilon+i\psi)}\times\\[0.2cm]
\sum^{+\infty}_{\lambda=-\infty}
\sum^{\ld\frac{\lambda}{2}\rd}_{k=-\frac{\lambda}{2}}i^{m-k}
\sqrt{\Gamma(\tfrac{\lambda}{2}-m+1)
\Gamma(\tfrac{\lambda}{2}+m+1)\Gamma(\tfrac{\lambda}{2}-k+1)
\Gamma(\tfrac{\lambda}{2}+k+1)}\times\\
\cos^{\lambda}\frac{\theta}{2}\tg^{m-k}\frac{\theta}{2}\times\\[0.2cm]
\sum^{\ld\min\left(\frac{\lambda}{2}-m,\frac{\lambda}{2}+k\right)
\rd}_{j=\max(0,k-m)}
\frac{i^{2j}\tg^{2j}\dfrac{\theta}{2}}
{\Gamma(j+1)\Gamma(\tfrac{\lambda}{2}-m-j+1)
\Gamma(\tfrac{\lambda}{2}+k-j+1)\Gamma(m-k+j+1)}\times\\[0.2cm]
\sqrt{\Gamma(\tfrac{1}{2}+i\rho-n)
\Gamma(\tfrac{1}{2}+i\rho+n)\Gamma(\tfrac{1}{2}+i\rho-k)
\Gamma(\tfrac{1}{2}+i\rho+k)}
\ch^{-1+2i\rho}\frac{\tau}{2}\tnh^{n-k}\frac{\tau}{2}\times\\[0.2cm]
\sum^{\ld\min\left(\frac{\lambda}{2}-n,\frac{\lambda}{2}+k\right)
\rd}_{s=\max(0,k-n)}
\frac{\tnh^{2s}\dfrac{\tau}{2}}
{\Gamma(s+1)\Gamma(\frac{1}{2}+i\rho-n-s)
\Gamma(\frac{1}{2}+i\rho+k-s)\Gamma(n-k+s+1)}.\label{MPrincip}
\end{multline}
Thus, the matrix elements of the principal unitary series representations 
of the group $SL(2,\C)$ are expressed via the function
\begin{equation}\label{MEPS}
\fM^{-\frac{1}{2}+i\rho}_{mn}(\fg)=e^{-m(\epsilon+i\varphi)}
Z^{-\frac{1}{2}+i\rho}_{mn}(\cos\theta^c)e^{-n(\varepsilon+i\psi)},
\end{equation}
where
\[
Z^{-\frac{1}{2}+i\rho}_{mn}(\cos\theta^c)=\sum^{+\infty}_{\lambda=-\infty}
\sum^{\ld\frac{\lambda}{2}\rd}_{k=-\frac{\lambda}{2}}
P^{\frac{\lambda}{2}}_{mk}(\cos\theta)\fP^{-\frac{1}{2}+i\rho}_{kn}(\ch\tau).
\]
In the case of associated functions ($n=0$) we obtain
\begin{equation}\label{AHPS}
Z^m_{-\frac{1}{2}+i\rho}(\cos\theta^c)=\sum^{+\infty}_{\lambda=-\infty}
\sum^{\ld\frac{\lambda}{2}\rd}_{k=-\frac{\lambda}{2}}
P^{\frac{\lambda}{2}}_{mk}(\cos\theta)\fP^k_{-\frac{1}{2}+i\rho}(\ch\tau),
\end{equation}
where $\fP^k_{-\frac{1}{2}+i\rho}(\ch\tau)$ are {\it conical functions}
(see \cite{Bat}). In this case our result agrees with the paper \cite{FNR66},
where matrix elements (eigenfunctions of Casimir operators) of noncompact
rotation groups are expressed in terms of conical and spherical functions
(see also \cite{Vil68}).

Further, at $\lambda=0$ and $\rho=i\sigma$ from (\ref{Principal}) it follows
that
\[
T^\alpha f(z)=|a_{12}z+a_{22}|^{-2-\sigma}f
\left(\frac{a_{11}z+a_{21}}{a_{12}z+a_{22}}\right).
\]
This formula defines an unitary representation $a\rightarrow T^\alpha$ of
supplementary series
$\fD_\sigma$ of the group $SL(2,\C)$. At this point, for the supplementary
series the relations
\[
l_0=0,\quad l_1=\pm\frac{\sigma}{2}
\]
hold. In turn, the representation $S_l$ of the group $SU(2)$ is contained
in the representation $\fD_\sigma$ of supplementary series when
$l$ is an integer number. In this case $S_l$ is contained in $\fD_\sigma$
exactly one time and the number $\frac{\lambda}{2}=0$ is one from the set
$-l,-l+1,\ldots,l$ \cite{Nai58}.

Thus, matrix elements of supplementary series appear as a particular case
of the matrix elements of the principal series at $\lambda=0$
and $\rho=i\sigma$:
\begin{multline}
t^{-\frac{1}{2}-\sigma}_{mn}(\mathfrak{g})=
e^{-m(\epsilon+i\varphi)-n(\varepsilon+i\psi)}
Z^{-\frac{1}{2}-\sigma}_{mn}=
e^{-m(\epsilon+i\varphi)-n(\varepsilon+i\psi)}\times\\[0.2cm]
\sqrt{\Gamma(\tfrac{1}{2}-\sigma-n)
\Gamma(\tfrac{1}{2}-\sigma+n)\Gamma(\tfrac{1}{2}-\sigma-k)
\Gamma(\tfrac{1}{2}-\sigma+k)}
\ch^{-1-2\sigma}\frac{\tau}{2}\tnh^{n-k}\frac{\tau}{2}\times\\[0.2cm]
\sum^{\infty}_{s=\max(0,k-n)}
\frac{\tnh^{2s}\dfrac{\tau}{2}}
{\Gamma(s+1)\Gamma(\tfrac{1}{2}-\sigma-n-s)
\Gamma(\tfrac{1}{2}-\sigma+k-s)\Gamma(n-k+s+1)}.\label{MSupl}
\end{multline}
Or
\[
\fM^{-\frac{1}{2}-\sigma}_{mn}(\fg)=e^{-m(\epsilon+i\varphi)}
\fP^{-\frac{1}{2}-\sigma}_{mn}(\ch\tau)e^{-n(\varepsilon+i\psi)},
\]
that is, the hyperspherical function $Z^{-\frac{1}{2}+i\rho}_{mn}(\cos\theta^c)$
in the case of supplementary series is degenerated to the Jacobi function
$\fP^{-\frac{1}{2}-\sigma}_{mn}(\ch\tau)$. For the associated functions of
supplementary series we obtain
\[
\fP^m_{-\frac{1}{2}-\sigma}(\fg)=e^{-m(\epsilon+i\varphi)}
\fP^m_{-\frac{1}{2}-\sigma}(\ch\tau).
\]

\section{Harmonic analysis on the group $SL(2,\C)$}
First of all, on the group $SL(2,\C)$
there exists an
invariant measure $d\fg$, that is, such a
measure that for any finite continuous function
$f(\fg)$ on $SL(2,\C)$ the following equality
\[
\int f(\fg)d\fg=\int f(\fg_0\fg)d\fg=\int f(\fg\fg_0)d\fg=\int f(\fg^{-1})d\fg
\]
holds.
Applying the equations (\ref{X1})--(\ref{Y3}), we express
the Haar measure (left or right) in terms of the parameters (\ref{CEA}):
\begin{equation}\label{HA1}
d\fg=\sin\theta^c\sin\dot{\theta}^cd\theta d\varphi d\psi d\tau d\epsilon
d\varepsilon.
\end{equation}
Thus, an invariant integration on the group $SL(2,\C)$
is defined by the formula
\[
\int\limits_{SL(2,\C)}f(g)d\fg=\frac{1}{32\pi^4}
\int\limits^{+\infty}_{-\infty}
\int\limits^{+\infty}_{-\infty}
\int\limits^{+\infty}_{-\infty}
\int\limits^{2\pi}_{-2\pi}
\int\limits^{2\pi}_0
\int\limits^\pi_0
f(\theta,\varphi,\psi,\tau,\epsilon,\varepsilon)
\sin\theta^c\sin\dot{\theta}^cd\theta d\varphi d\psi d\tau d\epsilon
d\varepsilon.
\]
When we consider finite-dimensional (spinor) representations of
$SL(2,\C)$, we come naturally to a local isomorphism
$SU(2)\otimes SU(2)\simeq SL(2,\C)$ considered by many authors
\cite{HS70,Ryd85}.
Since a dimension of the spinor representation $T_{\fg}$ of
$SU(2)\otimes SU(2)$ is equal to $(2l+1)(2\dot{l}+1)$, then the functions
$\sqrt{(2l+1)(2\dot{l}+1)}t^l_{mn}(\fg)$ form a full orthogonal normalized
system on this group with respect to the invariant measure $d\fg$. At this point,
the index $l$ runs all possible integer or half-integer non-negative values,
and the indices $m$ and $n$ run the values
$-l,-l+1,\ldots,l-1,l$. In virtue of (\ref{HS}) the matrix elements
$t^l_{mn}$ are expressed via the generalized hyperspherical function
$t^l_{mn}(\fg)=\fM^l_{mn}(\varphi,\epsilon,\theta,\tau,\psi,\varepsilon)$.
Therefore,
\begin{equation}\label{HA2}
\int\limits_{SU(2)\otimes SU(2)}\fM^l_{mn}(\fg)
\overline{\fM^l_{mn}(\fg)}d\fg=\frac{32\pi^4}{(2l+1)(2\dot{l}+1)}
\delta(\fg^\prime-\fg),
\end{equation}
where $\delta(\fg^\prime-\fg)$ is a $\delta$-function on the group
$SU(2)\otimes SU(2)$. An explicit form of $\delta$-function is
\begin{multline}
\delta(\fg^\prime-\fg)=\delta(\varphi^\prime-\varphi)
\delta(\epsilon^\prime-\epsilon)\delta(\cos\theta^\prime\ch\tau^\prime-
\cos\theta\ch\tau)\times\\
\times\delta(\sin\theta^\prime\sh\tau^\prime-\sin\theta\sh\tau)
\delta(\psi^\prime-\psi)\delta(\varepsilon^\prime-\varepsilon).\nonumber
\end{multline}
Substituting into (\ref{HA2}) the expression
\[
\fM^l_{mn}(\fg)=e^{-m(\epsilon+i\varphi)}Z^l_{mn}e^{-n(\varepsilon+i\psi)}
\]
and taking into account (\ref{HA1}), we obtain
\begin{multline}
\int\limits_{SU(2)\otimes SU(2)}
Z^l_{mn}\overline{Z^s_{pq}}e^{-(m+p)\epsilon}e^{-i(m-p)\varphi}\times\\
\times e^{-(n+q)\varepsilon}e^{-i(n-q)\psi}
\sin\theta^c\sin\dot{\theta}^cd\theta d\varphi d\psi d\tau d\epsilon
d\varepsilon=\frac{32\pi^4\delta_{ls}\delta_{mp}\delta_{nq}
\delta(\fg^\prime-\fg)}{(2l+1)(2\dot{l}+1)}.\nonumber
\end{multline}

Thus, any square integrable function $f(\varphi^c,\theta^c,\psi^c)$ on the
group $SU(2)\otimes SU(2)$, such that
\[
\int\limits_{SU(2)\otimes SU(2)}
|f(\varphi^c,\theta^c,\psi^c)|^2
\sin\theta^c\sin\dot{\theta}^cd\theta d\varphi d\psi d\tau d\epsilon
d\varepsilon
<+\infty,
\]
is expanded into a convergent (on an average) Fourier series on
$SU(2)\otimes SU(2)$,
\begin{equation}\label{CFS}
f(\varphi^c,\theta^c,\psi^c)=\sum^\infty_{l=0}\sum^l_{m=-l}\sum^l_{n=-l}
\alpha^l_{mn}e^{-m(\epsilon+i\varphi)}Z^l_{mn}(\cos\theta^c)
e^{-n(\varepsilon+i\psi)},
\end{equation}
where
\begin{multline}
\alpha^l_{mn}=\frac{(-1)^{m-n}(2l+1)(2\dot{l}+1)}{32\pi^4}\times\\
\int\limits_{SU(2)\otimes SU(2)}
f(\varphi^c,\theta^c,\psi^c)e^{i(m\varphi^c+n\psi^c)}
Z^l_{mn}(\cos\theta^c)
\sin\theta^c\sin\dot{\theta}^cd\theta d\varphi d\psi d\tau d\epsilon
d\varepsilon.
\nonumber
\end{multline}
The Parseval equality for the case of $SU(2)\otimes SU(2)$ is defined
as follows
\begin{multline}
\sum^\infty_{l=0}\sum^l_{m=-l}\sum^l_{n=-l}|\alpha^l_{mn}|^2=\\
\frac{(2l+1)(2\dot{l}+1)}{32\pi^4}
\int\limits_{SU(2)\otimes SU(2)}
|f(\varphi^c,\theta^c,\psi^c)|^2
\sin\theta^c\sin\dot{\theta}^cd\theta d\varphi d\psi d\tau d\epsilon
d\varepsilon.
\nonumber
\end{multline}
About convergence of Fourier series of the type (\ref{CFS}) see \cite{BD70}.

In like manner we can define Fourier series on the two-dimensional
complex sphere via the associated hyperspherical functions.
An expansion of the functions on the surface of the two-dimensional
sphere has an important meaning for the subsequent physical
applications.

So, let $f(\varphi^c,\theta^c)$ be a function on the complex two-sphere
$\dS^2$, such that
\[
\int\limits_{\dS^2}
|f(\varphi^c,\theta^c)|^2
\sin\theta^c\sin\dot{\theta}^cd\theta d\varphi d\tau d\epsilon <+\infty,
\]
then $f(\varphi^c,\theta^c)$ is expanded into a convergent Fourier series
on $\dS^2$,
\[
f(\varphi^c,\theta^c)=\sum^\infty_{l=0}\sum^l_{m=-l}
\alpha^l_me^{-m(\epsilon+i\varphi)}Z^m_l(\cos\theta^c),
\]
where
\[
\alpha^l_m=\frac{(-1)^m(2l+1)(2\dot{l}+1)}{32\pi^4}
\int\limits_{\dS^2}
f(\varphi^c,\theta^c)e^{im\varphi^c}Z^m_l(\cos\theta^c)
\sin\theta^c\sin\dot{\theta}^cd\theta d\varphi d\tau d\epsilon,
\]
and $Z^m_l(\cos\theta^c)$ is an associated hyperspherical function,
$d\fg=\sin\theta^c\sin\dot{\theta}^cd\theta d\varphi d\tau d\epsilon$ is
a Haar measure on the sphere $\dS^2$. Correspondingly, the Parseval
equality on $\dS^2$ has a form
\[
\sum^\infty_{l=0}\sum^l_{m=-l}|\alpha^l_m|^2=
\frac{(2l+1)(2\dot{l}+1)}{32\pi^4}
\int\limits_{\dS^2}
|f(\varphi^c,\theta^c)|^2
\sin\theta^c\sin\dot{\theta}^cd\theta d\varphi d\tau d\epsilon.
\]

In the case of infinite-dimensional representations of $SL(2,\C)$ we come
to the Fourier integrals on the Lorentz group. Let $f(\fg)$ be a square
integrable function on $SL(2,\C)$ and let
\begin{equation}\label{FTPS}
T^\alpha(f)=\int\limits_{SL(2,\C)}f(\fg)T^\alpha(\fg)d\fg
\end{equation}
be a Fourier transform for all representations of the principal unitary
series, then the inverse Fourier transform is defined by a formula
\begin{equation}\label{IFTPS}
f(\fg)=\frac{1}{16\pi^4}\sum^{+\infty}_{\lambda=-\infty}
\int\limits^{+\infty}_0\Tr\ld T^{\rho,\lambda}(f)
(T^{\rho,\lambda}(\fg))^\ast\rd(\lambda^2+\rho^2)d\rho,
\end{equation}
where the integral in the right part is convergent on an average, and
the representation $T^\alpha=T^{\rho,\lambda}$ is defined by the formula
(\ref{Principal}). There is an analog of the Plancherel formula,
\begin{equation}\label{IFPS}
\int\limits_{SL(2,\C)}|f(\fg)|^2d\fg=\frac{1}{16\pi^4}
\sum^{+\infty}_{\lambda=-\infty}\int\limits^{+\infty}_0
\Tr\ld(T^{\rho,\lambda}(f))^\ast T^{\rho,\lambda}(f)\rd
(\lambda^2+\rho^2)d\rho,
\end{equation}
where $d\fg$ is the Haar measure on $SL(2,\C)$. The integrals and series
in the right part of (\ref{IFPS}) are convergent absolutely.
Fourier integrals on the Lorentz group were studied by many authors
\cite{Nai58,Sha56,Sha62,CZ58,Pop59,GGV62,VS64,WF65,KS67,LSS68,KLMS69,Hai69,Hai71,Ruh70,War72,Zhe74,ZS83}.
It is obvious that the formulae (\ref{FTPS})--(\ref{IFPS}) can be expressed
via the matrix elements (\ref{MPrincip}).

\section{Fields on the Poincar\'{e} group}
Fields on the Poincar\'{e} group present itself a natural generalization
of the concept of wave function. These fields (generalized wave functions)
were introduced independently by several authors 
\cite{GT47,BW48,Yuk50,Shi51} mainly in connection with constructing
relativistic wave equations (a so called $Z$-description of the
relativistic spin \cite{GS03}). The following logical step was done by
Finkelstein \cite{Fin55}, he suggested to consider the wave function
depending both the coordinates on the Minkowski spacetime and some
continuous variables corresponding to spin degrees of freedom
(internal space). In essence, this generalization consists in replacing
the Minkowski space by a larger space on which the Poincar\'{e} group
acts. If this action is to be transitive, one is lead to consider the
homogeneous spaces of the Poincar\'{e} group. All the homogeneous spaces
of this type were listed by Finkelstein \cite{Fin55} and by Bacry and
Kihlberg \cite{BK69} and the fields on these spaces were considered
in the works \cite{Lur64,BN67,NB67,Kih68,Kih70,Tol96,GS01}.

A homogeneous space $\cM$ of a group $G$ has the following properties:\\
a) It is a topological space on which the group $G$ acts continuously,
that is, let $y$ be a point in $\cM$, then $gy$ is defined and is again
a point in $\cM$ ($g\in G$).\\
b) This action is transitive, that is, for any two points $y_1$ and
$y_2$ in $\cM$ it is always possible to find a group element $g\in G$
such that $y_2=gy_1$.\\
There is a one-to-one correspondence between the homogeneous spaces of
$G$ and the coset spaces of $G$. Let $H_0$ be a maximal subgroup of
$G$ which leaves the point $y_0$ invariant, $gy_0=y_0$, $g\in H_0$, then
$H_0$ is called the stabilizer of $y_0$. Representing now any group
element of $G$ in the form $g=g_cg_0$, where $g_0\in H_0$ and
$g_c\in G/H_0$, we see that, by virtie of the transitivity property,
any point $y\in\cM$ can be given by $y=g_cg_0y_0=g_cy$. Hence it follows
that the elements $g_c$ of the coset space give a parametrization
of $\cM$. The mapping $\cM\leftrightarrow G/H_0$ is continuous since
the group multiplication is continuous and the action on $\cM$ is
continuous by definition. The stabilizers $H$ and $H_0$ of two different
points $y$ and $y_0$ are conjugate, since from $H_0g_0=g_0$,
$y_0=g^{-1}y$, it follows that $gH_0g^{-1}y=y$, that is, $H=gH_0g^{-1}$.

Coming back to the Poincar\'{e} group $\cP$, we see that the enumeration
of the different homogeneous spaces $\cM$ of $\cP$ amounts to an
enumeration of the subgroups of $\cP$ up to a conjugation. Following to
Finkelstein, we require that $\cM$ always contains the Minkowski space
$\R^{1,3}$ which means that four parameters of $\cM$ can be denoted by
$x\,(x^\mu)$. This means that the stabilizer $H$ of a given point in $\cM$
can never contain an element of the translation subgroup of $\cP$.
Thus, the stabilizer must be a subgroup of the homogeneous Lorentz
group $\fG_+$.

In such a way, studying different subgroups of $\fG_+$, we obtain a full
list of homogeneous spaces $\cM=\cP/H$ of the Poincar\'{e} group.
In the present paper we restrict ourselves by a consideration of the
following four homogeneous spaces:
\begin{eqnarray}
\cM_{10}&=&\R^{1,3}\times\fL_6,\quad H=0;\nonumber\\
\cM_8&=&\R^{1,3}\times\dS^2,\quad H=\Omega^c_\psi;\nonumber\\
\cM_7&=&\R^{1,3}\times\BH^3,\quad H=SU(2);\nonumber\\
\cM_6&=&\R^{1,3}\times S^2,\quad H=\{\Omega^c_\psi,\Omega_\tau,\Omega_\epsilon\}.
\nonumber
\end{eqnarray}
Hence it follows that a group manifold of the Poincar\'{e} group,
$\cM_{10}=\R^{1,3}\times\fL_6$, is a maximal homogeneous space of $\cP$,
$\fL_6$ is a group manifold of the Lorentz group. The fields on the
manifold $\cM_{10}$ were considered by Lur\c{c}at \cite{Lur64}.
These fields depend on all the ten parameters of $\cP$:
\[
\boldsymbol{\psi}(x,\fg)=\psi(x)\psi(\fg)=\psi(x_0,x_1,x_2,x_3)
\psi(\fg_1,\fg_2,\fg_3,\fg_4,\fg_5,\fg_6),
\]
where an explicit form of $\psi(x)$ is given by the exponentials, and the
functions $\psi(\fg)$ are expressed via the generalized hyperspherical
functions $\fM^l_{mn}(\fg)$ (see (\ref{HS2})) in the case of finite
dimensional representations and via the function (\ref{MPrincip}) in the
case of principal series of unitary representations.

The following eight-dimensional homogeneous space 
$\cM_8=\R^{1,3}\times\dS^2$ is a direct product of the Minkowski space
$\R^{1,3}$ and the complex two-sphere $\dS^2$. In this case the stabilizer
$H$ consists of the subgroup $\Omega^c_\psi$ of the diagonal matrices
$\begin{pmatrix}
e^{\frac{i\psi^c}{2}} & 0\\
0 & e^{-\frac{i\psi^c}{2}}
\end{pmatrix}$. Bacry and Kihlberg \cite{BK69} claimed that the space
$\cM_8$ is the most suitable for a description of both half-integer and
integer spins. The fields, defined in $\cM_8$, depend on the eight
parameters of $\cP$:
\begin{equation}\label{WF}
\boldsymbol{\psi}(x,\varphi^c,\theta^c)=\psi(x)\psi(\varphi^c,\theta^c)=
\psi(x_0,x_1,x_2,x_3)\psi(\varphi,\epsilon,\theta,\tau),
\end{equation}
where the functions $\psi(\varphi^c,\theta^c)$ are expressed via the
associated hyperspherical functions (\ref{Associated}) defined on the
surface of the complex two-sphere $\dS^2$.

In turn, a seven-dimensional homogeneous space 
$\cM_7=\R^{1,3}\times\BH^3$ is a direct product of $\R^{1,3}$ and a
three-dimensional timelike (two sheeted) hyperboloid $\BH^3$. The stabilizer
$H$ consists of the subgroup of three-dimensional rotations, $SU(2)$.
Quantum field theory on the space $\cM_7$ was studied by Boyer and
Fleming \cite{BF74}. They showed that the fields built over $\cM_7$ are
in general nonlocal and become local only when the finite dimensional
representations of the Lorentz group are used. It is easy to see that
the fields $\boldsymbol{\psi}\in\cM_7$ depend on the seven parameters of
the Poincar\'{e} group:
\[
\boldsymbol{\psi}(x,\tau,\epsilon,\varepsilon)=\psi(x_0,x_1,x_2,x_3)
\psi(\tau,\epsilon,\varepsilon),
\]
where the functions $\psi(\tau,\epsilon,\varepsilon)$ are expressed via
$e^{-(m\epsilon+n\varepsilon)}\fP^l_{mn}(\ch\tau)$ in the case of finite
dimensional representations (the function $\fP^l_{mn}(\ch\tau)$ is of the
type (\ref{Jacobi})), and also via 
$e^{-(m\epsilon+n\varepsilon)}\fP^{-\frac{1}{2}+i\rho}_{mn}(\ch\tau)$
in the case of principal series of unitary representations, and via
$e^{-(m\epsilon+n\varepsilon)}\fP^{l,\pm}_{mn}(\ch\tau)$ in the case of
the discrete series.

Further, a six-dimensional space $\cM_6=\R^{1,3}\times S^2$ is a minimal
homogeneous space of the Poincar\'{e} group, since the real two-sphere
$S^2$ has a minimal possible dimension among the homogeneous spaces of
the Lorentz group. In this case, the stabilizer $H$ consists of the
subgroup $\Omega^c_\psi$ and the subgroups $\Omega_\tau$ and 
$\Omega_\epsilon$ formed by the matrices 
$\begin{pmatrix}
\ch\frac{\tau}{2} & \sh\frac{\tau}{2}\\
\sh\frac{\tau}{2} & \ch\frac{\tau}{2}
\end{pmatrix}$ and
$\begin{pmatrix}
e^{\frac{\epsilon}{2}} & 0\\
0 & e^{-\frac{\epsilon}{2}}
\end{pmatrix}$, respectively. It is not hard to see that the
two-dimensional real sphere coincides with a well known Penrose's
celestial sphere $S^-$ (or anti-celestial sphere $S^+$) \cite{PR84}, and
by this reason we can define the two types of $\cM_6$, namely,
$\cM^{\pm}_6=\R^{1,3}\times S^{\pm}$. Field models on the homogeneous
space $\cM_6$ have been considered in recent works
\cite{KLS95,LSS96,Dre97}. In the paper \cite{Dre97} Drechsler considered
the real two-sphere as a `spin shell' $S^2_{r=2s}$ of radius $r=2s$,
where $s=0,\frac{1}{2},1,\frac{3}{2},\ldots$. The fields, defined in $\cM_6$,
depend on the six parameters of $\cP$:
\[
\boldsymbol{\psi}(x,\varphi,\theta)=\psi(x_0,x_1,x_2,x_3)
\psi(\varphi,\theta),
\]
where the functions $\psi(\varphi,\theta)$ are expressed via the generalized
spherical functions of the type $e^{-im\varphi}P^l_{m0}(\cos\theta)$ or
via the Wigner $D$-functions.

\subsection{Harmonic analysis on $SU(2)\otimes SU(2)\odot T_4$}
In this subsection we will study Fourier series on the Poincar\'{e} group
$\cP$. First of all, the group $\cP$ has the same number of connected
components as with the Lorentz group. Later on we will consider only the
component $\cP^\uparrow_+$ corresponding the connected component
$L^\uparrow_+$ (so called special Lorentz group, see \cite{RF}).
As is known, an universal covering $\overline{\cP^\uparrow_+}$ of the group
$\cP^\uparrow_+$ is defined by a semidirect product
$\overline{\cP^\uparrow_+}=SL(2,\C)\odot T_4\simeq\spin_+(1,3)\odot T_4$,
where $T_4$ is a subgroup of four-dimensional translations.
Since the Poincar\'{e} group is a 10-parameter group, then an invariant
measure on this group has a form
\[
d^{10}\balpha=d^6\fg d^4x,
\]
where $d^6\fg$ is the Haar measure on the Lorentz group. Or, taking
into account (\ref{HA1}), we obtain
\begin{equation}\label{HMP}
d\balpha=\sin\theta^c\sin\dot{\theta}^cd\theta d\varphi d\psi d\tau
d\epsilon d\varepsilon dx_1 dx_2 dx_3 dx_4,
\end{equation}
where $x_i\in T_4$.

Thus, an invariant integration on the group $SL(2,\C)\odot T_4$ is defined
by a formula
\[
\int\limits_{SL(2,\C)\odot T_4}f(\balpha)d^{10}\balpha=
\int\limits_{SL(2,\C)}\int\limits_{T_4}f(x,\fg)d^4xd^6\fg,
\]
where $f(\balpha)$ is a finite continuous function on $SL(2,\C)\odot T_4$.

In the case of finite-dimensional representations we come again to a local
isomorphism $SL(2,\C)\odot T_4\simeq SU(2)\otimes SU(2)\odot T_4$.
In this case basis representation functions of the Poincar\'{e} group
are defined by symmetric polynomials of the form 
\begin{gather}
p(x,z,\bar{z})=\sum_{\substack{(\alpha_1,\ldots,\alpha_k)\\
(\dot{\alpha}_1,\ldots,\dot{\alpha}_r)}}\frac{1}{k!\,r!}
a^{\alpha_1\cdots\alpha_k\dot{\alpha}_1\cdots\dot{\alpha}_r}(x)
z_{\alpha_1}\cdots z_{\alpha_k}\bar{z}_{\dot{\alpha}_1}\cdots
\bar{z}_{\dot{\alpha}_r},\label{SF2}\\
(\alpha_i,\dot{\alpha}_i=0,1)\nonumber
\end{gather}
where the coefficients
$a^{\alpha_1\cdots\alpha_k\dot{\alpha}_1\cdots\dot{\alpha}_r}$ 
depend on the variables $x^\alpha$ ($\alpha=0,1,2,3$) 
(the parameters of $T_4$).
The functions (\ref{SF2}) should be considered as {\it the functions on
the Poincar\'{e} group}.\index{function!on the Poincar\'{e} group}
Some applications of these functions contained
in \cite{GS01}. 
The group $T_4$ is an Abelian group formed by a direct
product of the four one-dimensional translation groups, $T_1$, where
$T_1$ is isomorphic to an additive group of real numbers $\R$ (usual
Fourier analysis is formulated in terms of the group $\R$). Hence it
follows that all irreducible representations of $T_4$ are one-dimensional
and expressed via the exponentials. Thus, the basis functions
(matrix elements) of the finite-dimensional representations of $\cP$
have the form
\begin{equation}\label{BFP}
t^l_{mn}(\balpha)=e^{-ipx}\fM^l_{mn}(\fg),
\end{equation}
where $x=(x_1,x_2,x_3,x_4)$, and $\fM^l_{mn}(\fg)$ is the 
generalized hyperspherical
function (\ref{HS2}).

Let us consider now the configuration space $\cM_8=\R^{1,3}\times \dS^2$.
In this case the Fourier series on $\cM_8$ can be defined as follows
\begin{equation}\label{FM8}
f(\balpha)=\sum^{+\infty}_{p=-\infty}e^{ipx}\sum^\infty_{l=0}
\sum^l_{m=-l}\alpha^l_m\fM^m_l(\varphi,\epsilon,\theta,\tau,0,0),
\end{equation}
where
\[
\alpha^l_m=\frac{(-1)^m(2l+1)(2\dot{l}+1)}{32\pi^4}
\int\limits_{\dS^2}\int\limits_{T_4}f(\balpha)e^{-ipx}
\fM^m_l(\varphi,\epsilon,\theta,\tau,0,0)d^4xd^4\fg,
\]
and $d^4\fg=\sin\theta^c\sin\dot{\theta}^cd\theta d\varphi d\tau d\epsilon$
is the Haar measure on $\dS^2$, $f(\balpha)$ is the square integrable
function on $\cM_8$, such that
\[
\int\limits_{\dS^2}\int\limits_{T_4}|f(\balpha)|^2d^4xd^4\fg<+\infty.
\]

Coming to the space $\cM_7=\R^{1,3}\times\BH^3$ and restricting by the
finite dimensional representations, we see that the Fourier series
on $\cM_7$ can be defined as
\[
f(x,\tau,\epsilon,\varepsilon)=\sum^{+\infty}_{p=-\infty}
e^{ipx}\sum^\infty_{l=0}\sum^l_{m,n=-l}\alpha^l_{mn}
e^{-(m\epsilon+n\varepsilon)}\fP^l_{mn}(\ch\tau),
\]
where
\[
\alpha^l_{mn}=\frac{(-1)^{m-n}(2l+1)}{16\pi^2}
\int\limits_{SU(1,1)}\int\limits_{T_4}
f(x,\tau,\epsilon,\varepsilon)e^{-ipx}\fP^l_{mn}(\ch\tau)
e^{m\epsilon+n\varepsilon}d^4xd^3g,
\]
and $d^3g=\sh\tau d\epsilon d\tau d\varepsilon$ is the Haar measure on
$SU(1,1)$, $f(x,\tau,\epsilon,\varepsilon)$ is the square integrable
function on $\cM_7$. In the case of infinite dimensional representations
(principal and discrete series) we come to the Fourier integrals on
$\cM_7$, where the direct and inverse Fourier transforms are of the type
(\ref{Har1})--(\ref{Har5}). The consideration of the infinite dimensional
representations leads immediately to a harmonic analysis on the
hyperboloids \cite{Str73}. However, all the fields built in terms of the
Fourier integrals on $\cM_7$ are in general nonlocal. The same statement
holds also for the fields built in terms of the Fourier integrals
on $\cM_{10}$ and $\cM_8$, where the direct and inverse Fourier transforms
are defined by the formulae (\ref{FTPS})--(\ref{IFPS}).

Further, the Fourier series on the homogeneous space 
$\cM_6=\R^{1,3}\times S^2$ are defined like the series on the group
$SU(2)$ considered in the subsection \ref{HASU2}. For the physical
purposes it is useful to express the fields on $\cM_6$ via the Wigner
$D$-functions. Replacing the basis (\ref{Basis}) by
\begin{equation}\label{Basis2}
f_n(\fz)=i^n\psi_n(\fz),\quad n=-l,-l+1,\ldots,l,
\end{equation}
we obtain matrix elements $D^l_{mn}(\theta)$ of the operators
$T_l(\theta)$ related with the matrix elements $t^l_{mn}(\theta)$ by
a formula
\[
D^l_{mn}(\theta)=i^{n-m}t^l_{mn}(\theta)=i^{n-m}P^l_{mn}(\cos\theta).
\]
In contrast to $P^l_{mn}(\cos\theta)$, the functions $D^l_{mn}(\theta)$
are real. The full matrix element in the basis (\ref{Basis2}) has a form
\[
D^l_{mn}(u)=e^{-i(m\varphi+n\psi)}D^l_{mn}(\theta).
\]
Therefore, for any square integrable function $f(x,\varphi,\theta)$ on
$\cM_6$ we have
\[
f(x,\varphi,\theta)=\sum^{+\infty}_{p=-\infty}e^{ipx}\sum^\infty_{l=0}
\sum^l_{m=-l}\alpha^l_me^{-im\varphi}D^l_{m0}(\theta),
\]
where
\[
\alpha^l_m=\frac{(-1)^m(2l+1)}{4\pi}\int\limits_{S^2}\int\limits_{T_4}
e^{-ipx}f(x,\varphi,\theta)e^{im\varphi}D^l_{m0}(\theta)
d^4 d^2\boldsymbol{\xi},
\]
and $d^2\boldsymbol{\xi}=1/4\pi\sin\theta d\theta d\varphi$ is an invariant
measure on the sphere $S^2$. In such a way we come here to a Fourier
analysis on the sphere \cite{She75}. It should be noted that solutions
of relativistic wave equations and quantization procedures was studied
by Malin \cite{Mal75,Mal79} in terms of the functions over the group $SU(2)$.

\section{Lagrangian formalism and field equations on the\protect\newline
Poincar\'{e} group}
We will start with a more general homogeneous space of the group $\cP$,
$\cM_{10}=\R^{1,3}\times\fL_6$ (group manifold of the Poincar\'{e} group).
Let $\cL(\balpha)$ be a Lagrangian on the group manifold $\cM_{10}$ 
(in other words, $\cL(\balpha)$ is a 10-dimensional
point function), where $\balpha$ is the parameter set of this group.
Then an integral for $\cL(\balpha)$ on some 10-dimensional volume $\Omega$
of the group manifold we will call {\it an action on the Poincar\'{e}
group}:
\[
A=\int\limits_\Omega d\balpha\cL(\balpha),
\]
where $d\balpha$ is a Haar measure on the group $\cP$ (see (\ref{HMP})).

Let $\boldsymbol{\psi}(\balpha)$ be a function on 
the group manifold $\cM_{10}$ (now it is
sufficient to assume that $\boldsymbol{\psi}(\balpha)$ 
is a square integrable function
on the Poincar\'{e} group) and let
\begin{equation}\label{ELE}
\frac{\partial\cL}{\partial\boldsymbol{\psi}}-\frac{\partial}{\partial\balpha}
\frac{\partial\cL}{\partial\frac{\partial\boldsymbol{\psi}}{\partial\balpha}}=0
\end{equation}
be Euler-Lagrange equations on $\cM_{10}$ (more precisely speaking, the equations
(\ref{ELE}) act on the tangent bundle 
$T\cM_{10}=\underset{\balpha\in\cM_{10}}{\cup}T_{\balpha}\cM_{10}$ 
of the manifold $\cM_{10}$,
see \cite{Arn}). Let us introduce a Lagrangian $\cL(\balpha)$ depending on
the field function $\boldsymbol{\psi}(\balpha)$ as follows
\[
\cL(\balpha)=-\frac{1}{2}\left(\boldsymbol{\psi}^\ast(\balpha)B_\mu
\frac{\partial\boldsymbol{\psi}(\balpha)}{\partial\balpha_\mu}-
\frac{\partial\boldsymbol{\psi}^\ast(\balpha)}
{\partial\balpha_\mu}B_\mu\boldsymbol{\psi}(\balpha)\right)
-\kappa\boldsymbol{\psi}^\ast(\balpha)B_{11}\boldsymbol{\psi}(\balpha),
\]
where $B_\nu$ ($\nu=1,2,\ldots,10$) are square matrices. The number of
rows and columns in these matrices is equal to the number of components
of $\boldsymbol{\psi}(\balpha)$, $\kappa$ is a non-null real constant.

Further, if $B_{11}$ is non-singular, then we can introduce the matrices
\[
\Gamma_\mu=B^{-1}_{11}B_\mu,\quad \mu=1,2,\ldots,10,
\]
and represent the Lagrangian $\cL(\balpha)$ in the form
\begin{equation}\label{Lagrange}
\cL(\balpha)=-\frac{1}{2}\left(\overline{\boldsymbol{\psi}}(\balpha)\Gamma_\mu
\frac{\partial\boldsymbol{\psi}(\balpha)}{\partial\balpha_\mu}-
\frac{\overline{\boldsymbol{\psi}}(\balpha)}{\partial\balpha_\mu}\Gamma_\mu
\boldsymbol{\psi}(\balpha)\right)-
\kappa\overline{\boldsymbol{\psi}}(\balpha)\boldsymbol{\psi}(\balpha),
\end{equation}
where
\[
\overline{\boldsymbol{\psi}}(\balpha)=\boldsymbol{\psi}^\ast(\balpha)B_{11}.
\]

Varying independently $\psi(x)$ and $\overline{\psi}(x)$, we obtain from
(\ref{Lagrange}) in accordance with (\ref{ELE}) the following equations:
\begin{equation}\label{FET}
\begin{array}{ccc}
\Gamma_i\dfrac{\partial\psi(x)}{\partial x_i}+\kappa\psi(x)&=&0,\\
\Gamma^T_i\dfrac{\partial\overline{\psi}(x)}{\partial x_i}-
\kappa\overline{\psi}(x)&=&0.
\end{array}\quad(i=1,\ldots,4)
\end{equation}
Analogously, varying independently $\psi(\fg)$ and $\overline{\psi}(\fg)$
one gets
\begin{equation}\label{FEL}
\begin{array}{ccc}
\Gamma_k\dfrac{\partial\psi(\fg)}{\partial\fg_k}+\kappa\psi(\fg)&=&0,\\
\Gamma^T_k\dfrac{\overline{\psi}(\fg)}{\partial\fg_k}-
\kappa\overline{\psi}(\fg)&=&0,
\end{array}\quad(k=1,\ldots,6)
\end{equation}
where
\[
\psi(\fg)=\begin{pmatrix}
\psi(\fg)\\
\dot{\psi}(\fg)
\end{pmatrix},\quad
\Gamma_k=\begin{pmatrix}
0 & \Lambda^\ast_k\\
\Lambda_k & 0
\end{pmatrix}.
\]
The doubling of representations, described by a bispinor
$\psi(\fg)=(\psi(\fg),\dot{\psi}(\fg))^T$, is the well known feature of
the Lorentz group representations \cite{GMS,Nai58}.  
Since an universal covering $SL(2,\C)$ of the
proper orthochronous Lorentz group is a complexification of the group
$SU(2)$ (see the section \ref{SL2C}), then it is more
convenient to express six parameters $\fg_k$ of the Lorentz group via
three parameters $a_1$, $a_2$, $a_3$ of the group $SU(2)$. It is obvious that
$\fg_1=a_1$, $\fg_2=a_2$, $\fg_3=a_3$, $\fg_4=ia_1$, $\fg_5=ia_2$,
$\fg_6=ia_3$. Then the first equation from (\ref{FEL}) can be written as
\begin{eqnarray}
\sum^3_{j=1}\Lambda^\ast_j\frac{\partial\dot{\psi}}
{\partial\widetilde{a}_j}+i\sum^3_{j=1}\Lambda^\ast_j
\frac{\partial\dot{\psi}}{\partial\widetilde{a}^\ast_j}+
\kappa^c\psi&=&0,\nonumber\\
\sum^3_{j=1}\Lambda_j\frac{\partial\psi}{\partial a_j}-
i\sum^3_{j=1}\Lambda_j\frac{\partial\psi}{\partial a^\ast_j}+
\dot{\kappa}^c\dot{\psi}&=&0,\label{Complex}
\end{eqnarray}
where $a^\ast_1=-i\fg_4$, $a^\ast_2=-i\fg_5$, $a^\ast_3=-i\fg_6$, and
$\widetilde{a}_j$, $\widetilde{a}^\ast_j$ are the parameters corresponding
the dual basis. In essence, the equations (\ref{Complex}) are defined
in a three-dimensional complex space $\C^3$. In turn, the space $\C^3$
is isometric to a 6-dimensional bivector space $\R^6$ (a parameter space
of the Lorentz group \cite{Kag26,Pet69}). The bivector space $\R^6$ is
a tangent space of the group manifold $\fL_6$ of the Lorentz group, that is,
the manifold $\fL_6$ in the each its point is equivalent locally to the
space $\R^6$. Thus, for all $\fg\in\fL_6$ we have $T_{\fg}\fL_6\simeq\R^6$.
There exists a close relationship between the metrics of the
Minkowski spacetime $\R^{1,3}$ and the metrics of $\R^6$ defined by the
formulae (see \cite{Pet69})
\begin{equation}\label{Metric}
g_{ab}\longrightarrow g_{\alpha\beta\gamma\delta}\equiv
g_{\alpha\gamma}g_{\beta\delta}-g_{\alpha\delta}g_{\beta\gamma},
\end{equation}
where $g_{\alpha\beta}$ is a metric tensor of the spacetime $\R^{1,3}$, and
collective indices are skewsymmetric pairs
$\alpha\beta\rightarrow a$, $\gamma\delta\rightarrow b$. In more detail, if
\[\ar
g_{\alpha\beta}=\begin{pmatrix}
-1 & 0 & 0 & 0\\
0  & -1& 0 & 0\\
0  & 0 & -1& 0\\
0  & 0 & 0 & 1
\end{pmatrix},
\]
then in virtue of (\ref{Metric}) for the metric tensor of $\R^6$ we obtain
\begin{equation}\label{MetB}
g_{ab}=\ar\begin{pmatrix}
-1& 0 & 0 & 0 & 0 & 0\\
0 & -1& 0 & 0 & 0 & 0\\
0 & 0 & -1& 0 & 0 & 0\\
0 & 0 & 0 & 1 & 0 & 0\\
0 & 0 & 0 & 0 & 1 & 0\\
0 & 0 & 0 & 0 & 0 & 1
\end{pmatrix},
\end{equation}
where the order of collective indices in $\R^6$ is
$23\rightarrow 0$, $10\rightarrow 1$, $20\rightarrow 2$,
$30\rightarrow 3$, $31\rightarrow 4$, $12\rightarrow 5$.
As it is shown in \cite{Kag26}, the Lorentz transformations can be represented
by linear transformations of the space $\R^6$. Let us write an
invariance condition of the system (\ref{Complex}). Let 
$\fg: a^\prime=\fg^{-1} a$ be a transformation of the bivector space $\R^6$,
that is, $a^\prime=\sum^6_{b=1}g_{ba} a_b$, where
$a=(a_1,a_2,a_3,a^\ast_1,a^\ast_2,a^\ast_3)$ and
$g_{ba}$ is the metric tensor (\ref{MetB}). We can write the tensor
(\ref{MetB}) in the form
$g_{ab}=\ar\begin{pmatrix}
g^-_{ik} & \\
& g^+_{ik}
\end{pmatrix}$,
then $a^\prime=\sum^3_{k=1}g^-_{ki}a_k$,
$a^\ast{}^\prime=\sum^3_{k=1}g^+_{ki}a^\ast_k$.
Replacing $\psi$ via $T^{-1}_{\fg}\psi^\prime$, and differentiation on
$a_k$ ($a^\ast_k$) by differentiation on $a^\prime_k$
($a^\ast_k{}^\prime$) via the formulae
\[
\frac{\partial}{\partial a_k}=\sum g^-_{ik}
\frac{\partial}{\partial a^\prime_i},\quad
\frac{\partial}{\partial a^\ast_k}=\sum g^+_{ik}
\frac{\partial}{\partial a^\ast_i{}^\prime},
\]
we obtain
\begin{multline}
\sum_i\left[
g^-_{i1}\sL_1T^{-1}_{\fg}\frac{\partial\psi^\prime}{\partial a^\prime_i}+
g^-_{i2}\sL_2T^{-1}_{\fg}\frac{\partial\psi^\prime}{\partial a^\prime_i}+
g^-_{i3}\sL_3T^{-1}_{\fg}\frac{\partial\psi^\prime}{\partial a^\prime_i}-
\right.\\
\shoveright{\left.
-ig^-_{i1}\sL_1T^{-1}_{\fg}\frac{\partial\psi^\prime}
{\partial a^\ast_i{}^\prime}-
ig^-_{i2}\sL_2T^{-1}_{\fg}\frac{\partial\psi^\prime}
{\partial a^\ast_i{}^\prime}-
ig^-_{i3}\sL_3T^{-1}_{\fg}\frac{\partial\psi^\prime}
{\partial a^\ast_i{}^\prime}\right]+
\kappa^cT^{-1}_{\fg}\psi^\prime=0,}\\
\shoveleft{
\sum_i\left[
g^+_{i1}\sL^\ast_1\overset{\ast}{T}_{\fg}\!\!{}^{-1}
\frac{\partial\dot{\psi}^\prime}{\partial\widetilde{a}^\prime_i}+
g^+_{i2}\sL^\ast_2\overset{\ast}{T}_{\fg}\!\!{}^{-1}
\frac{\partial\dot{\psi}^\prime}{\partial\widetilde{a}^\prime_i}+
g^+_{i3}\sL^\ast_3\overset{\ast}{T}_{\fg}\!\!{}^{-1}
\frac{\partial\dot{\psi}^\prime}{\partial\widetilde{a}^\prime_i}+\right.}\\
\left.
+ig^+_{i1}\sL^\ast_1\overset{\ast}{T}_{\fg}\!\!{}^{-1}
\frac{\partial\dot{\psi}^\prime}{\partial\widetilde{a^\ast}_i{}^\prime}+
ig^+_{i2}\sL^\ast_2\overset{\ast}{T}_{\fg}\!\!{}^{-1}
\frac{\partial\dot{\psi}^\prime}{\partial\widetilde{a^\ast}_i{}^\prime}+
ig^+_{i3}\sL^\ast_3\overset{\ast}{T}_{\fg}\!\!{}^{-1}
\frac{\partial\dot{\psi}^\prime}{\partial\widetilde{a^\ast}_i{}^\prime}\right]+
\dot{\kappa}^c\overset{\ast}{T}_{\fg}\!\!{}^{-1}\dot{\psi}^\prime=0.\nonumber
\end{multline}
For coincidence of the latter system with (\ref{Complex}) we must multiply
this system by $T_{\fg}$ ($\overset{\ast}{T}_{\fg}$) from the left:
\begin{eqnarray}
\sum_i\sum_k g^-_{ik}T_{\fg}\sL_kT^{-1}_{\fg}
\frac{\partial\psi^\prime}{\partial a^\prime_i}
-i\sum_i\sum_k g^-_{ik}T_{\fg}\sL_kT^{-1}_{\fg}
\frac{\partial\psi^\prime}{\partial a^\ast_i{}^\prime}+
\kappa^c\psi^\prime&=&0,\nonumber\\
\sum_i\sum_k g^+_{ik}\overset{\ast}{T}_{\fg}\sL^\ast_k
\overset{\ast}{T}_{\fg}\!\!{}^{-1}\frac{\partial\dot{\psi}^\prime}
{\partial\widetilde{a}^\prime_i}+
i\sum_i\sum_k g^+_{ik}\overset{\ast}{T}_{\fg}\sL^\ast_k
\overset{\ast}{T}_{\fg}\!\!{}^{-1}\frac{\partial\dot{\psi}^\prime}
{\partial\widetilde{a^\ast}_i{}^\prime}+
\dot{\kappa}^c\dot{\psi}^\prime&=&0.\nonumber
\end{eqnarray}
The requirement of invariance means that for any transformation $\fg$
between the matrices $\sL_k$ ($\sL^\ast_k$) we must have the relations
\begin{eqnarray}
\sum_k g^-_{ik}T_{\fg}\sL_kT^{-1}_{\fg}&=&\sL_i,\nonumber\\
\sum_k g^+_{ik}\overset{\ast}{T}_{\fg}\sL^\ast_k
\overset{\ast}{T}_{\fg}\!\!\!{}^{-1}&=&\sL^\ast_i,\label{IC}
\end{eqnarray}
where $\sL^\ast_i$ are the matrices of the equations in the dual
representation space, $\kappa^c$ is a complex number,
$\partial/\partial\widetilde{a}_i$ mean covariant derivatives in the dual
space.

Let us find commutation relations between the matrices $\sL_i$, $\sL^\ast_i$
and infinitesimal operators (\ref{OpA}), (\ref{OpB}), (\ref{DopA}),
(\ref{DopB}) defined in the helicity basis. 
First of all, let us present transformations $T_{\fg}$
($\overset{\ast}{T}_{\fg}$) in the infinitesimal form,
$\sI+\sA_i\xi+\ldots$, $\sI+\sB_i\xi+\ldots$, $\sI+\widetilde{\sA}_i\xi+
\ldots$, $\sI+\widetilde{\sB}_i\xi+\ldots$. Substituting these transformations
into invariance conditions (\ref{IC}), we obtain with an accuracy of the
terms of second order the following commutation relations
\begin{equation}\label{AL}
\begin{array}{rcl}
\ld\sA_1,\sL_1\rd &=& 0,\\
\ld\sA_2,\sL_1\rd &=&-\sL_3,\\
\ld\sA_3,\sL_1\rd &=& \sL_2,
\end{array}\;\;\;
\begin{array}{rcl}
\ld\sA_1,\sL_2\rd &=& \sL_3,\\
\ld\sA_2,\sL_2\rd &=& 0,\\
\ld\sA_3,\sL_2\rd &=&-\sL_1,
\end{array}\;\;\;
\begin{array}{rcl}
\ld\sA_1,\sL_3\rd &=&-\sL_2,\\
\ld\sA_2,\sL_3\rd &=& \sL_1,\\
\ld\sA_3,\sL_3\rd &=& 0.
\end{array}
\end{equation}
\begin{equation}\label{BL}
\begin{array}{rcl}
\ld\sB_1,\sL_1\rd &=& 0,\\
\ld\sB_2,\sL_1\rd &=&i\sL_3,\\
\ld\sB_3,\sL_1\rd &=& -i\sL_2,
\end{array}\;\;\;
\begin{array}{rcl}
\ld\sB_1,\sL_2\rd &=& -i\sL_3,\\
\ld\sB_2,\sL_2\rd &=& 0,\\
\ld\sB_3,\sL_2\rd &=&i\sL_1,
\end{array}\;\;\;
\begin{array}{rcl}
\ld\sB_1,\sL_3\rd &=&i\sL_2,\\
\ld\sB_2,\sL_3\rd &=& -i\sL_1,\\
\ld\sB_3,\sL_3\rd &=& 0.
\end{array}
\end{equation}
\begin{equation}\label{DAL}
{\renewcommand{\arraystretch}{1.6}
\begin{array}{rcl}
\ld\widetilde{\sA}_1,\sL^\ast_1\rd &=& 0,\\
\ld\widetilde{\sA}_2,\sL^\ast_1\rd &=& -\sL^\ast_3,\\
\ld\widetilde{\sA}_3,\sL^\ast_1\rd &=&\sL^\ast_2,
\end{array}\;\;\;
\begin{array}{rcl}
\ld\widetilde{\sA}_1,\sL^\ast_2\rd &=&\sL^\ast_3,\\
\ld\widetilde{\sA}_2,\sL^\ast_2\rd &=& 0,\\
\ld\widetilde{\sA}_3,\sL^\ast_2\rd &=& -\sL^\ast_1,
\end{array}\;\;\;
\begin{array}{rcl}
\ld\widetilde{\sA}_1,\sL^\ast_3\rd &=& -\sL^\ast_2,\\
\ld\widetilde{\sA}_2,\sL^\ast_3\rd &=&\sL^\ast_1,\\
\ld\widetilde{\sA}_3,\sL^\ast_3\rd &=& 0.
\end{array}}
\end{equation}
\begin{equation}\label{DBL}
{\renewcommand{\arraystretch}{1.6}
\begin{array}{rcl}
\ld\widetilde{\sB}_1,\sL^\ast_1\rd &=& 0,\\
\ld\widetilde{\sB}_2,\sL^\ast_1\rd &=&-i\sL^\ast_3,\\
\ld\widetilde{\sB}_3,\sL^\ast_1\rd &=&i\sL^\ast_2,
\end{array}\;\;\;
\begin{array}{rcl}
\ld\widetilde{\sB}_1,\sL^\ast_2\rd &=&i\sL^\ast_3,\\
\ld\widetilde{\sB}_2,\sL^\ast_2\rd &=& 0,\\
\ld\widetilde{\sB}_3,\sL^\ast_2\rd &=&-i\sL^\ast_1,
\end{array}\;\;\;
\begin{array}{rcl}
\ld\widetilde{\sB}_1,\sL^\ast_3\rd &=&-i\sL^\ast_2,\\
\ld\widetilde{\sB}_2,\sL^\ast_3\rd &=&i\sL^\ast_1,\\
\ld\widetilde{\sB}_3,\sL^\ast_3\rd &=& 0.
\end{array}}
\end{equation} 
\begin{sloppypar}
Further, using the latter relations and taking into account (\ref{SL25}),
it is easy to establish commutation relations between $\sL_3$, $\sL^\ast_3$ and
generators $\sY_\pm$, $\sY_3$, $\sX_\pm$, $\sX_3$:
\end{sloppypar}
\begin{equation}\label{LX}
{\renewcommand{\arraystretch}{1.5}
\left.\begin{array}{l}
\ld\ld\sL_3,\sX_-\rd,\sX_+\rd = 2\sL_3,\\
\ld\sL_3,\sX_3\rd =0,
\end{array}\right.
}
{\renewcommand{\arraystretch}{1.5}
\left.\begin{array}{l}
\ld\ld\sL^\ast_3,\sY_-\rd,\sY_+\rd =2\sL^\ast_3,\\
\ld\sL^\ast_3,\sY_3\rd =0,
\end{array}\right.
}
\end{equation}
Using the relations (\ref{LX}), we will find an explicit form of the matrices
$\sL_3$ and $\sL^\ast_3$, and after this we will find
$\sL_1$, $\sL_2$ and $\sL^\ast_1$,
$\sL^\ast_2$. The wave function $\boldsymbol{\psi}$ is transformed within
some representation $T_{\fg}$ of the group $SL(2,\C)$. We assume that $T_{\fg}$
is decomposed into irreducible representations.
The components of the
function $\boldsymbol{\psi}$ we will numerate by the indices
$l$ and $m$, where $l$ is a weight of irreducible representation,
$m$ is a number of the components in the representation of the weight $l$.
In the case when a representation with one and the same weight $l$ at the
decomposition of $\boldsymbol{\psi}$ occurs more than one time, then with
the aim to distinguish these representations we will add the index
$k$, which indicates a number of the representations of the weight $l$. 
Denoting
$\zeta_{lm;\dot{l}\dot{m}}=\mid lm;\dot{l}\dot{m}\rangle$ and coming to the
helicity basis, we obtain a following decomposition for the wave function:
\[
\boldsymbol{\psi}(a_1,a_2,a_3,a^\ast_1,a^\ast_2,a^\ast_3)=
\sum_{l,m,k}\psi^k_{lm;\dot{l}\dot{m}}(a_1,a_2,a_3,a^\ast_1,a^\ast_2,a^\ast_3)
\zeta^k_{lm;\dot{l};\dot{m}},
\]
where $a_1,a_2,a_3,a^\ast_1,a^\ast_2,a^\ast_3$ 
are the coordinates of the complex space
$\C^3\simeq\R^6$
(parameters of $SL(2,\C)$)\footnote{Recall that the wave function
$\boldsymbol{\psi}(a_j,a^\ast_j)$ is defined on the group manifold
$\fL_6$, that is, $\boldsymbol{\psi}$ is a function on the Lorentz group.}.
Analogously, for the dual representation we have
\[
\dot{\boldsymbol{\psi}}(\widetilde{a}_1,\widetilde{a}_2,\widetilde{a}_3,
\widetilde{a^\ast}_1,\widetilde{a^\ast}_2,\widetilde{a^\ast}_3)=
\sum_{\dot{l},\dot{m},\dot{k}}\psi^{\dot{k}}_{\dot{l}\dot{m};lm}
(\widetilde{a}_1,\widetilde{a}_2,\widetilde{a}_3,\widetilde{a^\ast}_1,
\widetilde{a^\ast}_2,\widetilde{a^\ast}_3)
\zeta^{\dot{k}}_{\dot{l}\dot{m};lm}.
\]
The transformation $\sL_3$ in the helicity basis has a form
\[
\sL_3\zeta^k_{lm;\dot{l}\dot{m}}=\sum_{l^\prime,m^\prime,k^\prime}
c^{k^\prime k}_{l^\prime l,m^\prime m}
\zeta^{k^\prime}_{l^\prime m^\prime;\dot{l}\dot{m}}.
\]
Calculating the commutators $\ld\sL_3,\sX_3\rd$,
$\ld\ld\sL_3,\sX_-\rd,\sX_+\rd$, we find the numbers
$c^{k^\prime k}_{l^\prime l,m^\prime m}$:
\begin{equation}\label{L3}
{\renewcommand{\arraystretch}{1.25}
\sL_3:\quad\left\{\begin{array}{ccc}
c^{k^\prime k}_{l-1,l,m}&=&
c^{k^\prime k}_{l-1,l}\sqrt{l^2-m^2},\\
c^{k^\prime k}_{l,l,m}&=&c^{k^\prime k}_{ll}m,\\
c^{k^\prime k}_{l+1,l,m}&=&
c^{k^\prime k}_{l+1,l}\sqrt{(l+1)^2-m^2}.
\end{array}\right.}
\end{equation}
All other elements of the matrix $\sL_3$ are equal to zero. Let us define now
elements of the matrices $\sL_1$ and $\sL_2$. For the transformations
$\sL_1$ and $\sL_2$ in the helicity basis we have
\begin{eqnarray}
\sL_1\zeta^k_{lm;\dot{l}\dot{m}}&=&\sum_{l^\prime,m^\prime,k^\prime}
a^{k^\prime k}_{l^\prime l,m^\prime m}
\zeta^{k^\prime}_{l^\prime m^\prime;\dot{l}\dot{m}},\nonumber\\
\sL_2\zeta^k_{lm;\dot{l}\dot{m}}&=&\sum_{l^\prime,m^\prime,k^\prime}
b^{k^\prime k}_{l^\prime l,m^\prime m}
\zeta^{k^\prime}_{l^\prime m^\prime;\dot{l}\dot{m}}.\nonumber
\end{eqnarray}
Using the relations $\sL_1=\ld\sA_2,\sL_3\rd$ (or $\sL_1=i\ld\sB_2,\sL_3\rd$)
and (\ref{OpA}) (or (\ref{OpB})), and also (\ref{L3}), we find the elements
$a^{k^\prime k}_{l^\prime l,m^\prime m}$ of the matrix $\sL_1$. 
Analogously, from the relations
$\sL_2=-\ld\sA_1,\sL_3\rd$ (or $\sL_2=-i\ld\sB_1,\sL_3\rd$)
and (\ref{OpA}) (or (\ref{OpB})), (\ref{L3}) we obtain the elements
$b^{k^\prime k}_{l^\prime l,m^\prime m}$
of $\sL_2$. Thus,
\begin{equation}\label{L1}
{\renewcommand{\arraystretch}{1.25}
\sL_1:\quad\left\{\begin{array}{ccc}
a^{k^\prime k}_{l-1,l,m-1,m}&=&
-\frac{c_{l-1,l}}{2}\sqrt{(l+m)(l+m-1)},\\
a^{k^\prime k}_{l,l,m-1,m}&=&
\frac{c_{ll}}{2}\sqrt{(l+m)(l-m+1)},\\
a^{k^\prime k}_{l+1,l,m-1,m}&=&
\frac{c_{l+1,l}}{2}\sqrt{(l-m+1)(l-m+2)},\\
a^{k^\prime k}_{l-1,l,m+1,m}&=&
\frac{c_{l-1,l}}{2}\sqrt{(l-m)(l-m-1)},\\
a^{k^\prime k}_{l,l,m+1,m}&=&
\frac{c_{ll}}{2}\sqrt{(l+m+1)(l-m)},\\
a^{k^\prime k}_{l+1,l,m+1,m}&=&
-\frac{c_{l+1,l}}{2}\sqrt{(l+m+1)(l+m+2)}.
\end{array}\right.}
\end{equation}
\begin{equation}\label{L2}
{\renewcommand{\arraystretch}{1.25}
\sL_2:\quad\left\{\begin{array}{ccc}
b^{k^\prime k}_{l-1,l,m-1,m}&=&
-\frac{ic_{l-1,l}}{2}\sqrt{(l+m)(l+m-1)},\\
b^{k^\prime k}_{l,l,m-1,m}&=&
\frac{ic_{ll}}{2}\sqrt{(l+m)(l-m+1)},\\
b^{k^\prime k}_{l+1,l,m-1,m}&=&
\frac{ic_{l+1,l}}{2}\sqrt{(l-m+1)(l-m+2)},\\
b^{k^\prime k}_{l-1,l,m+1,m}&=&
-\frac{ic_{l-1,l}}{2}\sqrt{(l-m)(l-m-1)},\\
b^{k^\prime k}_{l,l,m+1,m}&=&
-\frac{ic_{ll}}{2}\sqrt{(l+m+1)(l-m)},\\
b^{k^\prime k}_{l+1,l,m+1,m}&=&
\frac{ic_{l+1,l}}{2}\sqrt{(l+m+1)(l+m+2)}.
\end{array}\right.}
\end{equation}
Coming to the dual representations, we find elements of the matrices
$\sL^\ast_1$,
$\sL^\ast_2$ and $\sL^\ast_3$. The transformations $\sL^\ast_i$ in the dual
helicity basis are
\begin{eqnarray}
\sL^\ast_1\zeta^{\dot{k}}_{\dot{l}\dot{m};lm}&=&
\sum_{\dot{l}^\prime,\dot{m}^\prime,\dot{k}^\prime}
d^{\dot{k}^\prime\dot{k}}_{\dot{l}^\prime\dot{l},\dot{m}^\prime\dot{m}}
\zeta^{\dot{k}^\prime}_{\dot{l}^\prime\dot{m}^\prime;lm},\nonumber\\
\sL^\ast_2\zeta^{\dot{k}}_{\dot{l}\dot{m};lm}&=&
\sum_{\dot{l}^\prime,\dot{m}^\prime,\dot{k}^\prime}
e^{\dot{k}^\prime\dot{k}}_{\dot{l}^\prime\dot{l},\dot{m}^\prime\dot{m}}
\zeta^{\dot{k}^\prime}_{\dot{l}^\prime\dot{m}^\prime;lm},\nonumber\\
\sL^\ast_3\zeta^{\dot{k}}_{\dot{l}\dot{m};lm}&=&
\sum_{\dot{l}^\prime,\dot{m}^\prime,\dot{k}^\prime}
f^{\dot{k}^\prime\dot{k}}_{\dot{l}^\prime\dot{l},\dot{m}^\prime\dot{m}}
\zeta^{\dot{k}^\prime}_{\dot{l}^\prime\dot{m}^\prime;lm}.\nonumber
\end{eqnarray}
\begin{sloppypar}\noindent
Calculating the commutators $\ld\sL^\ast_3,\sY_3\rd$,
$\ld\ld\sL^\ast_3,\sY_-\rd,\sY_+\rd$ with respect to the vectors
$\zeta^{\dot{k}}_{\dot{l}\dot{m};lm}$ of the dual basis, we find elements of
the matrix $\sL^\ast_3$. Using the relations
$\sL^\ast_1=\ld\widetilde{\sA}_2,\sL^\ast_3\rd$ (or
$\sL^\ast_1=-i\ld\widetilde{\sB}_2,\sL^\ast_3\rd$) and (\ref{DopA})
(or (\ref{DopB})), we find elements
$d^{\dot{k}^\prime\dot{k}}_{\dot{l}^\prime\dot{l},\dot{m}^\prime\dot{m}}$
of the matrix $\sL^\ast_1$. And also from the relations
$\sL^\ast_2=-\ld\widetilde{\sA}_1,\sL^\ast_3\rd$ (or
$\sL^\ast_2=i\ld\widetilde{\sB}_1,\sL^\ast_3\rd$) we obtain elements 
$e^{\dot{k}^\prime\dot{k}}_{\dot{l}^\prime\dot{l},\dot{m}^\prime\dot{m}}$ of
$\sL^\ast_2$. All calculations are analogous to the calculations presented
for the case of $\sL_i$. In the result we have \end{sloppypar}
\begin{equation}\label{L1'}
{\renewcommand{\arraystretch}{1.25}
\sL^\ast_1:\quad\left\{\begin{array}{ccc}
d^{\dot{k}^\prime\dot{k}}_{\dot{l}-1,\dot{l},\dot{m}-1,\dot{m}}&=&
-\frac{c_{\dot{l}-1,\dot{l}}}{2}
\sqrt{(\dot{l}+\dot{m})(\dot{l}-\dot{m}-1)},\\
d^{\dot{k}^\prime\dot{k}}_{\dot{l},\dot{l},\dot{m}-1,\dot{m}}&=&
\frac{c_{\dot{l}\dot{l}}}{2}
\sqrt{(\dot{l}+\dot{m})(\dot{l}-\dot{m}+1)},\\
d^{\dot{k}^\prime\dot{k}}_{\dot{l}+1,\dot{l},\dot{m}-1,\dot{m}}&=&
\frac{c_{\dot{l}+1,\dot{l}}}{2}
\sqrt{(\dot{l}-\dot{m}+1)(\dot{l}-\dot{m}+2)},\\
d^{\dot{k}^\prime\dot{k}}_{\dot{l}-1,\dot{l},\dot{m}+1,\dot{m}}&=&
\frac{c_{\dot{l}-1,\dot{l}}}{2}
\sqrt{(\dot{l}-\dot{m})(\dot{l}-\dot{m}-1)},\\
d^{\dot{k}^\prime\dot{k}}_{\dot{l},\dot{l},\dot{m}+1,\dot{m}}&=&
\frac{c_{\dot{l}\dot{l}}}{2}
\sqrt{(\dot{l}+\dot{m}+1)(\dot{l}-\dot{m})},\\
d^{\dot{k}^\prime\dot{k}}_{\dot{l}+1,\dot{l},\dot{m}+1,\dot{m}}&=&
-\frac{c_{\dot{l}+1,\dot{l}}}{2}
\sqrt{(\dot{l}+\dot{m}+1)(\dot{l}+\dot{m}+2)}.
\end{array}\right.}
\end{equation}
\begin{equation}\label{L2'}
{\renewcommand{\arraystretch}{1.25}
\sL^\ast_2:\quad\left\{\begin{array}{ccc}
e^{\dot{k}^\prime\dot{k}}_{\dot{l}-1,\dot{l},\dot{m}-1,\dot{m}}&=&
-\frac{ic_{\dot{l}-1,\dot{l}}}{2}
\sqrt{(\dot{l}+\dot{m})(\dot{l}-\dot{m}-1)},\\
e^{\dot{k}^\prime\dot{k}}_{\dot{l},\dot{l},\dot{m}-1,\dot{m}}&=&
\frac{ic_{\dot{l}\dot{l}}}{2}
\sqrt{(\dot{l}+\dot{m})(\dot{l}-\dot{m}+1)},\\
e^{\dot{k}^\prime\dot{k}}_{\dot{l}+1,\dot{l},\dot{m}-1,\dot{m}}&=&
\frac{ic_{\dot{l}+1,\dot{l}}}{2}
\sqrt{(\dot{l}-\dot{m}+1)(\dot{l}-\dot{m}+2)},\\
e^{\dot{k}^\prime\dot{k}}_{\dot{l}-1,\dot{l},\dot{m}+1,\dot{m}}&=&
\frac{-ic_{\dot{l}-1,\dot{l}}}{2}
\sqrt{(\dot{l}-\dot{m})(\dot{l}-\dot{m}-1)},\\
e^{\dot{k}^\prime\dot{k}}_{\dot{l},\dot{l},\dot{m}+1,\dot{m}}&=&
\frac{-ic_{\dot{l}\dot{l}}}{2}
\sqrt{(\dot{l}+\dot{m}+1)(\dot{l}-\dot{m})},\\
e^{\dot{k}^\prime\dot{k}}_{\dot{l}+1,\dot{l},\dot{m}+1,\dot{m}}&=&
-\frac{ic_{\dot{l}+1,\dot{l}}}{2}
\sqrt{(\dot{l}+\dot{m}+1)(\dot{l}+\dot{m}+2)}.
\end{array}\right.}
\end{equation}
\begin{equation}\label{L3'}
{\renewcommand{\arraystretch}{1.25}
\sL^\ast_3:\quad\left\{\begin{array}{ccc}
f^{\dot{k}^\prime\dot{k}}_{\dot{l}-1,\dot{l},\dot{m}}&=&
c^{\dot{k}^\prime\dot{k}}_{\dot{l}-1,\dot{l}}
\sqrt{\dot{l}^2-\dot{m}^2},\\
f^{\dot{k}^\prime\dot{k}}_{\dot{l}\dot{l},\dot{m}}&=&
c^{\dot{k}^\prime\dot{k}}_{\dot{l}\dot{l}}\dot{m},\\
f^{\dot{k}^\prime\dot{k}}_{\dot{l}+1,\dot{l},\dot{m}}&=&
c^{\dot{k}^\prime\dot{k}}_{\dot{l}+1,\dot{l}}
\sqrt{(\dot{l}+1)^2-\dot{m}^2}.
\end{array}\right.}
\end{equation}

\subsection{Boundary value problem}
Following to the classical methods of mathematical physics \cite{CH31},
it is quite natural to set up {\it a boundary value problem for the
relativistic wave equation (relativistically invariant system)}.
It is well known that all the physically meaningful requirements,
which follow from the experience, are contained in the boundary value
problem.

We will set up a boundary value problem for the two-dimensional complex
sphere $\dS^2$ (this problem can be considered as a relativistic
generalization of the classical Dirichlet problem for the sphere $S^2$).\\
{\it Let $T$ be an unbounded region in $\C^3\simeq\R^6$
and let $\Sigma$ be a surface of the complex two-sphere (correspondingly,
$\dot{\Sigma}$, for the dual two-sphere), then it needs to find
a function $\boldsymbol{\psi}(\fg)=(\psi_m(\fg),\dot{\psi}_{\dot{m}}(\fg))^T$
satisfying the following conditions:\\
1) $\boldsymbol{\psi}(\fg)$ is a solution of the system
\begin{eqnarray}
\sum^3_{j=1}\Lambda_j\frac{\partial\psi}{\partial a_j}-
i\sum^3_{j=1}\Lambda_j\frac{\partial\psi}{\partial a^\ast_j}+
\dot{\kappa}^c\dot{\psi}&=&0,\\
\sum^3_{j=1}\Lambda^\ast_j\frac{\partial\dot{\psi}}
{\partial\widetilde{a}_j}+i\sum^3_{j=1}\Lambda^\ast_j
\frac{\partial\dot{\psi}}{\partial\widetilde{a}^\ast_j}+
\kappa^c\psi&=&0,\label{Complex2}
\end{eqnarray}
in the all region $T$;\\
2) $\boldsymbol{\psi}(\fg)$ is a continuous function (everywhere in $T$),
including the surfaces $\Sigma$ and $\dot{\Sigma}$;\\
3) $\left.\phantom{\frac{x}{x}}\psi_m(\fg)\right|_\Sigma=F_m(\fg)$,
$\left.\phantom{\frac{x}{x}}\dot{\psi}_{\dot{m}}(\fg)\right|_{\dot{\Sigma}}=
\dot{F}_{\dot{m}}(\fg)$, where $F_m(\fg)$ and 
$\dot{F}_{\dot{m}}(\fg)$ are square integrable
functions defined on the surfaces $\Sigma$ and $\dot{\Sigma}$, respectively.}

In particular, boundary conditions can be represented by constants,
\[
\left.\phantom{\frac{x}{x}}\psi(\fg)\right|_\Sigma=\text{const}=F_0,\quad
\left.\phantom{\frac{x}{x}}\dot{\psi}(\fg)\right|_{\dot{\Sigma}}=\text{const}=
\dot{F}_0.
\]
It is obvious that an explicit form of the boundary conditions follows
from the experience. For example, they can describe a distribution of
energy in the experiment.

With the aim to solve the boundary value problem we come to the complex
Euler angles (\ref{CEA}) and represent the function
$\boldsymbol{\psi}(r,\theta^c,\varphi^c)=(\psi_m(r,\theta^c,\varphi^c),
\dot{\psi}_{\dot{m}}(r^\ast,\dot{\theta}^c,\dot{\varphi}^c))^T$ in the
form of following series
\begin{eqnarray}
\psi_m(r,\theta^c,\varphi^c)&=&\sum^\infty_{l=0}\sum_k
\boldsymbol{f}_{lmk}(r)\sum^l_{n=-l}\alpha^m_{l,n}
\fM^l_{mn}(\varphi,\epsilon,\theta,\tau,0,0),\label{Fourier1}\\
\dot{\psi}_{\dot{m}}(r^\ast,\dot{\theta}^c,\dot{\varphi}^c)&=&
\sum^\infty_{\dot{l}=0}\sum_{\dot{k}}
\boldsymbol{f}_{\dot{l}\dot{m}\dot{k}}(r^\ast)
\sum^{\dot{l}}_{\dot{n}=-\dot{l}}\alpha^{\dot{m}}_{\dot{l},\dot{n}}
\fM^{\dot{l}}_{\dot{m}\dot{n}}(\varphi,\epsilon,\theta,\tau,0,0),
\label{Fourier2}
\end{eqnarray}
where
\begin{eqnarray}
\alpha^m_{l,n}&=&\frac{(-1)^n(2l+1)(2\dot{l}+1)}{32\pi^4}
\int\limits_{\dS^2}
F_m(\theta^c,\varphi^c)\fM_l^{n}(\varphi,\epsilon,\theta,\tau,0,0)
\sin\theta^c\sin\dot{\theta}^cd\theta d\varphi d\tau d\epsilon,\nonumber\\
\alpha^{\dot{m}}_{\dot{l},\dot{n}}&=
&\frac{(-1)^{\dot{n}}(2l+1)(2\dot{l}+1)}{32\pi^4}
\int\limits_{\dS^2}
\dot{F}_{\dot{m}}(\dot{\theta}^c,\dot{\varphi}^c)
\fM^{\dot{n}}_{\dot{l}}(\varphi,\epsilon,\theta,\tau,0,0)
\sin\theta^c\sin\dot{\theta}^cd\theta d\varphi d\tau d\epsilon,\nonumber
\end{eqnarray}
The index $k$ numerates equivalent representations.
$\fM^n_l(\varphi,\epsilon,\theta,\tau,0,0)$
($\fM_{\dot{l}}^{\dot{n}}(\varphi,\epsilon,\theta,\tau,0,0)$) are
associated hyperspherical functions defined on 
the surface $\Sigma$ ($\dot{\Sigma}$) of the 
two-dimensional complex sphere of the radius $r$ ($r^\ast$), 
$\boldsymbol{f}_{lmk}(r)$ and
$\boldsymbol{f}_{\dot{l}\dot{m}\dot{k}}(r^\ast)$ are radial
functions. It is easy to see that we come here to the harmonic analysis
on the complex two-sphere, since the series (\ref{Fourier1}) and
(\ref{Fourier2}) have the structure of the Fourier series on $\dS^2$.

General solutions of the system (\ref{Complex}) have been found in the
work \cite{Var031}
on the tangent bundle
$T\fL_6=\underset{\fg\in\fL_6}{\cup}T_{\fg}\fL_6$ of the group manifold $\fL_6$.
A separation of variables in
(\ref{Complex}) is realized via the following factorization
\begin{eqnarray}
\psi^k_{lm;\dot{l}\dot{m}}&=&\boldsymbol{f}^{l_0}_{lmk}(r)
\fM^{l_0}_{mn}(\varphi,\epsilon,\theta,\tau,0,0),\nonumber\\
\psi^{\dot{k}}_{\dot{l}\dot{m};lm}&=&
\boldsymbol{f}^{\dot{l}_0}_{\dot{l}\dot{m}\dot{k}}(r^\ast)
\fM^{\dot{l}_0}_{\dot{m}\dot{n}}(\varphi,\epsilon,\theta,\tau,0,0),\label{F}
\end{eqnarray}
where $l_0\ge l$, $-l_0\le m$, $n\le l_0$ and $\dot{l}_0\ge\dot{l}$,
$-\dot{l}_0\le\dot{m}$, $\dot{n}\le \dot{l}_0$. In the result of separation
of variables  
the relativistically invariant system (\ref{Complex}) is
reduced to a system of ordinary differential equations 
(for more details see \cite{Var031}):
\begin{multline}
\sum_{k^\prime}c^{kk^\prime}_{l,l-1}\left[
2\sqrt{l^2-m^2}
\frac{d\boldsymbol{f}^{l_0}_{l-1,m,k^\prime}(r)}{d r}
-\frac{1}{r}(l+1)\sqrt{l^2-m^2}\boldsymbol{f}^{l_0}_{l-1,m,k^\prime}(r)+\right.\\
+\frac{1}{r}\sqrt{(l+m)(l+m-1)}
\sqrt{(l_0+m)(l_0-m+1)}
\boldsymbol{f}^{l_0}_{l-1,m-1,k^\prime}(r)+\\
%\shoveright{
\left.+\frac{1}{r}\sqrt{(l-m)(l-m-1)}
\sqrt{(l_0+m+1)(l_0-m)}
\boldsymbol{f}^{l_0}_{l-1,m+1,k^\prime}(r)\right]+
%}
\nonumber
\end{multline}
\begin{multline}
%\shoveleft{
\sum_{k^\prime}c^{kk^\prime}_{ll}\left[
2m\frac{d\boldsymbol{f}^{l_0}_{l,m,k^\prime}(r)}{d r}-
\frac{1}{r}m\boldsymbol{f}^{l_0}_{l,m,k^\prime}(r)-\right.
%}
\\
-\frac{1}{r}\sqrt{(l+m)(l-m+1)}
\sqrt{(l_0+m)(l_0-m+1)}
\boldsymbol{f}^{l_0}_{l,m-1,k^\prime}(r)+\\
%\shoveright{
\left.+\frac{1}{r}\sqrt{(l+m+1)(l-m)}
\sqrt{(l_0+m+1)(l_0-m)}
\boldsymbol{f}^{l_0}_{l,m+1,k^\prime}(r)\right]+
%}
\nonumber
\end{multline}
\begin{multline}
%\shoveleft{
\sum_{k^\prime}c^{kk^\prime}_{l,l+1}\left[
2\sqrt{(l+1)^2-m^2}
\frac{d\boldsymbol{f}^{l_0}_{l+1,m,k^\prime}(r)}{d r}
%}
+\frac{1}{r}l\sqrt{(l+1)^2-m^2}
\boldsymbol{f}^{l_0}_{l+1,m,k^\prime}(r)-\right.\\
-\frac{1}{r}\sqrt{(l-m+1)(l-m+2)}
\sqrt{(l_0+m)(l_0-m+1)}
\boldsymbol{f}^{l_0}_{l+1,m-1,k^\prime}(r)-\\
%\shoveright{
\left.-\frac{1}{r}\sqrt{(l+m+1)(l+m+2)}
\sqrt{(l_0+m+1)(l_0-m)}
\boldsymbol{f}^{l_0}_{l+1,m+1,k^\prime}(r)\right]+
%}
\\
+\kappa^c\boldsymbol{f}^{l_0}_{lmk}(r)=0,\nonumber
\end{multline}
\begin{multline}
\sum_{\dot{k}^\prime}c^{\dot{k}\dot{k}^\prime}_{\dot{l},\dot{l}-1}\left[
2\sqrt{\dot{l}^2-\dot{m}^2}
\frac{d\boldsymbol{f}^{\dot{l}_0}_{\dot{l}-1,\dot{m},\dot{k}^\prime}(r^\ast)}
{d r^\ast}
-\frac{1}{r^\ast}(\dot{l}+1)\sqrt{\dot{l}^2-\dot{m}^2}
\boldsymbol{f}^{\dot{l}_0}_{\dot{l}-1,\dot{m},\dot{k}^\prime}(r^\ast)+\right.\\
+\frac{1}{r^\ast}\sqrt{(\dot{l}+\dot{m})(\dot{l}+\dot{m}-1)}
\sqrt{(\dot{l}_0+\dot{m})(\dot{l}_0-\dot{m}+1)}
\boldsymbol{f}^{\dot{l}_0}_{\dot{l}-1,\dot{m}-1,\dot{k}^\prime}(r^\ast)+\\
%\shoveright{
\left.
+\frac{1}{r^\ast}\sqrt{(\dot{l}-\dot{m})(\dot{l}-\dot{m}-1)}
\sqrt{(\dot{l}_0+\dot{m}+1)(\dot{l}_0-\dot{m})}
\boldsymbol{f}^{\dot{l}_0}_{\dot{l}-1,\dot{m}+1,\dot{k}^\prime}(r^\ast)\right]+
%}
\nonumber
\end{multline}
\begin{multline}
%\shoveleft{
\sum_{\dot{k}^\prime}c^{\dot{k}\dot{k}^\prime}_{\dot{l},\dot{l}}\left[
2\dot{m}
\frac{d\boldsymbol{f}^{\dot{l}_0}_{\dot{l},\dot{m},\dot{k}^\prime}(r^\ast)}
{d r^\ast}
-\frac{1}{r^\ast}\dot{m}
\boldsymbol{f}^{\dot{l}_0}_{\dot{l},\dot{m},\dot{k}^\prime}(r^\ast)-\right.
%}
\\
-\frac{1}{r^\ast}\sqrt{(\dot{l}+\dot{m})(\dot{l}-\dot{m}-1)}
\sqrt{(\dot{l}_0+\dot{m})(\dot{l}_0-\dot{m}+1)}
\boldsymbol{f}^{\dot{l}_0}_{\dot{l},\dot{m}-1,\dot{k}^\prime}(r^\ast)+\\
%\shoveright{
\left.
+\frac{1}{r^\ast}\sqrt{(\dot{l}+\dot{m}+1)(\dot{l}-\dot{m})}
\sqrt{(\dot{l}_0+\dot{m}+1)(\dot{l}_0-\dot{m})}
\boldsymbol{f}^{\dot{l}_0}_{\dot{l},\dot{m}+1,\dot{k}^\prime}(r^\ast)\right]+
%}
\nonumber
\end{multline}
\begin{multline}
%\shoveleft{
\sum_{\dot{k}^\prime}c^{\dot{k}\dot{k}^\prime}_{\dot{l},\dot{l}+1}\left[
2\sqrt{(\dot{l}+1)^2-\dot{m}^2}
\frac{d\boldsymbol{f}^{\dot{l}_0}_{\dot{l}+1,\dot{m},\dot{k}^\prime}(r^\ast)}
{d r^\ast}
%}
+\frac{1}{r^\ast}\dot{l}\sqrt{(\dot{l}+1)^2-\dot{m}^2}
\boldsymbol{f}^{\dot{l}_0}_{\dot{l}+1,\dot{m},\dot{k}^\prime}(r^\ast)-\right.\\
-\frac{1}{r^\ast}\sqrt{(\dot{l}-\dot{m}+1)(\dot{l}-\dot{m}+2)}
\sqrt{(\dot{l}_0+\dot{m})(\dot{l}_0-\dot{m}+1)}
\boldsymbol{f}^{\dot{l}_0}_{\dot{l}+1,\dot{m}-1,\dot{k}^\prime}(r^\ast)-\\
%\shoveright{
\left.
-\frac{1}{r^\ast}\sqrt{(\dot{l}+\dot{m}+1)(\dot{l}+\dot{m}+2)}
\sqrt{(\dot{l}_0+\dot{m}+1)(\dot{l}_0-\dot{m})}
\boldsymbol{f}^{\dot{l}_0}_{\dot{l}+1,\dot{m}+1,\dot{k}^\prime}(r^\ast)\right]+
%}
\\
+\dot{\kappa}^c
\boldsymbol{f}^{\dot{l}_0}_{\dot{l}\dot{m}\dot{k}^\prime}(r^\ast)=0.
\label{RFS}
\end{multline}
Substituting solutions of this system into the series (\ref{Fourier1}) and
(\ref{Fourier2}), we obtain a solution of the boundary value problem.
It is easy to see that a boundary value problem of the same type can be
defined on the homogeneous spaces $\cM_8$ and $\cM_6$. The analogous
problem on the homogeneous space $\cM_7$ leads to relativistic wave
equations in $2+1$ dimensions \cite{JN91,Ply92,CP94,GS97} (or
relativistically invariant system on the group $SU(1,1)$). The boundary
value problem for the hyperboloid $\BH^3$ comes beyond the framework
of this paper and will be studied in a future work.
\section{The Dirac field}
\label{Sec:Dir}
In this section we will consider a boundary value problem for the Dirac
field $(1/2,0)\oplus(0,1/2)$ (electron-positron field) defined on the
homogeneous space $\cM_8$. Solution of this problem allows us to construct
field operators and further to define a quantization procedure for the
Dirac field on the space $\cM_8$.

We start with the Lagrangian (\ref{Lagrange}) on the group manifold $\cM_{10}$:
\begin{multline}\label{LagDir}
\cL(\balpha)=-\frac{1}{2}\left(\overline{\boldsymbol{\psi}}(\balpha)\Gamma_\mu
\frac{\partial\boldsymbol{\psi}(\balpha)}{\partial x_\mu}-
\frac{\partial\overline{\boldsymbol{\psi}}(\balpha)}{\partial x_\mu}
\Gamma_\mu\boldsymbol{\psi}(\balpha)\right)-\\
-\frac{1}{2}\left(\overline{\boldsymbol{\psi}}(\balpha)\Upsilon_\nu
\frac{\partial\boldsymbol{\psi}(\balpha)}{\partial\fg_\nu}-
\frac{\partial\overline{\boldsymbol{\psi}}(\balpha)}{\partial\fg_\nu}
\Upsilon_\nu\boldsymbol{\psi}(\balpha)\right)-
\kappa\overline{\boldsymbol{\psi}}(\balpha)\boldsymbol{\psi}(\balpha),
\end{multline}
where $\boldsymbol{\psi}(\balpha)=\psi(x)\psi(\fg)$ 
($\mu=0,1,2,3,\;\nu=1,\ldots,6$),
and
\begin{equation}\label{Gamma1}
\gamma_0=\begin{pmatrix}
\sigma_0 & 0\\
0 & -\sigma_0
\end{pmatrix},\;\;\gamma_1=\begin{pmatrix}
0 & \sigma_1\\
-\sigma_1 & 0
\end{pmatrix},\;\;\gamma_2=\begin{pmatrix}
0 & \sigma_2\\
-\sigma_2 & 0
\end{pmatrix},\;\;\gamma_3=\begin{pmatrix}
0 & \sigma_3\\
-\sigma_3 & 0
\end{pmatrix},
\end{equation}
\begin{equation}\label{Upsilon1}
\Upsilon_1=\begin{pmatrix}
0 & \Lambda^\ast_1\\
\Lambda_1 & 0
\end{pmatrix},\quad\Upsilon_2=\begin{pmatrix}
0 & \Lambda^\ast_2\\
\Lambda_2 & 0
\end{pmatrix},\quad\Upsilon_3=\begin{pmatrix}
0 & \Lambda^\ast_3\\
\Lambda_3 & 0
\end{pmatrix},
\end{equation}
\begin{equation}\label{Upsilon2}
\Upsilon_4=\begin{pmatrix}
0 & i\Lambda^\ast_1\\
i\Lambda_1 & 0
\end{pmatrix},\quad\Upsilon_5=\begin{pmatrix}
0 & i\Lambda^\ast_2\\
i\Lambda_2 & 0
\end{pmatrix},\quad\Upsilon_6=\begin{pmatrix}
0 & i\Lambda^\ast_3\\
i\Lambda_3 & 0
\end{pmatrix},
\end{equation}
where $\sigma_i$ are the Pauli matrices, and the matrices $\Lambda_j$ and
$\Lambda^\ast_j$ are derived from (\ref{L1})--(\ref{L3}) and 
(\ref{L1'})--(\ref{L3'}) at $l=1/2$:
\begin{eqnarray}
&&\sL_1=\frac{1}{2}c_{\frac{1}{2}\frac{1}{2}}\begin{pmatrix}
0 & 1\\
1 & 0
\end{pmatrix},\quad
\sL_2=\frac{1}{2}c_{\frac{1}{2}\frac{1}{2}}\begin{pmatrix}
0 & -i\\
i & 0
\end{pmatrix},\quad
\sL_3=\frac{1}{2}c_{\frac{1}{2}\frac{1}{2}}\begin{pmatrix}
1 & 0\\
0 & -1
\end{pmatrix},\nonumber\\
&&\sL^\ast_1=\frac{1}{2}\dot{c}_{\frac{1}{2}\frac{1}{2}}\begin{pmatrix}
0 & 1\\
1 & 0
\end{pmatrix},\quad
\sL^\ast_2=\frac{1}{2}\dot{c}_{\frac{1}{2}\frac{1}{2}}\begin{pmatrix}
0 & -i\\
i & 0
\end{pmatrix},\quad
\sL^\ast_3=\frac{1}{2}\dot{c}_{\frac{1}{2}\frac{1}{2}}\begin{pmatrix}
1 & 0\\
0 & -1
\end{pmatrix}.\label{LDir}
\end{eqnarray}
It is easy to see that these matrices coincide with the Pauli matrices
$\sigma_i$ when $c_{\frac{1}{2}\frac{1}{2}}=2$. 

Varying independently $\psi(x)$ and $\overline{\psi}(x)$ in the
Lagrangian (\ref{LagDir}), and then $\psi(\fg)$ and $\overline{\psi}(\fg)$,
we come to the following equations:
\begin{equation}\label{FET2}
\begin{array}{ccc}
\gamma_i\dfrac{\partial\psi(x)}{\partial x_i}+\kappa\psi(x)&=&0,\\
\gamma^T_i\dfrac{\partial\overline{\psi}(x)}{\partial x_i}-
\kappa\overline{\psi}(x)&=&0.
\end{array}\quad(i=1,\ldots,4)
\end{equation}
\begin{equation}\label{FEL2}
\begin{array}{ccc}
\Upsilon_k\dfrac{\partial\psi(\fg)}{\partial\fg_k}+\kappa\psi(\fg)&=&0,\\
\Upsilon^T_k\dfrac{\overline{\psi}(\fg)}{\partial\fg_k}-
\kappa\overline{\psi}(\fg)&=&0,
\end{array}\quad(k=1,\ldots,6)
\end{equation}
Now we can formulate the boundary value problem. {\it Let $T$ be an 
unbounded region in $\cM_8=\R^{1,3}\times\dS^2$ and let $\Sigma$ 
($\dot{\Sigma}$) be a surface of the complex two-sphere, then it needs to
find the function $\boldsymbol{\psi}(\balpha)=(\psi_1(\balpha),\psi_2(\balpha),
\dot{\psi}_1(\balpha),\dot{\psi}_2(\balpha))^T$, such that\\
1) $\boldsymbol{\psi}(\balpha)$ satisfies the equations (\ref{FET2}) and
(\ref{FEL2}) in the all region $T$.\\
2) $\boldsymbol{\psi}(\balpha)$ is a continuous function everywhere in $T$.\\
3) $\left.\phantom{\frac{x}{x}}\psi_m(\balpha)\right|_\Sigma=F_m(\balpha)$,
$\left.\phantom{\frac{x}{x}}\dot{\psi}_m(\balpha)\right|_{\dot{\Sigma}}=
\dot{F}_m(\balpha)$, where $F_m(\balpha)$ and $\dot{F}_m(\balpha)$ are
square integrable and infinitely differentiable finctions in $\cM_8$,
$m=1,2$}.

The first equation from (\ref{FET2}) coincides with the Dirac equation, and
the second equation coincides with the Dirac equation for antiparticle.
As is known, solutions of these equations are found in the
plane-wave approximation, that is, 
in the form \cite{BD64,Ryd85}\footnote{This form is a direct consequence
of the $T_4$-structure of the field equations (\ref{FET2}), since the
variables $x_i$ are parameters of $T_4$ and all the irreducible
representations of $T_4$ are expressed via the exponentials.}:
\begin{eqnarray}
\psi^+(x)&=&u(\bp)e^{-ipx},\nonumber\\
\psi^-(x)&=&v(\bp)e^{ipx},\nonumber
\end{eqnarray}
where the solutions $\psi^+(x)$ and $\psi^-(x)$ correspond to positive
and negative energy, respectively, and the amplitudes $u(\bp)$ and
$v(\bp)$ have the following components
\begin{gather}
u_1(\bp)=\left(\frac{E+m}{2m}\right)^{1/2}\begin{pmatrix}
1\\
0\\
\frac{p_z}{E+m}\\
\frac{p_+}{E+m}
\end{pmatrix},\quad
u_2(\bp)=\left(\frac{E+m}{2m}\right)^{1/2}\begin{pmatrix}
0\\
1\\
\frac{p_-}{E+m}\\
\frac{-p_z}{E+m}
\end{pmatrix},\nonumber\\
v_1(\bp)=\left(\frac{E+m}{2m}\right)^{1/2}\begin{pmatrix}
\frac{p_z}{E+m}\\
\frac{p_+}{E+m}\\
1\\
0
\end{pmatrix},\quad
v_2(\bp)=\left(\frac{E+m}{2m}\right)^{1/2}\begin{pmatrix}
\frac{p_-}{E+m}\\
\frac{-p_z}{E+m}\\
0\\
1
\end{pmatrix},\nonumber
\end{gather}
where $p_\pm=p_x\pm ip_y$.

Further, the first equation from (\ref{FEL2}) can be written as
\begin{eqnarray}
\sum^3_{j=1}\Lambda_j\frac{\partial\psi}{\partial a_j}-
i\sum^3_{j=1}\Lambda_j\frac{\partial\psi}{\partial a^\ast_j}+
\dot{\kappa}^c\dot{\psi}&=&0,\\
\sum^3_{j=1}\Lambda^\ast_j\frac{\partial\dot{\psi}}
{\partial\widetilde{a}_j}+i\sum^3_{j=1}\Lambda^\ast_j
\frac{\partial\dot{\psi}}{\partial\widetilde{a}^\ast_j}+
\kappa^c\psi&=&0,\label{Complex3}
\end{eqnarray}
Or, taking into account the explicit form of the matrices $\Lambda_j$
($\Lambda^\ast_j$) given by (\ref{LDir}), we obtain
\begin{eqnarray}
&&-\frac{1}{2}\frac{\partial\dot{\psi}_2}{\partial\widetilde{a}_1}+
\frac{i}{2}\frac{\partial\dot{\psi}_2}{\partial\widetilde{a}_2}-
\frac{1}{2}\frac{\partial\dot{\psi}_1}{\partial\widetilde{a}_3}-
\frac{i}{2}\frac{\partial\dot{\psi}_2}{\partial\widetilde{a}^\ast_1}-
\frac{1}{2}\frac{\partial\dot{\psi}_2}{\partial\widetilde{a}^\ast_2}-
\frac{i}{2}\frac{\partial\dot{\psi}_1}{\partial\widetilde{a}^\ast_3}-
\kappa^c\psi_1=0,\nonumber\\
&&-\frac{1}{2}\frac{\partial\dot{\psi}_1}{\partial\widetilde{a}_1}-
\frac{i}{2}\frac{\partial\dot{\psi}_1}{\partial\widetilde{a}_2}+
\frac{1}{2}\frac{\partial\dot{\psi}_2}{\partial\widetilde{a}_3}-
\frac{i}{2}\frac{\partial\dot{\psi}_1}{\partial\widetilde{a}^\ast_1}+
\frac{1}{2}\frac{\partial\dot{\psi}_1}{\partial\widetilde{a}^\ast_2}+
\frac{i}{2}\frac{\partial\dot{\psi}_2}{\partial\widetilde{a}^\ast_3}-
\kappa^c\psi_2=0,\nonumber\\
&&\phantom{-}\frac{1}{2}\frac{\partial\psi_2}{\partial a_1}-
\frac{i}{2}\frac{\partial\psi_2}{\partial a_2}+
\frac{1}{2}\frac{\partial\psi_1}{\partial a_3}-
\frac{i}{2}\frac{\partial\psi_2}{\partial a^\ast_1}-
\frac{1}{2}\frac{\partial\psi_2}{\partial a^\ast_2}-
\frac{i}{2}\frac{\partial\psi_1}{\partial a^\ast_3}-
\dot{\kappa}^c\dot{\psi}_1=0,\nonumber\\
&&\phantom{-}\frac{1}{2}\frac{\partial\psi_1}{\partial a_1}+
\frac{i}{2}\frac{\partial\psi_1}{\partial a_2}-
\frac{1}{2}\frac{\partial\psi_2}{\partial a_3}-
\frac{i}{2}\frac{\partial\psi_1}{\partial a^\ast_1}+
\frac{1}{2}\frac{\partial\psi_1}{\partial a^\ast_2}+
\frac{i}{2}\frac{\partial\psi_2}{\partial a^\ast_3}-
\dot{\kappa}^c\dot{\psi}_2=0,\label{Dirac2}
\end{eqnarray}
The latter system acts on the tangent bundle $T\fL_6$ of the group
manifold $\fL_6$. 
Coming to the helicity basis,
we will find solutions of the
system (\ref{Dirac2}), that is, we will present components of the Dirac
bispinor $\psi=(\psi_1,\psi_2,\dot{\psi}_1,\dot{\psi}_2)^T$
in terms of the functions on the two-dimensional complex sphere
(the indices $k$ and $\dot{k}$ we can omit, since representations
$\boldsymbol{\tau}_{\frac{1}{2},0}$ and
$\boldsymbol{\tau}_{0,\frac{1}{2}}$ occur only one time):
\begin{eqnarray}
\psi_1&=&\psi_{\frac{1}{2},\frac{1}{2};\frac{1}{2},\frac{1}{2}}=
\boldsymbol{f}^l_{\frac{1}{2},\frac{1}{2}}(r)
\fM^l_{\frac{1}{2},n}(\varphi,\epsilon,\theta,\tau,0,0),\nonumber\\
\psi_2&=&\psi_{\frac{1}{2},-\frac{1}{2};\frac{1}{2},\frac{1}{2}}=
\boldsymbol{f}^l_{\frac{1}{2},-\frac{1}{2}}(r)
\fM^l_{-\frac{1}{2},n}(\varphi,\epsilon,\theta,\tau,0,0),\nonumber\\
\dot{\psi}_1&=&\dot{\psi}_{\frac{1}{2},\frac{1}{2};\frac{1}{2},\frac{1}{2}}=
\boldsymbol{f}^{\dot{l}}_{\frac{1}{2},\frac{1}{2}}(r^\ast)
\fM^{\dot{l}}_{\frac{1}{2},\dot{n}}(\varphi,\epsilon,\theta,\tau,0,0),\nonumber\\
\dot{\psi}_2&=&\dot{\psi}_{\frac{1}{2},\frac{1}{2};\frac{1}{2},-\frac{1}{2}}=
\boldsymbol{f}^{\dot{l}}_{\frac{1}{2},-\frac{1}{2}}(r^\ast)
\fM^{\dot{l}}_{-\frac{1}{2},\dot{n}}(\varphi,\epsilon,\theta,\tau,0,0),\nonumber
\end{eqnarray}
Substituting these functions into (\ref{Dirac2}) and separating the variables
with the aid of recurrence relations between hyperspherical functions,
we come to the following system of ordinary differential 
equations (the system (\ref{RFS}) at $l=1/2$):
\begin{eqnarray}
&&-\frac{1}{2}\frac{d\boldsymbol{f}^{\dot{l}}_{\frac{1}{2},\frac{1}{2}}(r^\ast)}
{d r^\ast}+\frac{1}{4r^\ast}\boldsymbol{f}^{\dot{l}}_{\frac{1}{2},\frac{1}{2}}
(r^\ast)+\frac{\dot{l}+\frac{1}{2}}{2r^\ast}
\boldsymbol{f}^{\dot{l}}_{\frac{1}{2},-\frac{1}{2}}(r^\ast)-
\kappa^c\boldsymbol{f}^l_{\frac{1}{2},\frac{1}{2}}(r)=0,\nonumber\\
&&\phantom{-}\frac{1}{2}\frac{d\boldsymbol{f}^{\dot{l}}_{\frac{1}{2},-\frac{1}{2}}(r^\ast)}
{d r^\ast}-\frac{1}{4r^\ast}\boldsymbol{f}^{\dot{l}}_{\frac{1}{2},-\frac{1}{2}}
(r^\ast)-\frac{\dot{l}+\frac{1}{2}}{2r^\ast}
\boldsymbol{f}^{\dot{l}}_{\frac{1}{2},\frac{1}{2}}(r^\ast)-
\kappa^c\boldsymbol{f}^l_{\frac{1}{2},-\frac{1}{2}}(r)=0,\nonumber\\
&&\phantom{-}\frac{1}{2}\frac{d\boldsymbol{f}^l_{\frac{1}{2},\frac{1}{2}}(r)}{dr}-
\frac{1}{4r}\boldsymbol{f}^l_{\frac{1}{2},\frac{1}{2}}(r)-
\frac{l+\frac{1}{2}}{2r}\boldsymbol{f}^l_{\frac{1}{2},-\frac{1}{2}}-
\dot{\kappa}^c\boldsymbol{f}^{\dot{l}}_{\frac{1}{2},\frac{1}{2}}(r^\ast)=0,
\nonumber\\
&&-\frac{1}{2}\frac{d\boldsymbol{f}^l_{\frac{1}{2},-\frac{1}{2}}(r)}{dr}+
\frac{1}{4r}\boldsymbol{f}^l_{\frac{1}{2},-\frac{1}{2}}(r)+
\frac{l+\frac{1}{2}}{2r}\boldsymbol{f}^l_{\frac{1}{2},\frac{1}{2}}-
\dot{\kappa}^c\boldsymbol{f}^{\dot{l}}_{\frac{1}{2},-\frac{1}{2}}(r^\ast)=0,
\nonumber
\end{eqnarray}
For the brevity of exposition we suppose 
$\boldsymbol{f}_1=\boldsymbol{f}^l_{\frac{1}{2},\frac{1}{2}}(r)$,
$\boldsymbol{f}_2=\boldsymbol{f}^l_{\frac{1}{2},-\frac{1}{2}}(r)$,
$\boldsymbol{f}_3=\boldsymbol{f}^{\dot{l}}_{\frac{1}{2},\frac{1}{2}}(r^\ast)$,
$\boldsymbol{f}_4=\boldsymbol{f}^{\dot{l}}_{\frac{1}{2},-\frac{1}{2}}(r^\ast)$.
Then
\begin{eqnarray}
&&-2\frac{d\boldsymbol{f}_3}{dr^\ast}+\frac{1}{r^\ast}\boldsymbol{f}_3+
\frac{2\left(\dot{l}+\frac{1}{2}\right)}{r^\ast}\boldsymbol{f}_4-
4\kappa^c\boldsymbol{f}_1=0,\nonumber\\
&&\phantom{-}2\frac{d\boldsymbol{f}_4}{dr^\ast}-\frac{1}{r^\ast}\boldsymbol{f}_4-
\frac{2\left(\dot{l}+\frac{1}{2}\right)}{r^\ast}\boldsymbol{f}_3-
4\kappa^c\boldsymbol{f}_2=0,\nonumber\\
&&\phantom{-}2\frac{d\boldsymbol{f}_1}{dr}-\frac{1}{r}\boldsymbol{f}_1-
\frac{2\left(l+\frac{1}{2}\right)}{r}\boldsymbol{f}_2-
4\dot{\kappa}^c\boldsymbol{f}_3=0,\nonumber\\
&&-2\frac{d\boldsymbol{f}_2}{dr}+\frac{1}{r}\boldsymbol{f}_2+
\frac{2\left(l+\frac{1}{2}\right)}{r}\boldsymbol{f}_1-
4\dot{\kappa}^c\boldsymbol{f}_4=0.\nonumber
\end{eqnarray}
Let us assume that $\boldsymbol{f}_3=\mp\boldsymbol{f}_4$ and
$\boldsymbol{f}_2=\pm\boldsymbol{f}_1$, then the first equation coincides
with the second, and the third equations coincides with the fourth.
Therefore,
\begin{eqnarray}
&&\frac{d\boldsymbol{f}_4}{dr^\ast}+\frac{\dot{l}}{r^\ast}\boldsymbol{f}_4-
2\kappa^c\boldsymbol{f}_1=0,\nonumber\\
&&\frac{d\boldsymbol{f}_1}{dr}-\frac{l+1}{r}\boldsymbol{f}_1+
2\dot{\kappa}^c\boldsymbol{f}_4=0.\nonumber
\end{eqnarray}
Let us consider a real part $\re r$ of the radius of complex two-sphere.
It is obvious that $\re r=\re r^\ast$. Denoting $\sz=\re r=\re r^\ast$ and
excluding the function $\boldsymbol{f}_4$ at $l=\dot{l}$, 
we come to the following differential equation:
\begin{equation}\label{Bess}
\sz^2\frac{d^2\boldsymbol{f}_1}{d\sz^2}-\sz\frac{d\boldsymbol{f}_1}{d\sz}-
(l^2-1-4\kappa^c\dot{\kappa}^c\sz^2)\boldsymbol{f}_1=0.
\end{equation}
The latter equation is solvable in the Bessel functions of half-integer
order:
\[
\boldsymbol{f}_1(\sz)=C_1\sqrt{\kappa^c\dot{\kappa}^c}\sz 
J_l\left(\sqrt{\kappa^c\dot{\kappa}^c}\sz\right)+
C_2\sqrt{\kappa^c\dot{\kappa}^c}\sz
J_{-l}\left(\sqrt{\kappa^c\dot{\kappa}^c}\sz\right).
\]
Further, using recurrence relations between Bessel functions, we find
\begin{multline}
\boldsymbol{f}_4(\sz)=\frac{1}{2\kappa^c}
\left(\frac{l+1}{\sz}\boldsymbol{f}_1(\sz)-
\frac{d\boldsymbol{f}_1}{d\sz}\right)=\\
=\frac{C_1}{2}\sqrt{\frac{\dot{\kappa}^c}{\kappa^c}}\sz
J_{l+1}\left(\sqrt{\kappa^c\dot{\kappa}^c}\sz\right)-
\frac{C_2}{2}\sqrt{\frac{\dot{\kappa}^c}{\kappa^c}}\sz
J_{-l-1}\left(\sqrt{\kappa^c\dot{\kappa}^c}\sz\right).\nonumber
\end{multline}
Therefore,
\[
\boldsymbol{f}^l_{\frac{1}{2},\frac{1}{2}}(\re r)=
C_1\sqrt{\kappa^c\dot{\kappa}^c}\re r 
J_l\left(\sqrt{\kappa^c\dot{\kappa}^c}\re r\right)+
C_2\sqrt{\kappa^c\dot{\kappa}^c}\re r 
J_{-l}\left(\sqrt{\kappa^c\dot{\kappa}^c}\re r\right),
\]
\[
\boldsymbol{f}^{\dot{l}}_{\frac{1}{2},-\frac{1}{2}}(\re r^\ast)=
\frac{C_1}{2}\sqrt{\frac{\dot{\kappa}^c}{\kappa^c}}\re r^\ast
J_{l+1}\left(\sqrt{\kappa^c\dot{\kappa}^c}\re r^\ast\right)
-\frac{C_2}{2}\sqrt{\frac{\dot{\kappa}^c}{\kappa^c}}\re r^\ast
J_{-l-1}\left(\sqrt{\kappa^c\dot{\kappa}^c}\re r^\ast\right).
\]
%where $\nu=-(l-1)$, $\dot{\nu}=-(\dot{l}-1)$, 
%$l=\frac{2s+1}{2}$, $\dot{l}=\frac{2\dot{s}+1}{2}$,
%$s,\dot{s}=0,1,2,\ldots$ and
%\begin{multline}
%J_{\frac{2s+1}{2}}(z)=\sqrt{\frac{2}{\pi z}}\left[
%\sin\left(z-\frac{s\pi}{2}\right)\sum^{\frac{s}{2}}_{k=0}
%\frac{(-1)^k(s+2k)!}{(2k)!(s-2k)!(2z)^{2k}}+\right.\\
%\left.
%+\cos\left(z-\frac{s\pi}{2}\right)\sum^{\left[\frac{s-1}{2}\right]}_{k=0}
%\frac{(-1)^k(s+2k+1)!}{(2k+1)!(s-2k-1)!(2z)^{2k+1}}\right].\label{Bessel}
%\end{multline}
%is the Bessel function of half-integer order. 
In such a way, solutions of the system (\ref{Dirac2})
are defined by the following functions:
\begin{eqnarray}
\psi_1(r,\varphi^c,\theta^c)&=&\boldsymbol{f}^l_{\frac{1}{2},\frac{1}{2}}
(\re r)\fM^l_{\frac{1}{2},n}(\varphi,\epsilon,\theta,\tau,0,0),\nonumber\\
\psi_2(r,\varphi^c,\theta^c)&=&\pm\boldsymbol{f}^l_{\frac{1}{2},\frac{1}{2}}
(\re r)\fM^l_{-\frac{1}{2},n}(\varphi,\epsilon,\theta,\tau,0,0),\nonumber\\
\dot{\psi}_1(r^\ast,\dot{\varphi}^c,\dot{\theta}^c)&=&
\mp\boldsymbol{f}^{\dot{l}}_{\frac{1}{2},-\frac{1}{2}}
(\re r^\ast)\fM^{\dot{l}}_{\frac{1}{2},\dot{n}}
(\varphi,\epsilon,\theta,\tau,0,0),\nonumber\\
\dot{\psi}_2(r^\ast,\dot{\varphi}^c,\dot{\theta}^c)&=&
\boldsymbol{f}^{\dot{l}}_{\frac{1}{2},-\frac{1}{2}}
(\re r^\ast)\fM^l_{-\frac{1}{2},\dot{n}}(\varphi,\epsilon,\theta,\tau,0,0),
\nonumber
\end{eqnarray}
where
\begin{eqnarray}
&&l=\frac{1}{2},\;\frac{3}{2},\;\frac{5}{2},\ldots;\quad
n=-l,\;-l+1,\;\ldots,\; l;\nonumber\\
&&\dot{l}=\frac{1}{2},\;\frac{3}{2},\;\frac{5}{2},\ldots;\quad
\dot{n}=-\dot{l},\;-\dot{l}+1,\;\ldots,\; \dot{l},\nonumber
\end{eqnarray}
\[
\fM^l_{\pm\frac{1}{2},n}(\varphi,\epsilon,\theta,\tau,0,0)=
e^{\mp\frac{1}{2}(\epsilon+i\varphi)}Z^l_{\pm\frac{1}{2},n}(\theta,\tau),
\]
\begin{multline}
Z^l_{\pm\frac{1}{2},n}(\theta,\tau)=\cos^{2l}\frac{\theta}{2}
\ch^{2l}\frac{\tau}{2}\sum^l_{k=-l}i^{\pm\frac{1}{2}-k}
\tg^{\pm\frac{1}{2}-k}\frac{\theta}{2}\tnh^{n-k}\frac{\tau}{2}\times\\
\hypergeom{2}{1}{\pm\frac{1}{2}-l+1,1-l-k}{\pm\frac{1}{2}-k+1}
{i^2\tg^2\frac{\theta}{2}}
\hypergeom{2}{1}{n-l+1,1-l-k}{n-k+1}{\tnh^2\frac{\tau}{2}},\nonumber
\end{multline}
\[
\fM^{\dot{l}}_{\pm\frac{1}{2},\dot{n}}(\varphi,\epsilon,\theta,\tau,0,0)=
e^{\mp\frac{1}{2}(\epsilon-i\varphi)}
Z^{\dot{l}}_{\pm\frac{1}{2},\dot{n}}(\theta,\tau),
\]
\begin{multline}
Z^{\dot{l}}_{\pm\frac{1}{2},\dot{n}}(\theta,\tau)=
\cos^{2\dot{l}}\frac{\theta}{2}
\ch^{2\dot{l}}\frac{\tau}{2}
\sum^{\dot{l}}_{\dot{k}=-\dot{l}}i^{\pm\frac{1}{2}-\dot{k}}
\tg^{\pm\frac{1}{2}-\dot{k}}\frac{\theta}{2}
\tnh^{\dot{n}-\dot{k}}\frac{\tau}{2}\times\\
\hypergeom{2}{1}{\pm\frac{1}{2}-\dot{l}+1,1-\dot{l}-\dot{k}}
{\pm\frac{1}{2}-\dot{k}+1}
{i^2\tg^2\frac{\theta}{2}}
\hypergeom{2}{1}{\dot{n}-\dot{l}+1,1-\dot{l}-\dot{k}}
{\dot{n}-\dot{k}+1}{\tnh^2\frac{\tau}{2}}.\nonumber
\end{multline}

Therefore, in accordance with the factorization (\ref{WF}), an explicit
form of the particular solutions of the system (\ref{FET2})--(\ref{FEL2})
are given by expressions
\begin{eqnarray}
&&\psi^l_{1n}(\balpha)=\psi^+_1(x)\psi^l_{1n}(\fg)=
u_1(\bp)e^{-ipx}\boldsymbol{f}^l_{\frac{1}{2},\frac{1}{2}}
(\re r)\fM^l_{\frac{1}{2},n}(\varphi,\epsilon,\theta,\tau,0,0),\nonumber\\
&&\psi^l_{2n}(\balpha)=\psi^+_2(x)\psi^l_{2n}(\fg)=
\pm u_2(\bp)e^{-ipx}\boldsymbol{f}^l_{\frac{1}{2},\frac{1}{2}}
(\re r)\fM^l_{-\frac{1}{2},n}(\varphi,\epsilon,\theta,\tau,0,0),\nonumber\\
&&\dot{\psi}^{\dot{l}}_{1\dot{n}}(\balpha)=
\psi^-_1(x)\dot{\psi}^{\dot{l}}_{1\dot{n}}(\fg)=
\mp v_1(\bp)e^{ipx}\boldsymbol{f}^{\dot{l}}_{\frac{1}{2},-\frac{1}{2}}
(\re r^\ast)\fM^{\dot{l}}_{\frac{1}{2},\dot{n}}
(\varphi,\epsilon,\theta,\tau,0,0),\nonumber\\
&&\dot{\psi}^{\dot{l}}_{2\dot{n}}(\balpha)=
\psi^-_2(x)\dot{\psi}^{\dot{l}}_{2\dot{n}}(\fg)=
v_2(\bp)e^{ipx}\boldsymbol{f}^{\dot{l}}_{\frac{1}{2},-\frac{1}{2}}
(\re r^\ast)\fM^{\dot{l}}_{-\frac{1}{2},\dot{n}}
(\varphi,\epsilon,\theta,\tau,0,0).
\label{WF2}
\end{eqnarray}
The general solutions we obtain via an expansion in the particular solutions:
\begin{eqnarray}
\psi_1(\balpha)&=&\sum^{+\infty}_{p=-\infty}u_1(\bp)e^{ipx}
\sum^\infty_{l=\frac{1}{2}}\boldsymbol{f}^l_{\frac{1}{2},\frac{1}{2}}(\re r)
\sum^l_{n=-l}\alpha^{\frac{1}{2}}_{l,n}
\fM^l_{\frac{1}{2},n}(\varphi,\epsilon,\theta,\tau,0,0),\label{FSM81}\\
\psi_2(\balpha)&=&\pm\sum^{+\infty}_{p=-\infty}u_2(\bp)e^{ipx}
\sum^\infty_{l=\frac{1}{2}}\boldsymbol{f}^l_{\frac{1}{2},\frac{1}{2}}(\re r)
\sum^l_{n=-l}\alpha^{-\frac{1}{2}}_{l,n}
\fM^l_{-\frac{1}{2},n}(\varphi,\epsilon,\theta,\tau,0,0),\label{FSM82}\\
\dot{\psi}_1(\balpha)&=&\mp\sum^{+\infty}_{p=-\infty}v_1(\bp)e^{-ipx}
\sum^\infty_{\dot{l}=\frac{1}{2}}
\boldsymbol{f}^{\dot{l}}_{\frac{1}{2},-\frac{1}{2}}(\re r^\ast)
\sum^{\dot{l}}_{\dot{n}=-\dot{l}}\alpha^{\frac{1}{2}}_{\dot{l},\dot{n}}
\fM^{\dot{l}}_{\frac{1}{2},\dot{n}}
(\varphi,\epsilon,\theta,\tau,0,0),\label{FSM83}\\
\dot{\psi}_2(\balpha)&=&\sum^{+\infty}_{p=-\infty}v_2(\bp)e^{-ipx}
\sum^\infty_{\dot{l}=\frac{1}{2}}
\boldsymbol{f}^{\dot{l}}_{\frac{1}{2},-\frac{1}{2}}(\re r^\ast)
\sum^{\dot{l}}_{\dot{n}=-\dot{l}}\alpha^{-\frac{1}{2}}_{\dot{l},\dot{n}}
\fM^{\dot{l}}_{-\frac{1}{2},\dot{n}}
(\varphi,\epsilon,\theta,\tau,0,0),\label{FSM84}
\end{eqnarray}
where
\begin{eqnarray}
\alpha^{\pm\frac{1}{2}}_{l,n}&=&
\frac{(-1)^n(2l+1)(2\dot{l}+1)}{32\pi^4
\boldsymbol{f}^l_{\frac{1}{2},\frac{1}{2}}(\re a)}
\int\limits_{\dS^2}\int\limits_{T_4}
F_{\pm\frac{1}{2}}(\balpha)e^{-ipx}
\fM^l_{\pm\frac{1}{2},n}(\varphi,\epsilon,\theta,\tau,0,0)d^4xd^4\fg,\nonumber\\
\alpha^{\pm\frac{1}{2}}_{\dot{l},\dot{n}}&=&
\frac{(-1)^{\dot{n}}(2l+1)(2\dot{l}+1)}{32\pi^4
\boldsymbol{f}^{\dot{l}}_{\frac{1}{2},-\frac{1}{2}}(\re a^\ast)}
\int\limits_{\dS^2}\int\limits_{T_4}
\dot{F}_{\pm\frac{1}{2}}(\balpha)e^{ipx}
\fM^{\dot{l}}_{\pm\frac{1}{2},\dot{n}}
(\varphi,\epsilon,\theta,\tau,0,0)d^4xd^4\fg,\nonumber
\end{eqnarray}
These series give a solution of the boundary value problem for the Dirac
field.

\subsection{Quantization}
It is well known that the method of second quantization, 
introduced in the works
\cite{Dir27a,Dir27b,Foc32,Foc34,Jor27,JW28}, is a conceptual base of
quantum field theory. In connection with the problem of constructing
the quantum electrodynamics on the Poincar\'{e} group, it needs to transfer
the formalism of second quantization onto the group manifold $\cM_{10}$
($\cM_8$ or $\cM_6$). We consider briefly this question in case of $\cM_{10}$.

Let $\cH$ be a complex Hilbert space realized with the aid of functions
$f$ with summable square on the set $M$ endowed with a measure.
Let $f\rightarrow f^+$ be an involution on the space
$\cH$, that is, the mapping satisfying the following conditions:\\ 
1) $(f^+)^\ast=f$,\\
2) $(f_1+f_2)^+=f^+_1+f^+_2$,\\
3) $(\lambda f)^+=\overline{\lambda}f^+$,\\
4) $(f_1,f_2)^+=(f^+_2,f^+_1)$.\\
Let $M$ be a parameter set of the Poincar\'{e} group. When a physical
system consists of $n$ particles and the states of $k$-th particle are
described by a space $\cH_k$ of the functions with summable square on the
set $M_k$, then the states of the system are described by the functions
with summable square which depend on $n$ variables
$\balpha_1,\ldots,\balpha_n$, where $\balpha_k\in M_k$.
Let us denote this space as $\fH^n$. When the system consists of $n$
identical particles, the sets $M_k$ coincide with each other, and the
space $\fH^n$ in this case is called
{\it a configuration space}.

Let $\mid\Psi\rangle$ and $\mid\Phi\rangle$ be state vectors describing
the system of $n$ identical particles. The scalar product of these vectors
in the configuration space has a following form
\begin{multline}
(\Psi,\Phi)=\int d\balpha_1\int d\balpha_2\cdots\int d\balpha_n
\langle\Psi\mid\balpha_1,\ldots,\balpha_n\rangle
\langle\balpha_1,\ldots,\balpha_n\mid\Phi\rangle=\\
=\int d\balpha_1\int d\balpha_2\cdots\int d\balpha_n
\overline{\Psi}(\balpha_1,\ldots,\balpha_n)
\Phi(\balpha_1,\ldots,\balpha_n),\label{FS0}
\end{multline}
where $d\balpha_i$ is a Haar measure on the Poincar\'{e} group
of the form  (\ref{HMP}).

It is well known that in dependence on the kind of particles the system
of $n$ identical particles is described either by a subspace
$\cH^n_F\subset\cH^n$ of antisymmetric functions or by a subspace
$\cH^n_B\subset\cH^n$ of symmetric functions. In the first case the
particles are called {\it fermions}, in the second case we have 
{\it bosons}.

The states of the system, consisting of a variable number of particles,
are described by vectors of a space $\fH$. The space $\fH$ is a direct
sum of all $\fH^n$ and a one-dimensional space $\fH^0$ which corresponds to
the absence of particles (vacuum). The system, consisting of a variable
number of fermions, is described by a subspace
$\cH_F\subset\fH$, where
$\cH_F=\sum\limits^\infty_{n=0}\oplus\cH^n_F$, $\cH^0_F=\fH^0$. In turn,
the system, consisting of a variable number of bosons, is described by
a subspace $\cH_B\subset\fH$, where
$\cH_B=\sum\limits^\infty_{n=0}\cH^n_B$, $\cH^0_B=\fH^0$. The elements of
the subspaces $\cH_F$ and $\cH_B$ describe the states of real physical
systems. By this reason,
$\cH_F$ and $\cH_B$ are called {\it state spaces}
(or {\it Fock spaces}).

The vectors $\mid\Psi\rangle$ of $\fH$ can be written in the following
form:
\begin{equation}\label{FS1}
\mid\Psi\rangle=\left[\begin{array}{l}
\langle 0\mid\Psi\rangle\\
\langle\balpha_1\mid\Psi\rangle\\
\langle\balpha_1,\balpha_2\mid\Psi\rangle\\
\phantom{\langle\balpha_1,}\vdots\\
\langle\balpha_1,\balpha_2,\ldots,\balpha_n;n\mid\Psi\rangle\\
\phantom{\langle\balpha_1,}\vdots
\end{array}\right]=\left[\begin{array}{l}
\Psi^0\\
\Psi^1(\balpha_1)\\
\Psi^2(\balpha_1,\balpha_2)\\
\phantom{\langle\balpha_1,}\vdots\\
\Psi^n(\balpha_1,\balpha_2,\ldots,\balpha_n)\\
\phantom{\langle\balpha_1,}\vdots
\end{array}\right]
\end{equation}
and
\begin{equation}\label{FS2}
(\Psi,\Psi)=\left|\Psi^0\right|^2+\sum^\infty_{n=1}\int\left|
\Psi^n(\balpha_1,\balpha_2,\ldots,\balpha_n)\right|^2d^n\balpha,
\end{equation}
where $d^n\balpha=d\balpha_1d\balpha_2\cdots d\balpha_n$. 
It is obvious that the vectors, which have only $n$-th nonzero components,
form the subspace $\fH^n\subset\fH$. In turn, the vectors, belonging to
the spaces $\cH_F$ and $\cH_B$, and also their scalar products in these
spaces, are defined by the formulae (\ref{FS1}) and (\ref{FS2}), but
in the first case all the functions
$\Psi^n(\balpha_1,\balpha_2,\ldots,\balpha_n)$ are antisymmetric, and
in the second case they are symmetric.

The action of annihilation and creation operators on the $n$-particle
states are defined by the standard formulae
\begin{eqnarray}
\sa_i\mid n_1,n_2,\ldots,n_i,\ldots\rangle&=&
\sqrt{n_i}\mid n_1,n_2,\ldots,n_i-1,\ldots\rangle,\nonumber\\
\sa^+_i\mid n_1,n_2,\ldots,n_i,\ldots\rangle&=&
\sqrt{n_i+1}\mid n_1,n_2,\ldots,n_i+1,\ldots\rangle\nonumber
\end{eqnarray}
in the case of Bose-Einstein statistics, and
\begin{eqnarray}
\sa_i\mid n_1,n_2,\ldots,n_i,\ldots\rangle&=&
(-1)^{s_i}n_i\mid n_1,n_2,\ldots,n_i-1,\ldots\rangle,\nonumber\\
\sa^+_i\mid n_1,n_2,\ldots,n_i,\ldots\rangle&=&
(-1)^{s_i}(1-n_i)\mid n_1,n_2,\ldots,n_i+1,\ldots\rangle\nonumber
\end{eqnarray}
in the case of Fermi-Dirac statistics.

In like manner, the formalism of second quantization can be transferred
onto the homogeneous spaces $\cM_8$, $\cM_7$ and $\cM_6$, and also
onto other homogeneous spaces of the Poincar\'{e} group contained
in the Finkelstein-Bacry-Kihlberg list \cite{Fin55,BK69} and endowed
with a measure.

Coming back to the electron-positron field, we see that
the following logical step consists in definition of the field operators.
As is known, in the standard quantum field theory the field operators are
defined in the form of Fourier integrals (or Fourier series) on functions,
which, in general, are solutions of some relativistic wave equation in the
plane wave approximation.

In our case the electron-positron field can be represented in the form
of following superpositions (field operators):
\begin{eqnarray}
\boldsymbol{\psi}(\balpha)&=&\sum^2_{s=1}\sa_s\psi_s(\balpha)+
\sum^2_{s=1}\sfb^+_s\dot{\psi}_s(\balpha),\nonumber\\
\overline{\boldsymbol{\psi}}(\balpha)&=&\sum^2_{s=1}\sa^+_s
\overline{\psi}_s(\balpha)+
\sum^2_{s=1}\sfb_s\overline{\dot{\psi}}_s(\balpha),\label{FOE}
\end{eqnarray}\begin{sloppypar}\noindent
where $\sa^+_s$ and $\sa_s$ are creation and annihilation operators of the
electron in a state $s$, $\sfb^+_s$ and $\sfb_s$ are creation and
annihilation operators of the positron in a state $s$, $\psi_s(\balpha)$
and $\dot{\psi}_s(\balpha)$ are Fourier series (\ref{FSM81})--(\ref{FSM82})
and (\ref{FSM83})--(\ref{FSM84}) which form a bispinor
$(\psi_1(\balpha),\psi_2(\balpha),\dot{\psi}_1(\balpha),
\dot{\psi}_2(\balpha))^T$ on the homogeneous space $\cM_8$.
The operators $\sa^+_s$, $\sa_s$, $\sfb^+_s$, $\sfb_s$ satisfy the following
anticommutation relations:\end{sloppypar}
\begin{equation}\label{AntiCom}
{\renewcommand{\arraystretch}{1.25}
\begin{array}{ccc}
\ld\sa_s,\sa^+_{s^\prime}\rd_+&=&\delta_{ss^\prime},\\
\ld\sa_s,\sa_{s^\prime}\rd_+&=&0,\\
\ld\sa^+_s,\sa^+_{s^\prime}\rd_+&=&0,
\end{array}\quad
\begin{array}{ccc}
\ld\sfb_s,\sfb^+_{s^\prime}\rd_+&=&\delta_{ss^\prime},\\
\ld\sfb_s,\sfb_{s^\prime}\rd_+&=&0,\\
\ld\sfb^+_s,\sfb^+_{s^\prime}\rd_+&=&0,
\end{array}}
\end{equation}
\[
\ld\sa_s,\sfb_{s^\prime}\rd_+=\ld\sa_s,\sfb^+_{s^\prime}\rd_+=
\ld\sa^+_s,\sfb_{s^\prime}\rd_+=\ld\sa^+_s,\sfb^+_{s^\prime}\rd_+=0.
\]

Using the relations (\ref{AntiCom}), we can calculate anticommutators of
the electron-positron field:
\begin{eqnarray}
\ld\boldsymbol{\psi}_\alpha(\balpha),
\boldsymbol{\psi}_\beta(\balpha^\prime)\rd_+&=&0,\nonumber\\
\ld\overline{\boldsymbol{\psi}}_\alpha(\balpha),
\overline{\boldsymbol{\psi}}_\beta(\balpha^\prime)\rd_+&=&0,\nonumber\\
\ld\boldsymbol{\psi}_\alpha(\balpha),
\overline{\boldsymbol{\psi}}_\beta(\balpha^\prime)\rd_+&=&
S_{\alpha\beta}(\balpha,\balpha^\prime),\label{AntiCom2}
\end{eqnarray}
where
\begin{eqnarray}
S_{\alpha\beta}(\balpha,\balpha^\prime)&=&
S^+_{\alpha\beta}(\balpha,\balpha^\prime)+
S^-_{\alpha\beta}(\balpha,\balpha^\prime),\nonumber\\
S^+_{\alpha\beta}(\balpha,\balpha^\prime)&=&\sum^2_{s=1}
\psi_{s\alpha}(\balpha)\overline{\psi}_{s\beta}(\balpha^\prime),\nonumber\\
S^-_{\alpha\beta}(\balpha,\balpha^\prime)&=&
\sum^2_{s=1}\dot{\psi}_{s\alpha}(\balpha)
\overline{\dot{\psi}}_{s\beta}(\balpha^\prime).\nonumber
\end{eqnarray}
In the absence of an external electromagnetic field the functions
$S^+_{\alpha\beta}(\balpha,\balpha^\prime)$ and
$S^-_{\alpha\beta}(\balpha,\balpha^\prime)$ can be written as follows
\begin{multline}
S^+_{\alpha\beta}(\balpha-\balpha^\prime)=\sum^2_{s=1}\sum_p
u_{\alpha s}(\bp)\overline{u}_{\beta s}(\bp)e^{ip(x-x^\prime)}\times\\
\sum^\infty_{l=\frac{1}{2}}
%\left[\sum^l_{k=\frac{1}{2}}
\boldsymbol{f}^l_{\frac{1}{2},\frac{1}{2}}(\re r)
\overline{\boldsymbol{f}^{l}_{\frac{1}{2},\frac{1}{2}}
(\re r)}\times\\
\sum^l_{n=-l}
%\alpha^{\frac{(-1)^{s-1}}{2}}_{l,n}
\fM^l_{\frac{(-1)^{s-1}}{2},n}(\varphi,\epsilon,\theta,\tau,0,0)
%\sum^{l-k+\frac{1}{2}}_{n=-(l-k+\frac{1}{2})}
%\overline{\alpha^{\frac{(-1)^{s-1}}{2}}_{l,n}}
\overline{\fM^l_{\frac{(-1)^{s-1}}{2},n}
(\varphi^\prime,\epsilon^\prime,\theta^\prime,\tau^\prime,0,0)},
\nonumber
\end{multline}
\begin{multline}
S^-_{\alpha\beta}(\balpha-\balpha^\prime)=\sum^2_{s=1}\sum_p
v_{\alpha s}(\bp)\overline{v}_{\beta s}(\bp)e^{-ip(x-x^\prime)}\times\\
\sum^\infty_{\dot{l}=\frac{1}{2}}
%\left[\sum^{\dot{l}}_{\dot{k}=\frac{1}{2}}
\boldsymbol{f}^{\dot{l}}_{\frac{1}{2},-\frac{1}{2}}(\re r^\ast)
\overline{
\boldsymbol{f}^{\dot{l}}_{\frac{1}{2},-\frac{1}{2}}
(\re r^\ast)}\times\\
\sum^{\dot{l}}_{\dot{n}=-\dot{l}}
%\alpha^{\frac{(-1)^{s-1}}{2}}_{\dot{l},\dot{n}}
\fM^{\dot{l}}_{\frac{(-1)^{s-1}}{2},\dot{n}}(\varphi,\epsilon,\theta,\tau,0,0)
%\sum^{\dot{l}-\dot{k}+\frac{1}{2}}_{\dot{n}=-(\dot{l}-\dot{k}+\frac{1}{2})}
%\overline{\alpha^{\frac{(-1)^{s-1}}{2}}_{\dot{l},\dot{n}}}
\overline{\fM^{\dot{l}}_{\frac{(-1)^{s-1}}{2},\dot{n}}
(\varphi^\prime,\epsilon^\prime,\theta^\prime,\tau^\prime,0,0)},
\nonumber
\end{multline}\begin{sloppypar}\noindent
Taking into account that $\fM^l_{mn}(\varphi,\epsilon,\theta,\tau,0,0)=
e^{-m\varphi^c}Z^l_{mn}(\cos\theta^c)$, we obtain
$\overline{\fM^l_{mn}(\varphi,\epsilon,\theta,\tau,0,0)}=
e^{im\varphi^c}\overline{Z^l_{mn}(\cos\theta^c)}=
(-1)^{n-m}e^{im\varphi^c}Z^l_{nm}(\cos\theta^c)$. Using the addition
theorem for hyperspherical functions (see \cite{Var042}), we obtain
\end{sloppypar}
\begin{multline}
S^+_{\alpha\beta}(\balpha-\balpha^\prime)=\sum^2_{s=1}\sum_p
u_{\alpha s}(\bp)\overline{u}_{\beta s}(\bp)e^{ip(x-x^\prime)}\times\\
e^{-\frac{i(-1)^{s-1}}{2}(\varphi^-\varphi^c{}^\prime)}
\sum^\infty_{l=\frac{1}{2}}
%\left[\sum^l_{k=\frac{1}{2}}
\boldsymbol{f}^l_{\frac{1}{2},\frac{1}{2}}(\re r)
\overline{\boldsymbol{f}^{l}_{\frac{1}{2},\frac{1}{2}}
(\re r)}
Z^l_{\frac{(-1)^{s-1}}{2},\frac{(-1)^{s-1}}{2}}
(\cos\theta^c{}^{\prime\prime}),\nonumber
\end{multline}
\begin{multline}
S^-_{\alpha\beta}(\balpha-\balpha^\prime)=\sum^2_{s=1}\sum_p
v_{\alpha s}(\bp)\overline{v}_{\beta s}(\bp)e^{-ip(x-x^\prime)}\times\\
e^{\frac{i(-1)^{s-1}}{2}(\varphi^c-\varphi^c{}^\prime)}
\sum^\infty_{\dot{l}=\frac{1}{2}}
%\left[\sum^{\dot{l}}_{\dot{k}=\frac{1}{2}}
\boldsymbol{f}^{\dot{l}}_{\frac{1}{2},-\frac{1}{2}}(\re r^\ast)
\overline{
\boldsymbol{f}^{\dot{l}}_{\frac{1}{2},-\frac{1}{2}}
(\re r^\ast)}
Z^{\dot{l}}_{\frac{(-1)^{s-1}}{2},\frac{(-1)^{s-1}}{2}}
(\cos\dot{\theta}^c{}^{\prime\prime}),\nonumber
\end{multline}
where $\cos\theta^c{}^{\prime\prime}=\cos\theta^c\cos\theta^c{}^\prime-
\sin\theta^c\sin\theta^c{}^\prime\cos\varphi^c{}^\prime$.

Let us consider now normal products of the operators (\ref{FOE}).
Supposing
\[
\boldsymbol{\psi}(\balpha)=\psi^{(+)}(\balpha)+\psi^{(-)}(\balpha),\quad
\overline{\boldsymbol{\psi}}(\balpha)=\overline{\psi}^{(+)}(\balpha)+
\overline{\psi}^{(-)}(\balpha),
\]
where $\psi^{(+)}(\balpha)$ are annihilation operators of electrons and
$\psi^{(-)}(\balpha)$ are creation operators of positrons, and
correspondingly $\overline{\psi}^{(+)}(\balpha)$ are creation operators
of electrons and $\overline{\psi}^{(-)}(\balpha)$ are annihilation
operators of positrons, we see that
\begin{eqnarray}
N(\psi^{(+)}(\balpha)\overline{\psi}^{(+)}(\balpha^\prime))&=&
-\overline{\psi}^{(+)}(\balpha^\prime)\psi^{(+)}(\balpha),\nonumber\\
N(\psi^{(-)}(\balpha)\overline{\psi}^{(-)}(\balpha^\prime))&=&
\psi^{(-)}(\balpha)\overline{\psi}^{(-)}(\balpha^\prime),\nonumber\\
N(\psi^{(+)}(\balpha)\overline{\psi}^{(-)}(\balpha^\prime))&=&
\psi^{(+)}(\balpha)\overline{\psi}^{(-)}(\balpha^\prime),\nonumber\\
N(\psi^{(-)}(\balpha)\overline{\psi}^{(+)}(\balpha^\prime))&=&
\psi^{(-)}(\balpha)\overline{\psi}^{(+)}(\balpha^\prime).\nonumber
\end{eqnarray}
Taking into account the latter $N$-products, we have
\begin{multline}
\boldsymbol{\psi}(\balpha)\overline{\boldsymbol{\psi}}(\balpha^\prime)=
(\psi^{(+)}(\balpha)+\psi^{(-)}(\balpha))
(\overline{\psi}^{(+)}(\balpha^\prime)+
\overline{\psi}^{(-)}(\balpha^\prime))=\\
=\psi^{(+)}(\balpha)\overline{\psi}^{(+)}(\balpha^\prime)+
\psi^{(-)}(\balpha)\overline{\psi}^{(+)}(\balpha^\prime)+
\psi^{(+)}(\balpha)\overline{\psi}^{(-)}(\balpha^\prime)+\\
+\psi^{(-)}(\balpha)\overline{\psi}^{(-)}(\balpha^\prime).\nonumber
\end{multline}
In accordance with (\ref{AntiCom2}), 
$\psi^{(+)}(\balpha)\overline{\psi}^{(+)}(\balpha^\prime)=
-\overline{\psi}^{(+)}(\balpha^\prime)\psi^{(+)}(\balpha)+
S^+(\balpha,\balpha^\prime)$, therefore,
\begin{eqnarray}
\boldsymbol{\psi}(\balpha)\overline{\boldsymbol{\psi}}(\balpha^\prime)&=&
N(\boldsymbol{\psi}(\balpha)\overline{\boldsymbol{\psi}}(\balpha^\prime))+
S^+(\balpha-\balpha^\prime)\nonumber\\
&=&N(\boldsymbol{\psi}(\balpha)\overline{\boldsymbol{\psi}}(\balpha^\prime))+
\coupone,\label{NP1}
\end{eqnarray}
where $\coupone=S^+(\balpha-\balpha^\prime)$ is an operator coupling.
Analogously,
\begin{eqnarray}
\overline{\boldsymbol{\psi}}(\balpha)\boldsymbol{\psi}(\balpha^\prime)&=&
N(\overline{\boldsymbol{\psi}}(\balpha)\boldsymbol{\psi}(\balpha^\prime))+
S^-(\balpha-\balpha^\prime)\nonumber\\
&=&N(\overline{\boldsymbol{\psi}}(\balpha)\boldsymbol{\psi}(\balpha^\prime))+
\couptwo,\nonumber
\end{eqnarray}
\[
\overline{\boldsymbol{\psi}}(\balpha)\overline{\boldsymbol{\psi}}
(\balpha^\prime)=N(\overline{\boldsymbol{\psi}}(\balpha)
\overline{\boldsymbol{\psi}}(\balpha^\prime)),\quad
\boldsymbol{\psi}(\balpha)\boldsymbol{\psi}(\balpha^\prime)=
N(\boldsymbol{\psi}(\balpha)\boldsymbol{\psi}(\balpha^\prime)).
\]
Hence it follows that couplings of distinct $N$-products of the operators
$\boldsymbol{\psi}$ and $\overline{\boldsymbol{\psi}}$ are
\begin{eqnarray}
\coupone&=&S^+(\balpha-\balpha^\prime);\nonumber\\
\couptwo&=&S^-(\balpha-\balpha^\prime);\nonumber
\end{eqnarray}
\[
\coupthree=0;\quad\coupfour=0.
\]
Thus, the vacuum expectation values of the operator products are defined as
\begin{eqnarray}
\langle 0\mid\boldsymbol{\psi}(\balpha)\overline{\boldsymbol{\psi}}
(\balpha^\prime)\mid 0\rangle&=&\coupone=S^+(\balpha-\balpha^\prime),\nonumber\\
\langle 0\mid\overline{\boldsymbol{\psi}}(\balpha)\boldsymbol{\psi}
(\balpha^\prime)\mid 0\rangle&=&\couptwo=S^-(\balpha-\balpha^\prime),\nonumber\\
\langle 0\mid\boldsymbol{\psi}(\balpha)\boldsymbol{\psi}(\balpha^\prime)
\mid 0\rangle&=&0,\nonumber\\
\langle 0\mid\overline{\boldsymbol{\psi}}(\balpha)
\overline{\boldsymbol{\psi}}(\balpha^\prime)\mid 0\rangle&=&0.\nonumber
\end{eqnarray}

Further, we can define time ordered products of the field operators
(since $t$ is a parameter of the Poincar\'{e} group):
\begin{equation}\label{TP1}
T(\boldsymbol{\psi}(\balpha)\overline{\boldsymbol{\psi}}(\balpha^\prime))=
\begin{cases}
\boldsymbol{\psi}(\balpha)\overline{\boldsymbol{\psi}}(\balpha^\prime),&
t>t^\prime;\\
-\overline{\boldsymbol{\psi}}(\balpha^\prime)\boldsymbol{\psi}(\balpha),&
t^\prime>t.
\end{cases}
\end{equation}
Using (\ref{NP1}), we can express (\ref{TP1}) via the $N$-products:
\begin{equation}\label{TP2}
T(\boldsymbol{\psi}(\balpha)\overline{\boldsymbol{\psi}}(\balpha^\prime))=
\begin{cases}
N(\boldsymbol{\psi}(\balpha)\overline{\boldsymbol{\psi}}(\balpha^\prime))+
\coupone,& t>t^\prime;\\
-N(\overline{\boldsymbol{\psi}}(\balpha^\prime)\boldsymbol{\psi}(\balpha))-
\couptwo,& t^\prime>t.
\end{cases}
\end{equation}
Since $-N(\overline{\boldsymbol{\psi}}(\balpha^\prime)\boldsymbol{\psi}
(\balpha))=N(\boldsymbol{\psi}(\balpha)\overline{\boldsymbol{\psi}}
(\balpha^\prime))$, then (\ref{TP2}) can be rewritten as
\[
T(\boldsymbol{\psi}(\balpha)\overline{\boldsymbol{\psi}}(\balpha^\prime))=
N(\boldsymbol{\psi}(\balpha)\overline{\boldsymbol{\psi}}(\balpha^\prime))+
\coupfife,
\]
where $\coupfife$ is a time ordered coupling of the field operators:
\[
\coupfife=\begin{cases}
\coupone=S^+(\balpha-\balpha^\prime),& t>t^\prime;\\
-\coupnine=-S^-(\balpha-\balpha^\prime),& t^\prime>t.
\end{cases}
\]
In like manner we find for other $T$-products the following expressions
\begin{eqnarray}
T(\overline{\boldsymbol{\psi}}(\balpha)\boldsymbol{\psi}(\balpha^\prime))&=&
-T(\boldsymbol{\psi}(\balpha^\prime)\overline{\boldsymbol{\psi}}(\balpha))
\nonumber\\
&=&-N(\boldsymbol{\psi}(\balpha^\prime)\overline{\boldsymbol{\psi}}(\balpha))-
\coupsix,\nonumber
\end{eqnarray}
\[
T(\overline{\boldsymbol{\psi}}(\balpha)\overline{\boldsymbol{\psi}}
(\balpha^\prime))=N(\overline{\boldsymbol{\psi}}(\balpha)
\overline{\boldsymbol{\psi}}(\balpha^\prime)),\quad
T(\boldsymbol{\psi}(\balpha)\boldsymbol{\psi}(\balpha^\prime))=
N(\boldsymbol{\psi}(\balpha)\boldsymbol{\psi}(\balpha^\prime)),
\]
whence it follows that
\[
\coupfife=-\coupten,\quad\coupseven=0,\quad\coupeight=0.
\]
Therefore, vacuum expectation values of the time ordered products are
defined by expressions
\begin{eqnarray}
\langle 0\mid T(\boldsymbol{\psi}(\balpha)\overline{\boldsymbol{\psi}}
(\balpha^\prime))\mid 0\rangle&=&\coupfife,\nonumber\\
\langle 0\mid T(\overline{\boldsymbol{\psi}}(\balpha)\boldsymbol{\psi}
(\balpha^\prime))\mid 0\rangle&=&-\coupsix,\nonumber\\
\langle 0\mid T(\boldsymbol{\psi}(\balpha)\boldsymbol{\psi}(\balpha^\prime))
\mid 0\rangle&=&0,\nonumber\\
\langle 0\mid T(\overline{\boldsymbol{\psi}}(\balpha)
\overline{\boldsymbol{\psi}}(\balpha^\prime))\mid 0\rangle&=&0.\nonumber
\end{eqnarray}

\section{The Maxwell field}
\label{Sec:Max}
In this section we will set up a boundary value problem for the Maxwell
field $(1,0)\oplus(0,1)$ (photon field) defined on the homogeneous
space $\cM_8$. At this point, electromagnetic field should be defined
in the Riemann-Silberstein representation \cite{Web01,Sil07,Bir96}.
The Riemann-Silberstein (Majorana-Oppenheimer) representation considered
during long time by many authors 
\cite{Maj,Opp31,Goo57,Mos59,SS62,MRB74,DaS79,Gia85}.
The interest to this formulation of electrodynamics has grown in recent
years \cite{Ina94,Sip95,Ger98,Esp98}. One of the main
advantages of this approach lies in the fact that Dirac and Maxwell fields
are derived similarly from the Dirac-like Lagrangians\footnote{Moreover,
in contrast to the Gupta-Bleuler method, where the non-observable
four-potential $A_\mu$ is quantized, the Majorana-Oppenheimer 
electrodynamics deals directly with observable quantities, such as the
electric and magnetic fields. It allows one to avoid non-physical degrees
of freedom, indefinite metrics and other difficulties connected with
the Gupta-Bleuler method.}. These fields have the analogous mathematical
structure, namely, they are the functions on the Poincar\'{e} group.
This circumstance allows us to consider the fields
$(1/2,0)\oplus(0,1/2)$ and $(1,0)\oplus(0,1)$ on an equal footing, from
the one group theoretical viewpoint\footnote{In this connection it is
interesting to note that in the gauge theories electromagnetic field
is understood as a `gauge field' that leads to a peculiar opposition with
other physical fields called by this reason as `matter fields'.}.

We start with the Lagrangian (\ref{Lagrange}) on the group manifold $\cM_{10}$.
Let us rewrite (\ref{Lagrange}) in the form
\begin{multline}\label{LagMax}
\cL(\balpha)=-\frac{1}{2}\left(\overline{\boldsymbol{\phi}}(\balpha)\Gamma_\mu
\frac{\partial\boldsymbol{\phi}(\balpha)}{\partial x_\mu}-
\frac{\partial\overline{\boldsymbol{\phi}}(\balpha)}{\partial x_\mu}
\Gamma_\mu\boldsymbol{\phi}(\balpha)\right)-\\
-\frac{1}{2}\left(\overline{\boldsymbol{\phi}}(\balpha)\Upsilon_\nu
\frac{\partial\boldsymbol{\phi}(\balpha)}{\partial\fg_\nu}-
\frac{\partial\overline{\boldsymbol{\phi}}(\balpha)}{\partial\fg_\nu}
\Upsilon_\nu\boldsymbol{\phi}(\balpha)\right),
\end{multline}
where $\boldsymbol{\phi}(\balpha)=\phi(x)\phi(\fg)$ 
($\mu=0,1,2,3,\;\nu=1,\ldots,6$),
and
\begin{equation}\label{Gamma2}
\Gamma_0=\begin{pmatrix}
0 & I\\
I & 0
\end{pmatrix},\;\;\Gamma_1=\begin{pmatrix}
0 & -\alpha_1\\
\alpha_1 & 0
\end{pmatrix},\;\;\Gamma_2=\begin{pmatrix}
0 & -\alpha_2\\
\alpha_2 & 0
\end{pmatrix},\;\;\Gamma_3=\begin{pmatrix}
0 & -\alpha_3\\
\alpha_3 & 0
\end{pmatrix},
\end{equation}
\begin{equation}\label{Upsilon3}
\Upsilon_1=\begin{pmatrix}
0 & \Lambda^\ast_1\\
\Lambda_1 & 0
\end{pmatrix},\quad\Upsilon_2=\begin{pmatrix}
0 & \Lambda^\ast_2\\
\Lambda_2 & 0
\end{pmatrix},\quad\Upsilon_3=\begin{pmatrix}
0 & \Lambda^\ast_3\\
\Lambda_3 & 0
\end{pmatrix},
\end{equation}
\begin{equation}\label{Upsilon4}
\Upsilon_4=\begin{pmatrix}
0 & i\Lambda^\ast_1\\
i\Lambda_1 & 0
\end{pmatrix},\quad\Upsilon_5=\begin{pmatrix}
0 & i\Lambda^\ast_2\\
i\Lambda_2 & 0
\end{pmatrix},\quad\Upsilon_6=\begin{pmatrix}
0 & i\Lambda^\ast_3\\
i\Lambda_3 & 0
\end{pmatrix},
\end{equation}
where
\begin{equation}\label{Alpha}
\alpha_1=\begin{pmatrix}
0 & 0 & 0\\
0 & 0 & i\\
0 & -i& 0
\end{pmatrix},\quad\alpha_2=\begin{pmatrix}
0 & 0 & -i\\
0 & 0 & 0\\
i & 0 & 0
\end{pmatrix},\quad\alpha_3=\begin{pmatrix}
0 & i & 0\\
-i& 0 & 0\\
0 & 0 & 0
\end{pmatrix},
\end{equation}
and the matrices $\Lambda_j$ and $\Lambda^\ast_j$ are derived from
(\ref{L1})--(\ref{L3}) and (\ref{L1'})--(\ref{L3'}) at $l=1$:
\begin{equation}\label{Lambda1}
\Lambda_1=\frac{c_{11}}{\sqrt{2}}\begin{pmatrix}
0 & 1 & 0\\
1 & 0 & 0\\
0 & 1 & 0
\end{pmatrix},\quad\Lambda_2=\frac{c_{11}}{\sqrt{2}}\begin{pmatrix}
0 & -i & 0\\
i & 0 & -i\\
0 & i & 0
\end{pmatrix},\quad\Lambda_3=c_{11}\begin{pmatrix}
1 & 0 & 0\\
0 & 0 & 0\\
0 & 0 &-1
\end{pmatrix}.
\end{equation}
\begin{equation}\label{Lambda2}
\sL^\ast_1=\frac{\sqrt{2}}{2}\dot{c}_{11}\begin{pmatrix}
0 & 1 & 0\\
1 & 0 & 1\\
0 & 1 & 0
\end{pmatrix},\quad
\sL^\ast_2=\frac{\sqrt{2}}{2}\dot{c}_{11}\begin{pmatrix}
0 & -i & 0\\
i & 0  & -i\\
0 & i & 0
\end{pmatrix},\quad
\sL^\ast_3=\dot{c}_{11}\begin{pmatrix}
1 & 0 & 0\\
0 & 0 & 0\\
0 & 0 & -1
\end{pmatrix}.
\end{equation}

Varying independently $\phi(x)$ and $\overline{\phi}(x)$ in the Lagrangian
(\ref{LagMax}), and then $\phi(\fg)$ and $\overline{\phi}(\fg)$,
we come to the following equations:
\begin{eqnarray}
\Gamma_\mu\frac{\partial\phi(x)}{\partial x_\mu}&=&0,\label{Real}\\
\Gamma^T_\mu\frac{\partial\overline{\phi}(x)}{\partial x_\mu}&=&0.\label{Anti}\\
\Upsilon_\nu\frac{\partial\phi(\fg)}{\partial\fg_\nu}&=&0,\label{FEL3}\\
\Upsilon^T_\nu\frac{\partial\overline{\phi}(\fg)}{\partial\fg_\nu}&=&0.
\label{FEL4}
\end{eqnarray}

Let us formulate the boundary value problem for the field $(1,0)\oplus(0,1)$.
{\it Let $T$ be an unbounded region in $\cM_8=\R^{1,3}\times\dS^2$ and
let $\Sigma$ ($\dot{\Sigma}$) be a surface of the complex (dual) two-sphere,
then it needs to find the functions
$\boldsymbol{\phi}(\balpha)=(\phi_1(\balpha),\phi_2(\balpha),\phi_3(\balpha),
\dot{\phi}_1(\balpha),\dot{\phi}_2(\balpha),\dot{\phi}_3(\balpha))^T$,
such that\\
1) $\boldsymbol{\phi}(\balpha)$ satisfies the equations 
(\ref{Real})--(\ref{Anti}) and (\ref{FEL3})--(\ref{FEL4}) in the all
region $T$.\\
2) $\boldsymbol{\phi}(\balpha)$ is a continuous function everywhere in $T$.\\
3) $\left.\phantom{\frac{x}{x}}\phi_m(\balpha)\right|_{\Sigma}=F_m(\balpha)$,
$\left.\phantom{\frac{x}{x}}\dot{\phi}_m(\balpha)\right|_{\dot{\Sigma}}=
\dot{F}_m(\balpha)$, where $F_m(\balpha)$ and $\dot{F}_m(\balpha)$ are
square integrable and infinitely differentiable functions 
on $\cM_8$, $m=1,2,3$.}

First of all, the equation (\ref{Real}) can be written as follows:
\begin{equation}\label{ME}
\left[\frac{i\hbar}{c}\frac{\partial}{\partial t}\begin{pmatrix}
0 & I\\
I & 0
\end{pmatrix}-i\hbar\frac{\partial}{\partial\bx}\begin{pmatrix}
0 & -\alpha_i\\
\alpha_i & 0
\end{pmatrix}\right]\begin{pmatrix}
\phi(x)\\
\dot{\phi}(x)
\end{pmatrix}=0,
\end{equation}
where
\[
\begin{pmatrix}
\phi(x)\\
\dot{\phi}(x)
\end{pmatrix}=\begin{pmatrix}
\bE-i\bB\\
\bE+i\bB
\end{pmatrix}=\begin{pmatrix}
E_1-iB_1\\
E_2-iB_2\\
E_3-iE_3\\
E_1+iB_1\\
E_2+iB_2\\
E_3+iB_3
\end{pmatrix}.
\]
From the equation (\ref{ME}) it follows that
\begin{eqnarray}
&&\left(\frac{i\hbar}{c}\frac{\partial}{\partial t}-
i\hbar\alpha_i\frac{\partial}{\partial\bx}\right)\phi(x)=0,\label{ME1}\\
&&\left(\frac{i\hbar}{c}\frac{\partial}{\partial t}+
i\hbar\alpha_i\frac{\partial}{\partial\bx}\right)\dot{\phi}(x)=0.\label{ME2}
\end{eqnarray}
The latter equations with allowance for transversality conditions
($\bp\cdot\phi=0$, $\bp\cdot\dot{\phi}=0$) coincide with the Maxwell equations.
Indeed, taking into account that 
$(\bp\cdot\alpha)\phi=\hbar\nabla\times\phi$, we obtain
\begin{eqnarray}
\frac{i\hbar}{c}\frac{\partial\phi}{\partial t}&=&-\hbar\nabla\times\phi,
\label{Tr1}\\
-i\hbar\nabla\cdot\phi&=&0.\label{Tr1'}
\end{eqnarray}
Whence
\begin{eqnarray}
\nabla\times(\bE-i\bB)&=&-\frac{i}{c}\frac{\partial(\bE-i\bB)}{\partial t},
\nonumber\\
\nabla\cdot(\bE-i\bB)&=&0\nonumber
\end{eqnarray}
(the constant $\hbar$ is cancelled). Separating the real and imaginary
parts, we obtain Maxwell equations
\begin{eqnarray}
\nabla\times\bE&=&-\frac{1}{c}\frac{\partial\bB}{\partial t},\nonumber\\
\nabla\times\bB&=&\frac{1}{c}\frac{\partial\bE}{\partial t},\nonumber\\
\nabla\cdot\bE&=&0,\nonumber\\
\nabla\cdot\bB&=&0.\nonumber
\end{eqnarray}
It is easy to verify that we come again to Maxwell equations starting from
the equations
\begin{eqnarray}
\left(\frac{i\hbar}{c}\frac{\partial}{\partial t}+
i\hbar\alpha_i\frac{\partial}{\partial\bx}\right)\dot{\phi}(x)&=&0,\label{Tr2}\\
-i\hbar\nabla\cdot\dot{\phi}(x)&=&0.\label{Tr2'}
\end{eqnarray}
In spite of the fact that equations (\ref{ME1}) and (\ref{ME2}) lead
to the same Maxwell equations, the physical interpretation of these equations
is different (see \cite{Bir96,Ger98}). Namely, 
the equations (\ref{ME1}) and (\ref{ME2})
are equations with negative and positive 
helicity, 
respectively.

As usual, the conjugated wavefunction 
$\overline{\phi}(x)=\overset{+}{\phi}(x)\Gamma_0=(\phi(x),\dot{\phi}(x))$
corresponds to antiparticle (it is a direct consequence of the
Dirac-like Lagrangian (\ref{LagMax}), $\phi(x)$ is a complex
function). Therefore, we come here to a very controversial conclusion that
the equations (\ref{Anti}) describe the antiparticle (antiphoton)
and, moreover, hence it follows that there exist the current and
charge for the photon field. At first glance, we come to a drastic
contradiction with the widely accepted fact that the photon is
truly neutral particle. However, it is easy to verify that
equations (\ref{Anti}) lead to the Maxwell equations also. It means that
the photon coincides with its ``antiparticle". Following to the standard
procedure given in many textbooks, we can define the ``charge" of the
photon by an expression
\begin{equation}\label{Charge}
Q\sim\int d\bx\overline{\phi}\Gamma_0\phi,
\end{equation}
where $\overline{\phi}\Gamma_0\phi=2(\bE^2+\bB^2)$. However, Newton and
Wigner \cite{NW49} showed that for the photon there exist no localized
states. Therefore, the integral in the right side of (\ref{Charge})
presents an indeterminable expression. Since the integral (\ref{Charge})
does not exist in general, then the ``charge" of the photon cannot be
considered as a constant magnitude (as it takes place for the electron
field which has localized states \cite{NW49} and a well-defined
constant charge). In a sense, one can say that the ``charge " of the
photon is equal to the energy $\bE^2+\bB^2$ of the $\gamma$-quantum.

We see that the equation (\ref{ME}) leads to the two Dirac-like
equations (\ref{ME1}) and (\ref{ME2}) which in combination with the
transversality conditions (\ref{Tr1'}) and (\ref{Tr2'}) are equivalent to
the Maxwell equations. Let us represent solutions of (\ref{ME1}) in 
a plane-wave form
\begin{equation}\label{PW}
\phi(x)=\varepsilon(\bk)\exp[i\hbar^{-1}(\bk\cdot\bx-\omega t)].
\end{equation}
After substitution of (\ref{PW}) into (\ref{ME1}) we come to the following
matrix eigenvalue problem
\[
-c\begin{pmatrix}
0 & ik_3 & -ik_2\\
-ik_3 & 0 & ik_1\\
ik_2 & -ik_1 & 0
\end{pmatrix}\begin{pmatrix}
\varepsilon_1\\
\varepsilon_2\\
\varepsilon_3
\end{pmatrix}=\omega\begin{pmatrix}
\varepsilon_1\\
\varepsilon_2\\
\varepsilon_3
\end{pmatrix}.
\]
It is easy to verify that we come to the same eigenvalue problem starting
from (\ref{ME2}). The secular equation has the solutions
$\omega=\pm c\bk,0$.

Therefore, solutions of (\ref{ME}) in the plane-wave 
approximation are expressed via the functions
\begin{eqnarray}
\phi_\pm(\bk;\bx,t)&=&\lf 2(2\pi)^3\rf^{-\frac{1}{2}}
\begin{pmatrix}
\varepsilon_\pm(\bk)\\
\varepsilon_\pm(\bk)
\end{pmatrix}
\exp[i(\bk\cdot\bx-\omega t)],\nonumber\\
\phi_0(\bk;\bx)&=&\lf 2(2\pi)^3\rf^{-\frac{1}{2}}
\begin{pmatrix}
\varepsilon_0(\bk)\\
\varepsilon_0(\bk)
\end{pmatrix}
\exp[i\bk\cdot\bx]\nonumber
\end{eqnarray}
and the complex conjugate functions $\dot{\phi}_+(\bk;\bx,t)$ and
$\dot{\phi}_0(\bk;\bx)$ ($\bE+i\bB$) corresponding to positive helicity,
here $\omega=c|\bk|$ and $\varepsilon_\lambda(\bk)$ ($\lambda=\pm,0$)
are the polarization vectors of a photon:
\begin{eqnarray}
\varepsilon_\pm(\bk)&=&\lf 2|\bk|^2(k^2_1+k^2_2)\rf^{-\frac{1}{2}}
\begin{bmatrix}
-k_1k_3\pm ik_2|\bk|\\
-k_2k_3\mp ik_1|\bk|\\
k^2_1+k^2_2
\end{bmatrix},\nonumber\\
\varepsilon_0(\bk)&=&|\bk|^{-1}
\begin{bmatrix}
k_1\\
k_2\\
k_3
\end{bmatrix}.\nonumber
\end{eqnarray}

Let us find now solutions of the $SL(2,\C)$-part equations
(\ref{FEL3})--(\ref{FEL4}).
Taking into account the structure of 
$\Upsilon_\nu$ given by (\ref{Upsilon1})--(\ref{Upsilon2}), we can rewrite
the equation (\ref{FEL3}) as follows
\begin{eqnarray}
\sum^3_{k=1}\Lambda_k\frac{\partial\phi}{\partial a_k}-
i\sum^3_{k=1}\Lambda_k\frac{\partial\phi}{\partial a^\ast_k}&=&0,\nonumber\\
\sum^3_{k=1}\Lambda^\ast_k\frac{\partial\dot{\phi}}{\partial\widetilde{a}_k}+
i\sum^3_{k=1}\Lambda^\ast_k\frac{\partial\dot{\phi}}
{\partial\widetilde{a}^\ast_k}&=&0.\nonumber
\end{eqnarray}
Or, using the explicit form of the matrices $\Lambda_j$ and $\Lambda^\ast_j$
given by (\ref{Lambda1})--(\ref{Lambda2}), we obtain
\begin{gather}
\frac{\sqrt{2}}{2}\frac{\partial\phi_2}{\partial a_1}-
i\frac{\sqrt{2}}{2}\frac{\partial\phi_2}{\partial a_2}+
\frac{\partial\phi_1}{\partial a_3}-
i\frac{\sqrt{2}}{2}\frac{\partial\phi_2}{\partial a^\ast_1}-
\frac{\sqrt{2}}{2}\frac{\partial\phi_2}{\partial a^\ast_2}-
i\frac{\partial\phi_1}{\partial a^\ast_3}=0,\nonumber\\
\frac{\partial\phi_1}{\partial a_1}+
\frac{\partial\phi_3}{\partial a_1}+
i\frac{\partial\phi_1}{\partial a_2}-
i\frac{\partial\phi_3}{\partial a_2}-
i\frac{\partial\phi_1}{\partial a^\ast_1}-
i\frac{\partial\phi_3}{\partial a^\ast_1}-
\frac{\partial\phi_1}{\partial a^\ast_2}+
\frac{\partial\phi_3}{\partial a^\ast_2}=0,\nonumber\\
\frac{\sqrt{2}}{2}\frac{\partial\phi_2}{\partial a_1}+
i\frac{\sqrt{2}}{2}\frac{\partial\phi_2}{\partial a_2}-
\frac{\partial\phi_3}{\partial a_3}-
i\frac{\sqrt{2}}{2}\frac{\partial\phi_2}{\partial a^\ast_1}+
\frac{\sqrt{2}}{2}\frac{\partial\phi_2}{\partial a^\ast_2}+
i\frac{\partial\phi_3}{\partial a^\ast_3}=0,\nonumber\\
\frac{\sqrt{2}}{2}\frac{\partial\dot{\phi}_2}{\partial\widetilde{a}_1}-
i\frac{\sqrt{2}}{2}\frac{\partial\dot{\phi}_2}{\partial\widetilde{a}_2}+
\frac{\partial\dot{\phi}_1}{\partial\widetilde{a}_3}+
i\frac{\sqrt{2}}{2}\frac{\partial\dot{\phi}_2}{\partial\widetilde{a}^\ast_1}+
\frac{\sqrt{2}}{2}\frac{\partial\dot{\phi}_2}{\partial\widetilde{a}^\ast_2}+
i\frac{\partial\dot{\phi}_1}{\partial\widetilde{a}^\ast_3}=0,\nonumber\\
\frac{\partial\dot{\phi}_1}{\partial\widetilde{a}_1}+
\frac{\partial\dot{\phi}_3}{\partial\widetilde{a}_1}+
i\frac{\partial\dot{\phi}_1}{\partial\widetilde{a}_2}-
i\frac{\partial\dot{\phi}_3}{\partial\widetilde{a}_2}+
i\frac{\partial\dot{\phi}_1}{\partial\widetilde{a}^\ast_1}+
i\frac{\partial\dot{\phi}_3}{\partial\widetilde{a}^\ast_1}-
\frac{\partial\dot{\phi}_1}{\partial\widetilde{a}^\ast_2}+
\frac{\partial\dot{\phi}_3}{\partial\widetilde{a}^\ast_2}=0,\nonumber\\
\frac{\sqrt{2}}{2}\frac{\partial\dot{\phi}_2}{\partial\widetilde{a}_1}+
i\frac{\sqrt{2}}{2}\frac{\partial\dot{\phi}_2}{\partial\widetilde{a}_2}-
\frac{\partial\dot{\phi}_3}{\partial\widetilde{a}_3}+
i\frac{\sqrt{2}}{2}\frac{\partial\dot{\phi}_2}{\partial\widetilde{a}^\ast_1}-
\frac{\sqrt{2}}{2}\frac{\partial\dot{\phi}_2}{\partial\widetilde{a}^\ast_2}-
i\frac{\partial\dot{\phi}_3}{\partial\widetilde{a}^\ast_3}=0,\label{CompM}
\end{gather}
\begin{sloppypar}
Coming to the helicity basis, we will find solutions of the equations 
(\ref{CompM}), that is, we will present components of the Majorana--Oppenheimer
`bispinor'\index{bispinor!Majorana-Oppenheimer}
$\boldsymbol{\phi}=(\phi_1,\phi_2,\phi_3,\dot{\phi}_1,\dot{\phi}_2,
\dot{\phi}_3)^T$ in terms of the functions on the two-dimensional complex
sphere
(it is obvious that the indices $k$ and $\dot{k}$ can be omitted here):
\end{sloppypar}
\begin{eqnarray}
\phi_1&=&\phi_{1,1;1,1}=\boldsymbol{f}^l_{1,1}(r)
\fM^l_{1,n}(\varphi,\epsilon,\theta,\tau,0,0),\nonumber\\
\phi_2&=&\phi_{1,0;1,0}=\boldsymbol{f}^l_{1,0}(r)
\fM^l_{0,n}(\varphi,\epsilon,\theta,\tau,0,0),\nonumber\\
\phi_3&=&\phi_{1,-1;1,-1}=\boldsymbol{f}^l_{1,-1}(r)
\fM^l_{-1,n}(\varphi,\epsilon,\theta,\tau,0,0),\nonumber\\
\dot{\phi}_1&=&\dot{\phi}_{1,1;1,1}=
\boldsymbol{f}^{\dot{l}}_{1,1}(r^\ast)
\fM^{\dot{l}}_{1,\dot{n}}(\varphi,\epsilon,\theta,\tau,0,0),\nonumber\\
\dot{\phi}_2&=&\dot{\phi}_{1,0;1,0}=
\boldsymbol{f}^{\dot{l}}_{1,0}(r^\ast)
\fM^{\dot{l}}_{0,\dot{n}}(\varphi,\epsilon,\theta,\tau,0,0),\nonumber\\
\dot{\phi}_3&=&\dot{\phi}_{1,-1;1,-1}=
\boldsymbol{f}^{\dot{l}}_{1,-1}(r^\ast)
\fM^{\dot{l}}_{-1,\dot{n}}(\varphi,\epsilon,\theta,\tau,0,0),\nonumber
\end{eqnarray}
Substituting these functions into (\ref{CompM}) and separating the
variables with the aid of recurrence relations between hyperspherical
functions, we come to the following system of ordinary differential
equations (the system (\ref{RFS}) at $l=1$):
\begin{gather}
2\frac{d\boldsymbol{f}^l_{1,1}(r)}{dr}-
\frac{1}{r}\boldsymbol{f}^l_{1,1}(r)-
\frac{\sqrt{2l(l+1)}}{r}\boldsymbol{f}^l_{1,0}(r)=0,\nonumber\\
-\frac{\sqrt{2l(l+1)}}{r}\boldsymbol{f}^l_{1,-1}(r)+
\frac{\sqrt{2l(l+1)}}{r}\boldsymbol{f}^l_{1,1}(r)=0,\nonumber\\
-2\frac{d\boldsymbol{f}^l_{1,-1}(r)}{dr}+\frac{1}{r}\boldsymbol{f}^l_{1,-1}(r)+
\frac{\sqrt{2l(l+1)}}{r}\boldsymbol{f}^l_{1,0}(r)=0,\nonumber\\
2\frac{d\boldsymbol{f}^{\dot{l}}_{1,1}(r^\ast)}{dr^\ast}-\frac{1}{r^\ast}
\boldsymbol{f}^{\dot{l}}_{1,1}(r^\ast)-\frac{\sqrt{2\dot{l}(\dot{l}+1)}}{r^\ast}
\boldsymbol{f}^{\dot{l}}_{1,0}(r^\ast)=0,\nonumber\\
-\frac{i\sqrt{2\dot{l}(\dot{l}+1)}}{r^\ast}
\boldsymbol{f}^{\dot{l}}_{1,-1}(r^\ast)+\frac{\sqrt{2\dot{l}(\dot{l}+1)}}{r^\ast}
\boldsymbol{f}^{\dot{l}}_{1,1}(r^\ast)=0,\nonumber\\
-2\frac{d\boldsymbol{f}^{\dot{l}}_{1,-1}(r^\ast)}{dr^\ast}+\frac{1}{r^\ast}
\boldsymbol{f}^{\dot{l}}_{1,-1}(r^\ast)+\frac{\sqrt{2\dot{l}(\dot{l}+1)}}{r^\ast}
\boldsymbol{f}^{\dot{l}}_{1,0}(r^\ast)=0,\label{RFM}
\end{gather}
From the second and fifth equations it follows that
$\boldsymbol{f}^l_{1,-1}(r)=\boldsymbol{f}^l_{1,1}(r)$
and $\boldsymbol{f}^{\dot{l}}_{1,-1}(r^\ast)=
\boldsymbol{f}^{\dot{l}}_{1,1}(r^\ast)$. 
Taking into account these relations we can rewrite
the system (\ref{RFM}) as follows
\begin{eqnarray}
2\frac{d\boldsymbol{f}^l_{1,1}(r)}{dr}\;\;\;-\;\;
\frac{1}{r}\boldsymbol{f}^l_{1,1}(r)\;\;-\;\;
\frac{\sqrt{2l(l+1)}}{r}\boldsymbol{f}^l_{1,0}(r)
&=&0,\nonumber\\
-2\frac{d\boldsymbol{f}^l_{1,-1}(r)}{dr}\;\;+\,\,
\frac{1}{r}\boldsymbol{f}^l_{1,-1}(r)\;+\;
\frac{\sqrt{2l(l+1)}}{r}\boldsymbol{f}^l_{1,0}(r)
&=&0,\nonumber\\
2\frac{d\boldsymbol{f}^{\dot{l}}_{1,1}(r^\ast)}{dr^\ast}\,-\;\frac{1}{r^\ast}
\boldsymbol{f}^{\dot{l}}_{1,1}(r^\ast)\;-\;\frac{\sqrt{2\dot{l}(\dot{l}+1)}}{r^\ast}
\boldsymbol{f}^{\dot{l}}_{1,0}(r^\ast)&=&0,\nonumber\\
-2\frac{d\boldsymbol{f}^{\dot{l}}_{1,-1}(r^\ast)}{dr^\ast}+\frac{1}{r^\ast}
\boldsymbol{f}^{\dot{l}}_{1,-1}(r^\ast)+\frac{\sqrt{2\dot{l}(\dot{l}+1)}}{r^\ast}
\boldsymbol{f}^{\dot{l}}_{1,0}(r^\ast)&=&0,\label{RFM2}
\end{eqnarray}
It is easy to see that the first equation is equivalent to the second,
and third equation is equivalent to the fourth. Thus, we come to the
following inhomogeneous differential equations of the first order:
\begin{eqnarray}
&&2r\frac{d\boldsymbol{f}^l_{1,1}(r)}{dr}-\boldsymbol{f}^l_{1,1}(r)-
\sqrt{2l(l+1)}\boldsymbol{f}^l_{1,0}(r)=0,\nonumber\\
&&2r^\ast\frac{d\boldsymbol{f}^{\dot{l}}_{1,1}(r^\ast)}{dr^\ast}-
\boldsymbol{f}^{\dot{l}}_{1,1}(r^\ast)-
\sqrt{2\dot{l}(\dot{l}+1)}\boldsymbol{f}^{\dot{l}}_{1,0}(r^\ast)=0,\nonumber
\end{eqnarray}
where the functions $\boldsymbol{f}^l_{1,0}(r)$ and
$\boldsymbol{f}^{\dot{l}}_{1,0}(r^\ast)$ are understood as inhomogeneous
parts. Solutions of these equations are expressed via the elementary functions:
\begin{eqnarray}
\boldsymbol{f}^l_{1,1}(r)&=&C\sqrt{r}+\sqrt{2l(l+1)}r,\nonumber\\
\boldsymbol{f}^{\dot{l}}_{1,1}(r^\ast)&=&\dot{C}\sqrt{r^\ast}+
\sqrt{2\dot{l}(\dot{l}+1)}r^\ast.\nonumber
\end{eqnarray}
Therefore, solutions of the radial part have the form
\begin{eqnarray}
&&\boldsymbol{f}^l_{1,1}(r)=\boldsymbol{f}^l_{1,-1}(r)=
C\sqrt{r}+\sqrt{2l(l+1)}r,\nonumber\\
&&\boldsymbol{f}^l_{1,0}(r)=\sqrt{2l(l+1)}r,\nonumber\\
&&\boldsymbol{f}^{\dot{l}}_{1,1}(r^\ast)=
\boldsymbol{f}^{\dot{l}}_{1,-1}(r^\ast)=
\dot{C}\sqrt{r^\ast}+\sqrt{2\dot{l}(\dot{l}+1)}r^\ast,\nonumber\\
&&\boldsymbol{f}^{\dot{l}}_{1,0}(r^\ast)=
\sqrt{2\dot{l}(\dot{l}+1)}r^\ast.
\nonumber
\end{eqnarray}
In such a way, solutions of the system
(\ref{CompM}) are defined by the following functions:
\begin{eqnarray}
\phi_1(r,\varphi^c,\theta^c)&=&\boldsymbol{f}^l_{1,1}(r)
\fM^l_{1,n}(\varphi,\epsilon,\theta,\tau,0,0),\nonumber\\
\phi_2(r,\varphi^c,\theta^c)&=&\boldsymbol{f}^l_{1,0}(r)
\fM^l_{0,n}(\varphi,\epsilon,\theta,\tau,0,0),\nonumber\\
\phi_3(r,\varphi^c,\theta^c)&=&\boldsymbol{f}^l_{1,-1}(r)
\fM^l_{-1,n}(\varphi,\epsilon,\theta,\tau,0,0),\nonumber\\
\dot{\phi}_1(r^\ast,\dot{\varphi}^c,\dot{\theta}^c)&=&
\boldsymbol{f}^{\dot{l}}_{1,1}(r^\ast)
\fM^{\dot{l}}_{1,\dot{n}}(\varphi,\epsilon,\theta,\tau,0,0),\nonumber\\
\dot{\phi}_2(r^\ast,\dot{\varphi}^c,\dot{\theta}^c)&=&
\boldsymbol{f}^{\dot{l}}_{1,0}(r^\ast)
\fM^{\dot{l}}_{0,\dot{n}}(\varphi,\epsilon,\theta,\tau,0,0),\nonumber\\
\dot{\phi}_3(r^\ast,\dot{\varphi}^c,\dot{\theta}^c)&=&
\boldsymbol{f}^{\dot{l}}_{1,-1}(r^\ast)
\fM^{\dot{l}}_{-1,\dot{n}}(\varphi,\epsilon,\theta,\tau,0,0),\nonumber
\end{eqnarray}
where
\begin{eqnarray}
&&l=1,\;2,\;3,\;\ldots;\quad
n=-l,\;-l+1,\;\ldots,\; l;\nonumber\\
&&\dot{l}=1,\;2,\;3,\;\ldots;\quad
\dot{n}=-\dot{l},\;-\dot{l}+1,\;\ldots,\; \dot{l},\nonumber
\end{eqnarray}
\[
\fM^l_{\pm 1,n}(\varphi,\epsilon,\theta,\tau,0,0)=
e^{\mp(\epsilon+i\varphi)}Z^l_{\pm 1,n}(\theta,\tau),
\]
\begin{multline}
Z^l_{\pm 1,n}(\theta,\tau)=\cos^{2l}\frac{\theta}{2}
\ch^{2l}\frac{\tau}{2}\sum^l_{k=-l}i^{\pm 1-k}
\tg^{\pm 1-k}\frac{\theta}{2}\tnh^{n-k}\frac{\tau}{2}\times\\
\hypergeom{2}{1}{\pm 1-l+1,1-l-k}{\pm 1-k+1}
{i^2\tg^2\frac{\theta}{2}}
\hypergeom{2}{1}{n-l+1,1-l-k}{n-k+1}{\tnh^2\frac{\tau}{2}},\nonumber
\end{multline}
\[
\fM^l_{0,n}(\varphi,\epsilon,\theta,\tau,0,0)=
Z^l_{0,n}(\theta,\tau),
\]
\begin{multline}
Z^l_{0,n}(\theta,\tau)=\cos^{2l}\frac{\theta}{2}
\ch^{2l}\frac{\tau}{2}\sum^l_{k=-l}i^{-k}
\tg^{-k}\frac{\theta}{2}\tnh^{n-k}\frac{\tau}{2}\times\\
\hypergeom{2}{1}{-l+1,1-l-k}{-k+1}
{i^2\tg^2\frac{\theta}{2}}
\hypergeom{2}{1}{n-l+1,1-l-k}{n-k+1}{\tnh^2\frac{\tau}{2}},\nonumber
\end{multline}
\[
\fM^{\dot{l}}_{\pm 1,\dot{n}}(\varphi,\epsilon,\theta,\tau,0,0)=
e^{\mp(\epsilon-i\varphi)}
Z^{\dot{l}}_{\pm 1,\dot{n}}(\theta,\tau),
\]
\begin{multline}
Z^{\dot{l}}_{\pm 1,\dot{n}}(\theta,\tau)=
\cos^{2\dot{l}}\frac{\theta}{2}
\ch^{2\dot{l}}\frac{\tau}{2}
\sum^{\dot{l}}_{\dot{k}=-\dot{l}}i^{\pm 1-\dot{k}}
\tg^{\pm 1-\dot{k}}\frac{\theta}{2}
\tnh^{\dot{n}-\dot{k}}\frac{\tau}{2}\times\\
\hypergeom{2}{1}{\pm 1-\dot{l}+1,1-\dot{l}-\dot{k}}
{\pm 1-\dot{k}+1}
{i^2\tg^2\frac{\theta}{2}}
\hypergeom{2}{1}{\dot{n}-\dot{l}+1,1-\dot{l}-\dot{k}}
{\dot{n}-\dot{k}+1}{\tnh^2\frac{\tau}{2}},\nonumber
\end{multline}
\[
\fM^{\dot{l}}_{0,\dot{n}}(\varphi,\epsilon,\theta,\tau,0,0)=
Z^{\dot{l}}_{0,\dot{n}}(\theta,\tau),
\]
\begin{multline}
Z^{\dot{l}}_{0,\dot{n}}(\theta,\tau)=
\cos^{2\dot{l}}\frac{\theta}{2}
\ch^{2\dot{l}}\frac{\tau}{2}
\sum^{\dot{l}}_{\dot{k}=-\dot{l}}i^{-\dot{k}}
\tg^{-\dot{k}}\frac{\theta}{2}
\tnh^{\dot{n}-\dot{k}}\frac{\tau}{2}\times\\
\hypergeom{2}{1}{-\dot{l}+1,1-\dot{l}-\dot{k}}
{-\dot{k}+1}
{i^2\tg^2\frac{\theta}{2}}
\hypergeom{2}{1}{\dot{n}-\dot{l}+1,1-\dot{l}-\dot{k}}
{\dot{n}-\dot{k}+1}{\tnh^2\frac{\tau}{2}}.\nonumber
\end{multline}

Therefore, in accordance with the factorization (\ref{WF}), an explicit
form of the particular solutions of the system (\ref{Real})--(\ref{FEL4})
are given by expressions
\begin{multline}
\phi^l_{1,n}(\balpha)=\phi_+(\bk;\bx,t)\phi^l_{1,n}(\fg)=\\
\lf 2(2\pi)^3\rf^{-\frac{1}{2}}
\begin{pmatrix}\varepsilon_+(\bk)\\
\varepsilon_+(\bk)\end{pmatrix}
\exp[i(\bk\cdot\bx-\omega t)]\boldsymbol{f}^l_{1,1}(r)
\fM_{1,n}^l(\varphi,\epsilon,\theta,\tau,0,0),\nonumber
\end{multline}
\[
\phi^l_{0,n}(\balpha)=\phi_0(\bk;\bx)\phi^l_{0,n}(\fg)=
\lf 2(2\pi)^3\rf^{-\frac{1}{2}}
\begin{pmatrix}\varepsilon_0(\bk)\\
\varepsilon_0(\bk)\end{pmatrix}
\exp[i\bk\cdot\bx]\boldsymbol{f}^l_{1,0}(r)
\fM^l_{0,n}(0,0,\theta,\tau,0,0),
\]
\begin{multline}
\phi^l_{-1,n}(\balpha)=\phi_-(\bk;\bx,t)\phi^l_{-1,n}(\fg)=\\
\lf 2(2\pi)^3\rf^{-\frac{1}{2}}
\begin{pmatrix}\varepsilon_-(\bk)\\
\varepsilon_-(\bk)\end{pmatrix}
\exp[i(\bk\cdot\bx-\omega t)]\boldsymbol{f}^l_{1,-1}(r)
\fM_{-1,n}^l(\varphi,\epsilon,\theta,\tau,0,0),\nonumber
\end{multline}
\begin{multline}
\dot{\phi}^{\dot{l}}_{1,\dot{n}}(\balpha)=\phi^\ast_+(\bk;\bx,t)
\dot{\phi}^{\dot{l}}_{1,\dot{n}}(\fg)=\\
\lf 2(2\pi)^3\rf^{-\frac{1}{2}}
\begin{pmatrix}\varepsilon^\ast_+(\bk)\\
\varepsilon^\ast_+(\bk)\end{pmatrix}
\exp[-i(\bk\cdot\bx-\omega t)]\boldsymbol{f}^{\dot{l}}_{1,1}(r^\ast)
\fM_{1,\dot{n}}^{\dot{l}}(\varphi,\epsilon,\theta,\tau,0,0),\nonumber
\end{multline}
\[
\dot{\phi}^{\dot{l}}_{0,\dot{n}}(\balpha)=\phi^\ast_0(\bk;\bx)
\dot{\phi}^{\dot{l}}_{0,\dot{n}}(\fg)=
\lf 2(2\pi)^3\rf^{-\frac{1}{2}}
\begin{pmatrix}\varepsilon^\ast_0(\bk)\\
\varepsilon^\ast_0(\bk)\end{pmatrix}
\exp[-i\bk\cdot\bx]\boldsymbol{f}^{\dot{l}}_{1,0}(r^\ast)
\fM^{\dot{l}}_{0,\dot{n}}(0,0,\theta,\tau,0,0),
\]
\begin{multline}
\dot{\phi}^{\dot{l}}_{-1,\dot{n}}(\balpha)=
\phi^\ast_-(\bk;\bx,t)\dot{\phi}^{\dot{l}}_{-1,\dot{n}}(\fg)=\\
\lf 2(2\pi)^3\rf^{-\frac{1}{2}}
\begin{pmatrix}\varepsilon^\ast_-(\bk)\\
\varepsilon^\ast_-(\bk)\end{pmatrix}
\exp[-i(\bk\cdot\bx-\omega t)]\boldsymbol{f}^{\dot{l}}_{1,-1}(r^\ast)
\fM_{-1,\dot{n}}^{\dot{l}}(\varphi,\epsilon,\theta,\tau,0,0),\label{SME}
\end{multline}
Before we proceed to define a general solution of the boundary value
problem, let us note the following circumstance.
The set (\ref{SME}) consists of the transverse solutions 
$\phi^l_{\pm 1,n}(\balpha)$
(negative helicity), 
$\dot{\phi}^{\dot{l}}_{\pm 1,\dot{n}}(\balpha)$ (positive helicity) and
the zero-eigenvalue (longitudinal) solutions $\phi^l_{0,n}(\balpha)$ and
$\dot{\phi}^{\dot{l}}_{0,\dot{n}}(\balpha)$. 
%The negative energy solutions 
%$\dot{\psi}_{\pm 1}(\balpha)$ should be omitted, since photons have no
%antiparticles.
The longitudinal solutions $\phi^l_{0,n}(\balpha)$ and
$\dot{\phi}^{\dot{l}}_{0,\dot{n}}(\balpha)$ 
do not contribute to a real photon due to their
transversality conditions (\ref{Tr1'}) and (\ref{Tr2'}). Thus, any real
photon should be described by only $\phi^l_{\pm 1,n}(\balpha)$ and
$\dot{\phi}^{\dot{l}}_{\pm 1,\dot{n}}(\balpha)$:
\begin{multline}
\phi^l_{\pm 1,n}(\balpha)=\phi_{\pm}(\bk;\bx,t)\phi^l_{\pm 1,n}(\fg)=\\
\lf 2(2\pi)^3\rf^{-\frac{1}{2}}
\begin{pmatrix}\varepsilon_\pm(\bk)\\
\varepsilon_\pm(\bk)\end{pmatrix}
\exp[i(\bk\cdot\bx-\omega t)]\boldsymbol{f}^l_{1,\pm 1}(r)
\fM_{\pm 1,n}^l(\varphi,\epsilon,\theta,\tau,0,0),\nonumber
\end{multline}
\begin{multline}
\dot{\phi}^{\dot{l}}_{\pm 1,\dot{n}}(\balpha)=
\phi^\ast_\pm(\bk;\bx,t)\dot{\phi}^{\dot{l}}_{\pm 1,\dot{n}}(\fg)=\\
\lf 2(2\pi)^3\rf^{-\frac{1}{2}}
\begin{pmatrix}\varepsilon^\ast_\pm(\bk)\\
\varepsilon^\ast_\pm(\bk)\end{pmatrix}
\exp[-i(\bk\cdot\bx-\omega t)]\boldsymbol{f}^{\dot{l}}_{1,\pm 1}(r^\ast)
\fM_{\pm 1,\dot{n}}^{\dot{l}}(\varphi,\epsilon,\theta,\tau,0,0).\nonumber
\end{multline}
Taking into account only these physically meaningful solutions, we can define
a general solution of the boundary value problem by means of the
following expansions
\begin{eqnarray}
\boldsymbol{\phi}_{\pm 1}(\balpha)&=&
\lf 2(2\pi)^3\rf^{-\frac{1}{2}}\sum_k
\begin{pmatrix}\varepsilon_\pm(\bk)\\
\varepsilon_\pm(\bk)\end{pmatrix}
e^{ikx}\sum^\infty_{l=1}\boldsymbol{f}^l_{1,\pm 1}(r)\sum^l_{n=-l}
\alpha^{\pm 1}_{l,n}\fM_{\pm 1,n}^l(\varphi,\epsilon,\theta,\tau,0,0),
\label{FSM85}\\
\dot{\boldsymbol{\phi}}_{\pm 1}(\balpha)&=&
\lf 2(2\pi)^3\rf^{-\frac{1}{2}}\sum_k
\begin{pmatrix}\varepsilon^\ast_\pm(\bk)\\
\varepsilon^\ast_\pm(\bk)\end{pmatrix}
e^{-ikx}\sum^\infty_{\dot{l}=1}\boldsymbol{f}^{\dot{l}}_{1,\pm 1}(r^\ast)
\sum^{\dot{l}}_{\dot{n}=-\dot{l}}\alpha^{\pm 1}_{\dot{l},\dot{n}}
\fM_{\pm 1,\dot{n}}^{\dot{l}}(\varphi,\epsilon,\theta,\tau,0,0),
\label{FSM86}
\end{eqnarray}
where
\begin{eqnarray}
\alpha^{\pm 1}_{l,n}&=&\frac{(-1)^n(2l+1)(2\dot{l}+1)}{32\pi^4
\boldsymbol{f}^l_{1,\pm 1}(a)}
\int\limits_{\dS^2}\int\limits_{T_4}F_{\pm 1}(\balpha)e^{-ikx}
\fM^l_{\pm 1,n}(\varphi,\epsilon,\theta,\tau,0,0)d^4xd^4\fg,\nonumber\\
\alpha^{\pm 1}_{\dot{l},\dot{n}}&=&
\frac{(-1)^{\dot{n}}(2l+1)(2\dot{l}+1)}{32\pi^4
\boldsymbol{f}^{\dot{l}}_{1,\pm 1}(a^\ast)}
\int\limits_{\dS^2}\int\limits_{T_4}
\dot{F}_{\pm 1}(\balpha)e^{ikx}
\fM^{\dot{l}}_{\pm 1,\dot{n}}
(\varphi,\epsilon,\theta,\tau,0,0)d^4xd^4\fg.\nonumber
\end{eqnarray}

\subsection{Quantization}
In case of the photon field we define the field operator by a following
superposition
\begin{equation}\label{FOP}
\boldsymbol{\phi}(\balpha)=\sum^2_{s=1}\sfc_s\phi_s(\balpha)+
\sum^2_{s=1}\sfc^+_s\dot{\phi}_s(\balpha),
\end{equation}
where $\sfc^+_s$ and $\sfc_s$ are creation (emission) and annihilation
(absorption) operators of the photon in a state $s$,
$\phi_s(\balpha)$ and $\dot{\phi}_s(\balpha)$ are Fourier series
(\ref{FSM85}) (negative helicity) and (\ref{FSM86}) (positive helicity),
here we designate $\phi_1(\balpha)=\boldsymbol{\phi}_1(\balpha)$,
$\phi_2(\balpha)=\boldsymbol{\phi}_{-1}(\balpha)$,
$\dot{\phi}_1(\balpha)=\dot{\boldsymbol{\phi}}_1(\balpha)$,
$\dot{\phi}_2(\balpha)=\dot{\boldsymbol{\phi}}_{-1}(\balpha)$.
The operators $\sfc^+_s$ and $\sfc_s$ satisfy the following commutation
relations:
\begin{gather}
\ld\sfc_s,\sfc^+_{s^\prime}\rd_-=\delta_{ss^\prime},\nonumber\\
\ld\sfc_s,\sfc_{s^\prime}\rd_-=\ld\sfc^+_s,\sfc^+_{s^\prime}\rd_-=0,\quad
s,s^\prime=1,2.\nonumber
\end{gather}
Taking into account the latter relations, let us calculate a commutator
of the photon field:
\begin{equation}\label{ComPh}
\ld\boldsymbol{\phi}_\alpha(\balpha),
\boldsymbol{\phi}_\beta(\balpha^\prime)\rd_-=
D_{\alpha\beta}(\balpha,\balpha^\prime),
\end{equation}
where
\begin{eqnarray}
D_{\alpha\beta}(\balpha,\balpha^\prime)&=&
D^+_{\alpha\beta}(\balpha,\balpha^\prime)+
D^-_{\alpha\beta}(\balpha,\balpha^\prime),\nonumber\\
D^+_{\alpha\beta}(\balpha,\balpha^\prime)&=&
\sum^2_{s=1}\phi_{s\alpha}(\balpha)\dot{\phi}_{s\beta}(\balpha^\prime),
\nonumber\\
D^-_{\alpha\beta}(\balpha,\balpha^\prime)&=&
\sum^2_{s=1}\dot{\phi}_{s\alpha}(\balpha)\phi_{s\beta}(\balpha^\prime).
\nonumber
\end{eqnarray}
Or, using the formulae (\ref{FSM85}) and (\ref{FSM86}), we find an explicit
form of the functions $D^+_{\alpha\beta}(\balpha-\balpha^\prime)$ and
$D^-_{\alpha\beta}(\balpha-\balpha^\prime)$:
\begin{multline}
D^+_{\alpha\beta}(\balpha-\balpha^\prime)=\lf 2(2\pi)^3\rf^{-1}
\sum^2_{s=1}\sum_k\boldsymbol{\varepsilon}_{\alpha s}(\bk)
\boldsymbol{\varepsilon}^\ast_{\beta s}(\bk)e^{ik(x-x^\prime)}\times\\
e^{-i(-1)^{s-1}(\varphi^c-\varphi^c{}^\prime)}
\sum^\infty_{l=1}
%\left[\sum^l_{k=1}
\boldsymbol{f}^l_{1,(-1)^{s-1}}(r)
\boldsymbol{f}^{l}_{1,(-1)^{s-1}}(r^\ast)
%\sum^k_{n=-k}\alpha^{(-1)^{s-1}}_{l,n}
%\fM^l_{(-1)^{s-1},n}(\varphi,\epsilon,\theta,\tau,0,0)
%\sum^{l-k+1}_{\dot{n}=-(l-k+1)}
%\alpha^{(-1)^{s-1}}_{l,\dot{n}}\fM^l_{(-1)^{s-1},\dot{n}}
%(\varphi^\prime,\epsilon^\prime,\theta^\prime,\tau^\prime,0,0),\nonumber
Z^l_{(-1)^{s-1},(-1)^{s-1}}(\cos\theta^c{}^{\prime\prime}),\nonumber
\end{multline}
\begin{multline}
D^-_{\alpha\beta}(\balpha-\balpha^\prime)=\lf 2(2\pi)^3\rf^{-1}
\sum^2_{s=1}\sum_k\boldsymbol{\varepsilon}^\ast_{\alpha s}(\bk)
\boldsymbol{\varepsilon}_{\beta s}(\bk)e^{-ik(x-x^\prime)}\times\\
e^{i(-1)^{s-1}(\varphi^c-\varphi^c{}^\prime)}
\sum^\infty_{l=1}
\boldsymbol{f}^l_{1,(-1)^{s-1}}(r^\ast)
\boldsymbol{f}^{l}_{1,(-1)^{s-1}}(r)
%\sum^k_{\dot{n}=-k}\alpha^{(-1)^{s-1}}_{l,\dot{n}}
%\fM^l_{(-1)^{s-1},\dot{n}}(\varphi,\epsilon,\theta,\tau,0,0)
%\sum^{l-k+1}_{n=-(l-k+1)}
%\alpha^{(-1)^{s-1}}_{l,n}\fM^l_{(-1)^{s-1},n}
%(\varphi^\prime,\epsilon^\prime,\theta^\prime,\tau^\prime,0,0),\nonumber
Z^l_{(-1)^{s-1},(-1)^{s-1}}(\cos\theta^c{}^{\prime\prime}).\nonumber
\end{multline}

Let us define now normal and time ordered products of the field
operators (\ref{FOP}). Supposing
\[
\boldsymbol{\phi}(\balpha)=\phi^{(+)}(\balpha)+\phi^{(-)}(\balpha),
\]
where $\phi^{(+)}(\balpha)$ and $\phi^{(-)}(\balpha)$ are annihilation
and creation operators of the photons, we see that
\begin{eqnarray}
N(\phi^{(+)}(\balpha)\phi^{(-)}(\balpha^\prime))&=&
\phi^{(-)}(\balpha^\prime)\phi^{(+)}(\balpha),\nonumber\\
N(\phi^{(-)}(\balpha)\phi^{(+)}(\balpha^\prime))&=&
\phi^{(-)}(\balpha)\phi^{(+)}(\balpha^\prime),\nonumber\\
N(\phi^{(-)}(\balpha)\phi^{(-)}(\balpha^\prime))&=&
\phi^{(-)}(\balpha)\phi^{(-)}(\balpha^\prime),\nonumber\\
N(\phi^{(+)}(\balpha)\phi^{(+)}(\balpha^\prime))&=&
\phi^{(+)}(\balpha)\phi^{(+)}(\balpha^\prime).\nonumber
\end{eqnarray}
Taking into account the latter $N$-products, we have
\begin{multline}
\boldsymbol{\phi}(\balpha)\boldsymbol{\phi}(\balpha^\prime)=
(\phi^{(+)}(\balpha)+\phi^{(-)}(\balpha))(\phi^{(+)}(\balpha^\prime)+
\phi^{(-)}(\balpha^\prime))=\\
=\phi^{(+)}(\balpha)\phi^{(+)}(\balpha^\prime)+
\phi^{(-)}(\balpha)\phi^{(+)}(\balpha^\prime)+
\phi^{(+)}(\balpha)\phi^{(-)}(\balpha^\prime)+\\
+\phi^{(-)}(\balpha)\phi^{(-)}(\balpha^\prime).\nonumber
\end{multline}
In accordance with (\ref{ComPh}) 
$\phi^{(+)}(\balpha)\phi^{(-)}(\balpha^\prime)=
\phi^{(-)}(\balpha^\prime)\phi^{(+)}(\balpha)+D^-(\balpha-\balpha^\prime)$,
therefore,
\begin{eqnarray}
\boldsymbol{\phi}(\balpha)\boldsymbol{\phi}(\balpha^\prime)&=&
N(\boldsymbol{\phi}(\balpha)\boldsymbol{\phi}(\balpha^\prime))+
D^-(\balpha-\balpha^\prime)\nonumber\\
&=&N(\boldsymbol{\phi}(\balpha)\boldsymbol{\phi}(\balpha^\prime))+
\coupeleven,\nonumber
\end{eqnarray}
where
\[
\coupeleven=\langle 0\mid\boldsymbol{\phi}(\balpha)
\boldsymbol{\phi}(\balpha^\prime)\mid 0\rangle=D^-(\balpha-\balpha^\prime)
\]
is an operator coupling of the photon fields.

Further, the time ordered coupling of the field operators (\ref{FOP})
is defined as
\begin{multline}
\couptwelve=\langle 0\mid T(\boldsymbol{\phi}(\balpha)
\boldsymbol{\phi}(\balpha^\prime))\mid 0\rangle=\\
=\begin{cases}
\langle 0\mid\boldsymbol{\phi}(\balpha)\boldsymbol{\phi}(\balpha^\prime)
\mid 0\rangle=\coupeleven=D^-(\balpha-\balpha^\prime), & t>t^\prime;\\
\langle 0\mid\boldsymbol{\phi}(\balpha^\prime)\boldsymbol{\phi}(\balpha)
\mid 0\rangle=\coupthirteen=D^+(\balpha-\balpha^\prime), & t^\prime> t.
\end{cases}
\nonumber
\end{multline}
\section{Interacting fields}
\label{Sec:Int}
Up to now we analyze free Dirac and Maxwell fields. Let us consider
an interaction between these fields. As usual, interactions between the
fields are described by an interaction Lagrangian $\cL_I$. In our case
we take the following Lagrangian
\begin{equation}\label{LagInt}
\cL_I(\balpha)=\mu(\overline{\boldsymbol{\psi}}(\balpha)
\sigma^D_{(\mu\nu)^k}\boldsymbol{\psi}(\balpha))
(\Xi^M_{(\rho)k}\boldsymbol{\phi}(\balpha)),
\end{equation}
where $\sigma^D_{\mu\nu}=\frac{1}{2}(\Xi^D_\mu\Xi^D_\nu-\Xi^D_\nu\Xi^D_\mu)$
and $\Xi^D=(\Gamma^D_0,\Gamma^D_1,\Gamma^D_2,\Gamma^D_3,\Upsilon^D_1,
\Upsilon^D_2,\Upsilon^D_3,\Upsilon^D_4,\Upsilon^D_5,\Upsilon^D_6)$,
$\Xi^M=(\Gamma^M_0,\Gamma^M_1,\Gamma^M_2,\Gamma^M_3,\Upsilon^M_1,\Upsilon^M_2,
\Upsilon^M_3,\Upsilon^M_4,\Upsilon^M_5,\Upsilon^M_6)$, here $\Gamma^D$ and
$\Upsilon^D$ are the matrices (\ref{Gamma1}) and 
(\ref{Upsilon1})--(\ref{Upsilon2}), and $\Gamma^M$ and $\Upsilon^M$ are
the matrices (\ref{Gamma2}) and (\ref{Upsilon3})--(\ref{Upsilon4}).

The full Lagrangian of interacting Dirac and Maxwell fields equals to a sum
of the free field Lagrangians and the interaction Lagrangian:
\[
\cL(\balpha)=\cL_D(\balpha)+\cL_M(\balpha)+\cL_I(\balpha),
\]
where $\cL_D(\balpha)$ and $\cL_M(\balpha)$ are of the type (\ref{LagDir})
and (\ref{LagMax}), respectively. Or,
\begin{multline}
\cL(\balpha)=-\frac{1}{2}\left(\overline{\boldsymbol{\psi}}(\balpha)
\Xi^D_\mu\frac{\partial\boldsymbol{\psi}(\balpha)}{\partial\balpha_\mu}
-\frac{\partial\overline{\boldsymbol{\psi}}(\balpha)}
{\partial\balpha_\mu}\Xi^D_\mu\boldsymbol{\psi}(\balpha)\right)-\\
-\frac{1}{2}\left(\overline{\boldsymbol{\phi}}(\balpha)
\Xi^M_\mu\frac{\partial\boldsymbol{\phi}(\balpha)}{\partial\balpha_\mu}
-\frac{\partial\overline{\boldsymbol{\phi}}(\balpha)}
{\partial\balpha_\mu}\Xi^M_\mu\boldsymbol{\phi}(\balpha)\right)-\\
-\kappa\overline{\boldsymbol{\psi}}(\balpha)\boldsymbol{\psi}(\balpha)+
\mu(\overline{\boldsymbol{\psi}}(\balpha)
\sigma^D_{(\mu\nu)^k}\boldsymbol{\psi}(\balpha))
(\Xi^M_{(\rho)k}\boldsymbol{\phi}(\balpha)).\nonumber
\end{multline}
Since the Lagrangian (\ref{LagInt}) does not contain derivatives on the
field functions, then for a Hamiltonian density we have
$\cH_I(\balpha)=-\cL_I(\balpha)$.

As is known, in the standard quantum field theory the $S$-matrix is expressed
via the Dyson formula \cite{Sch61}
\begin{equation}\label{Dyson}
S=T\left[\exp\left(-\frac{i}{\hbar c}\int\limits^{+\infty}_{-\infty}
\cH_I(x)d^4x\right)\right],
\end{equation}
where $T$ is the time ordering operator. The Hamiltonian density
$\cH_I(x)$ has in general been assumed to have the form of an invariant
local products of fields.

In our case, the electron-positron and photon fields are defined on
the space $\cM_8=\R^{1,3}\times\dS^2$ which larger then the Minkowski
space $\R^{1,3}$. With a view to define a formula similar to the equation
(\ref{Dyson}) it is necessarily to replace $d^4x$ by the following
invariant measure on $\cM_8$:
\[
d^8\mu=d^4xd^4\fg,
\]
where
\[
d^4\fg=\sin\theta^c\sin\dot{\theta}^cd\theta d\tau d\varphi d\epsilon.
\]
Therefore, an analog of the Dyson formula (\ref{Dyson}) on the manifold
$\cM_8$ can be written as follows
\[
S=T\left[\exp\left(-\frac{i}{\hbar c}
\int\limits_{T_4}\int\limits_{\dS^2}
\cH_I(\balpha)d^4xd^4\fg\right)\right].
\]
Concrete calculations of the scattering amplitudes come beyond the
framework of the present paper and will be considered in a future work.

\section{Conclusion}
In this paper we have presented a general scheme of construction of
quantum electrodynamics on the Poincar\'{e} group $\cP$ (or, equally,
on the homogeneous spaces of $\cP$). Except the maximal homogeneous
space $\cM_{10}$, we consider here only three spaces $\cM_8$, $\cM_7$ and
$\cM_6$ (it is easy to see that $\cM_8$, $\cM_7$ and $\cM_6$ correspond
to the Wigner's little groups $SL(2,\C)$, $SU(1,1)$ and $SU(2)$,
respectively). However, the similar constructions are possible for all
other homogeneous spaces of $\cP$ contained in the
Finkelstein-Bacry-Kihlberg list \cite{Fin55,BK69} and endowed with
a measure.

It should be noted that discrete symmetries remain outside of our
consideration.
It is well-known that discrete transformations are of fundamental
importance for constructing relativistic wave equations
and for their analysis. An inclusion of discrete symmetries into the
framework of quantum field theory on the Poincar\'{e} group can be
obtained via an automorphism representation. It is known that
Gel'fand, Minlos and
Shapiro \cite{GMS} proposed to consider the discrete transformations as
outer involutory automorphisms of the Lorentz group (there are also other
realizations of the discrete symmetries via the outer automorphisms, see
\cite{Mic64,Kuo71,Sil92}). At present, the Gel'fand-Minlos-Shapiro ideas
have been found further development in the works of Buchbinder, Gitman and
Shelepin \cite{BGS00,GS01}, where the discrete symmetries are represented
by both outer and inner automorphisms of the Poincar\'{e} group.
It is pointed out by Shirokov \cite{Shi58,Shi60} that an universal
covering of the inhomogeneous Lorentz group has eight inequivalent
realizations. Later on, in the eighties this idea was applied to a general
orthogonal group $O(p,q)$ by D\c{a}browski \cite{Dab88}. It is well-known
that universal coverings of the groups $O(p,q)$ are completely formulated
within spinor groups \cite{Lou97}. In turn, the spinor group is an
intrinsic notion of the Clifford algebra \cite{Che55,Lou97}. By this reason
there exists a complete and consistent description of the discrete
transformations in terms of the Clifford algebra theory.
Such a description has been given in the works \cite{Var99,Var00,Var041},
where the discrete symmetries are represented by fundamental
automorphisms of the Clifford algebras. The fundamental automorphisms are
compared to elements of the finite group formed by the discrete
transformations. In like manner, charge conjugation is naturally included
into a general scheme by means of a complex conjugation
pseudoautomorphism \cite{Ras55,Ras57,Var041}. It allows us to define
64 universal coverings of $O(p,q)$ ($CPT$-structures) which include as
a particular case the eight $PT$-structures of Shirokov and
D\c{a}browski \cite{Var041}. Moreover, it allows us to incorporate this
algebraic description with the ideas presented in \cite{BGS00} and then
to apply it for analysis of relativistic wave equations on the
homogeneous spaces of the Poincar\'{e} group.

\section*{Acknowledgement}
I am deeply grateful to Prof. D. M. Gitman for many helpful conversations.

\end{document}